\documentclass{jfm}

\usepackage{graphicx}
\usepackage{natbib}

\ifCUPmtlplainloaded \else
  \checkfont{eurm10}
  \iffontfound
    \IfFileExists{upmath.sty}
      {\typeout{^^JFound AMS Euler Roman fonts on the system,
                   using the 'upmath' package.^^J}%
       \usepackage{upmath}}
       
      {\typeout{^^JFound AMS Euler Roman fonts on the system, but you
                   dont seem to have the}%
       \typeout{'upmath' package installed. JFM.cls can take advantage
                 of these fonts,^^Jif you use 'upmath' package.^^J}%
      }
  \else
  \fi
\fi

\ifCUPmtlplainloaded \else
  \checkfont{msam10}
  \iffontfound
    \IfFileExists{amssymb.sty}
      {\typeout{^^JFound AMS Symbol fonts on the system, using the
                'amssymb' package.^^J}%
       \usepackage{amssymb}%
         \let\leq=\leqslant
         \let\geq=\geqslant
      }{}
  \fi
\fi

\ifCUPmtlplainloaded \else
  \IfFileExists{amsbsy.sty}
    {\typeout{^^JFound the 'amsbsy' package on the system, using it.^^J}%
     \usepackage{amsbsy}}
    {\providecommand\boldsymbol[1]{\mbox{\boldmath $##1$}}}
\fi

\providecommand\bnabla{\boldsymbol{\nabla}}
\providecommand\bcdot{\boldsymbol{\cdot}}

\newcommand\Real{\mbox{Re}} 
\newcommand\Rey{\mbox{\textit{Re}}}  

\newsavebox{\astrutbox}
\sbox{\astrutbox}{\rule[-5pt]{0pt}{20pt}}

\newcommand\eg{e.g.\ }

\usepackage{amsmath}
\usepackage{color}
\usepackage{overpic} 
\usepackage{psfrag}

\definecolor{lightgray}{gray}{0.8}
\definecolor{darkgray}{gray}{0.5}
\definecolor{myyellow}{rgb}{1,0.5,0} 
\definecolor{mygreen}{rgb}{0,0.5,0}
\definecolor{mypurple}{rgb}{.75,0,.75}
\definecolor{myturquoise}{rgb}{0,.6,.6} 
\definecolor{mybrown}{rgb}{.75,.47,0}

\def\aa{\mathbf{a}}
\def\bb{\mathbf{b}}
\def\bdelta{\boldsymbol{\delta}}
\def\ff{\mathbf{f}}
\def\ex{\mathbf{e}_x}
\def\ey{\mathbf{e}_y}
\def\nn{\mathbf{n}}
\def\uu{\mathbf{u}}
\def\xx{\mathbf{x}}
\def\CC{\mathbf{C}}

\def\00{\mathbf{0}}

\def\UU{\mathbf{U}}

\def\uu{\mathbf{u}}			

\def\pa{p^\dag}
\def\uua{\mathbf{u}^\dag}

\def\Pa{P^\dag} 

\def\UUa{\mathbf{U}^\dag}

\def\nnu{\Rey^{-1}}

\newcommand\be{\begin{equation}}
\newcommand\ee{\end{equation}}

\DeclareMathOperator{\e}{e}
\DeclareMathOperator{\Tr}{Tr}

\title[Sensitivity of noise amplification]{Sensitivity and open-loop control of stochastic response in a 
noise amplifier flow:
 the backward-facing step}

\author[E. Boujo and F.  Gallaire]%
{E.\ns B\ls O\ls U\ls J\ls O$^1$%
  \thanks{Email address for correspondence: edouard.boujo@epfl.ch}
\and  F.\ns G\ls A\ls L\ls L\ls A\ls I\ls R\ls E$^1$}

\affiliation{$^1$LFMI, 
\'Ecole Polytechnique F\'ed\'erale de Lausanne,
CH-1015 Lausanne, Switzerland}

\pubyear{}
\volume{}
\pagerange{}
\date{?; revised ?; accepted ?. - To be entered by editorial office}
\begin{document}

\maketitle

\begin{abstract}
The two-dimensional backward-facing step flow is a canonical example of noise amplifier flow: global linear stability analysis predicts that it is stable, but perturbations can undergo large amplification in space and time as a result of non-normal effects. This amplification potential is best captured by optimal transient growth analysis, optimal harmonic forcing, or the response to sustained noise.
In view of reducing disturbance amplification in these globally stable open flows,
a variational technique is proposed to evaluate the sensitivity of stochastic amplification to steady control.
Existing sensitivity methods are extended in two ways to achieve a realistic representation of incoming noise: 
(i) perturbations are time-stochastic rather than time-harmonic,
(ii) perturbations are localised at the inlet rather than distributed in space.
This allows for the identification of regions where small-amplitude control is the most effective, without actually computing any controlled flows.
In particular,  passive control by means of a small cylinder  and active control by means of wall blowing/suction are analysed for Reynolds number $\Rey=500$ and step-to-outlet expansion ratio $\Gamma=0.5$.
Sensitivity maps for noise amplification appear largely similar to 
sensitivity maps for  optimal harmonic amplification at the most amplified frequency. This is observed at other values of $\Rey$ and $\Gamma$ too, and suggests that the design of steady control in this noise amplifier flow can be simplified by focusing on the most dangerous perturbation at the most dangerous frequency.
\end{abstract}

\begin{keywords}
Authors should not enter keywords on the manuscript, as these must be chosen by the author during the online submission process and will then be added during the typesetting process (see http://journals.cambridge.org/data/\linebreak[3]relatedlink/jfm-\linebreak[3]keywords.pdf for the full list)
\end{keywords}

\section{Introduction}

In his famous pipe flow experiment, \citet{Rey1883} observed transition to turbulence and showed that the critical value of a governing non-dimensional parameter, to be later coined \textit{Reynolds number}, was strongly dependent on the level of external noise.
However, linear stability theory predicts the Hagen-Poiseuille flow to be asymptotically stable for any value of $\Rey$ \citep{Sch01}. 
It is now well understood  that linear stability theory successfully captures bifurcations and instability mechanisms  for some flows (\eg Rayleigh--B\'enard convection, Taylor--Couette flow between rotating cylinders, or flow past a cylinder), but fails for other flows: 
the Navier--Stokes equations which govern fluid motion  constitute a non-normal system, 
able to amplify perturbations through non-modal mechanisms \citep{Tre93}; then, if amplification is large enough 
it may drive the system away from linearly stable solutions.
Non-normal stable systems can exhibit 
large  \textit{transient growth} 
$G(t)=||\uu(\xx,t)||/||\uu(\xx,0)||$,
i.e.  temporary amplification of initial conditions $\uu(\xx,0)$,
as well as large \textit{harmonic gain} 
$G(\omega)=||\uu(\xx)||/|| \ff(\xx)||$, 
i.e.  asymptotic amplification of the response 
$\uu(\xx)\cos(\omega t)$
to external harmonic forcing $\ff(\xx)\cos(\omega t)$.
Classical linear algebra techniques allow one to find the largest amplification possible together with \textit{optimal perturbations}, i.e. those specific structures  associated with
 maximal transient growth or harmonic gain.
Extensive literature exists about the calculation of transient growth \citep{But92,Cor00,bla08}
and harmonic gain \citep{Ake08,ali09,Gar13,Sipp13,Der13}.

A question of fundamental importance is whether these non-modal amplification mechanisms are sensitive:  
if they are significantly altered by small flow modifications,
one can design control strategies with a wide variety of applications:
enhance mixing, improve aerodynamic performance, and reduce noise and vibration.
To investigate this point, adjoint methods are particularly well suited, as they provide maps of sensitivity showing regions where a given quantity of interest is the most affected by small-amplitude flow modification or steady control. 
Since the actual modified flow need not be computed,  sensitivity analysis allows for a fast and systematic control design, 
without resorting to time-consuming parameter studies.
In the context of linearly unstable flows, \citet{Hill92AIAA} used such a variational technique to derive the gradient of the leading eigenvalue with respect to flow modification and to steady control, and successfully reproduced maps of vortex shedding suppression obtained experimentally by \cite{Stry90}.
Similar methods were later applied to eigenvalues in several parallel and non-parallel configurations \citep{Bot03,Marquet08cyl,Mel10}. 
For unstable flows,
\citet{Bra11} followed a similar technique to derive an expression for the sensitivity of the optimal harmonic gain $G(\omega)$ with respect to  flow modification and to steady control.
They applied their formula  to parallel and non-parallel 
 flat-plate boundary layers and discussed the sensitivity of 
$G(\omega)$ for Tollmien--Schlichting and lift-up instability mechanisms.
\citet{Bou13} used this method to identify sensitive regions in the separated flow past a wall-mounted bump, and designed a simple open-loop wall control able to delay noise-induced subcritical transition. 
They observed, however, that sensitivity to volume control was  dependent on frequency, indicating that a given control could reduce $G(\omega)$ at some frequencies but increase it at others.
In this case, the effect of control on the overall response of the stochastically driven flow is unclear, and control design is not easy.

In the present study, we extend sensitivity  methods for linearly stable flows in two ways.
First,  we compute the sensitivity of amplification when the flow is subject to  \textit{stochastic forcing} rather than 
harmonic forcing.
This step takes advantage of the relation between stochastic and harmonic amplification \citep{Far96a}. 
It allows us to consider the overall response of the flow to external noise, and to combine sensitivities at individual frequencies  into a single sensitivity.
Second, we derive an expression for the sensitivity of amplification when the flow is forced \textit{at the inlet} rather than in the whole domain,
with the aim  of dealing with a realistic model of incoming perturbations in convectively unstable open flows.
The method is illustrated with  the two-dimensional incompressible flow past a backward-facing step, a canonical noise amplifier flow.

This flow has been studied extensively experimentally 
and numerically for a wide variety of Reynolds numbers and expansion ratios.
For instance,
\cite{Armaly83} used Laser-Dopppler Velocimetry to measure the position of stagnation points in laminar and turbulent regimes, and discussed the appearance of three-dimensionality.
\cite{Beaudoin04} investigated in more detail three-dimensional effects with Particle Image Velocimetry.
\cite{Kai96} performed two-dimensional direct numerical simulations as well as local and global linear stability analyses, and studied the response to impulsive and harmonic forcing in the volume or at the inlet.

\cite{bar02} performed a three-dimensional linear  stability analysis for a step-to-outlet expansion ratio $\Gamma=0.5$ and showed the three-dimensional character of the first  globally unstable mode
at $\Rey_c=748$, the flow remaining globally stable to two-dimensional perturbations due to the convective nature of the shear layer instability.
\cite{Lan12-BF} found $\Rey_c=714$ for $\Gamma=0.5$,
and extended the stability analysis to smaller and larger expansion ratios.
\cite{bla08} studied convective instabilities for $\Gamma=0.5$.
Computing optimal transient growth, they observed three-dimensional values slightly larger than the two-dimensional one, for large wavelengths and with very similar flow structures.
Furthermore, white noise inlet velocity fluctuations in a
 direct numerical simulation below the global instability threshold ($\Rey=500$) resulted in ``predominantly two-dimensional wave packets whose properties are related to the optimal disturbances'' found in the transient growth analysis. 
Finally, \cite{Marquet10-BF1} investigated the optimal harmonic response for $\Gamma=0.5$, $\Rey=500$,  in the two-dimensional flow, and found values of the most amplified frequency and the streamwise location of maximum response consistent with results from the transient growth analysis and the perturbed direct numerical simulation of \cite{bla08}.

This paper is organised as follows.
Section~\ref{sec:pb} recalls how to characterise the response to harmonic and stochastic forcing in the volume and at the inlet, 
and presents how to compute the sensitivity of harmonic and stochastic gains.
Section~\ref{sec:num_valid} details the numerical method and its validation.
The harmonic response in the backward-facing step flow 
for $\Gamma=0.5$ and several values of $\Rey$ is presented in section~\ref{sec:harm_resp}, where a
connection with local stability analysis is also established.
Results from the sensitivity analysis of  harmonic and stochastic responses are given in section~\ref{sec:sensit} for $\Gamma=0.5$, $\Rey=500$.
Maps of sensitivity  to volume control and wall control are presented in sections~\ref{sec:volume} and \ref{sec:wall}, respectively.
The analysis shows the predominant role of the optimal harmonic response at the optimal frequency.
Validation against  non-linear calculations is also presented (section~\ref{sec:valid}), as well as an application to passive wall control (section~\ref{sec:passive_wall}).
Finally, section~\ref{sec:discussion} 
discusses the link between the sensitivities of noise amplification and recirculation length, 
and investigates other $\Gamma-\Rey$ configurations.
Conclusions are drawn in section~\ref{sec:conclu}.

\section{Problem formulation}
\label{sec:pb}

\subsection{Flow configuration}
\begin{figure}
  \centerline{
   	\begin{overpic}[width=8 cm,tics=10]{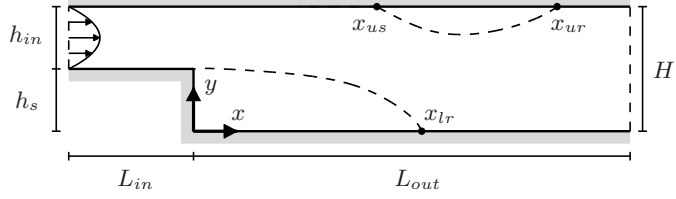}
	\put( 50,  22) {$x_{us}$}
	\put( 83,  22) {$x_{ur}$}
	\put( 62,   7) {$x_{lr}$}		
   	\put( 11,  -4) {$L_{in}$}
   	\put( 57,  -4) {$L_{out}$}
   	\put( -6,   9) {$h_s$}
   	\put( -7,  20) {$h_{in}$} 
   	\put(100,  14) {$H$}   	
   	\put( 30,   7) {$x$}
   	\put( 25.5,12) {$y$} 	
   	\end{overpic}
  }
  \vspace{0.2cm}
 \caption{Sketch of the geometry.
}
   \label{fig:sketch_step}
\end{figure}

We consider the two-dimensional flow over a backward-facing step, shown schematically in figure~\ref{fig:sketch_step}.
Geometrical parameters are the inlet   height $h_{in}$, 
the step height $h_s$, and the 
outlet height $H=h_s+h_{in}$,
which can be combined into a single governing parameter: the step-to-outlet expansion ratio $\Gamma=h_s/H$ or, equivalently, 
the outlet-to-inlet expansion ratio $e=H/h_{in}=1/(1-\Gamma)$.
Throughout this paper we will consider 
the classical geometry $\Gamma=0.5$ ($e=2$),
and a smaller step characterised by $\Gamma=0.3$ ($e\simeq1.43$).
The vertical wall and outlet lower wall define the $x=0$ and $y=0$ axes respectively.
The incoming flow is assumed to have a fully developed parabolic Poiseuille profile of maximum (centreline) velocity 
$U_\infty$
at the inlet  $\Gamma_{in}$ located at $x=-L_{in}$, while the 
outlet is at $x=L_{out}$.
The reference length is chosen as $L=H/2$ and the reference velocity as 
$U_\infty$.
The Reynolds number is consequently defined as 
$\Rey=L U_\infty/\nu$, where $\nu$ is the fluid kinematic viscosity.

The steady-state base flow  $(\UU_b,P_b)$ is solution of the stationary, incompressible Navier--Stokes equations in the domain $\Omega$, with no-slip boundary conditions at the walls $\Gamma_w$:
\begin{equation}
\begin{gathered}
	\bnabla \bcdot \UU_b = 0,
\\
\UU_b \bcdot \bnabla \UU_b   + \bnabla P_b - \nnu \bnabla^2 \UU_b = {\00}, 
\\
	 \UU_b=\00 
\quad \mbox{ on } \Gamma_w.
 \label{eq:NS}
\end{gathered}
\end{equation}
%

\subsection{Response to  forcing}
\label{sec:resp_forc}

In the following, we consider the response of a stable steady-state base flow to forcing. The focus of this paper is on \textit{stochastic inlet} forcing, but we 
mention  \textit{harmonic} and/or \textit{volume} forcing  to help understanding and highlight differences. 
More details can be found  for example in \cite{Far96a} and  \cite{Sch01}.
We first recall how the response to a small-amplitude harmonic  forcing 
$\ff'(\xx,t)=\ff(\xx) \e^{i\omega t}$ is characterised.
If the flow is linearly stable, the asymptotic  response is also harmonic at the same frequency. The forcing therefore introduces perturbations $(\uu',p')(\xx,t)=(\uu,p)(\xx) \e^{i\omega t}$ to the base flow  whose dynamics are governed by the  linearised equations
\begin{equation}
\begin{gathered}
	\bnabla \bcdot \uu_{vol} = 0,
\\
i \omega \uu_{vol} 
+ \bnabla \uu_{vol} \, \UU_b 
+ \bnabla \UU_b     \, \uu_{vol} 
+ \bnabla p_{vol} - \nnu \bnabla^2 \uu_{vol} = \ff_{vol}, 
\\
	 \uu_{vol}=\00 
\quad \mbox{ on } \Gamma_{in} \cup \Gamma_w
 \label{eq:LNharm_vol}
 \end{gathered}
\end{equation}
for volume forcing in $\Omega$,
and by
\begin{equation}
\begin{gathered}
	\bnabla \bcdot \uu_{in} = 0,
\\
i \omega \uu_{in} 
+ \bnabla \uu_{in} \, \UU_b
+ \bnabla \UU_b    \, \uu_{in} 
+ \bnabla p_{in} 
- \nnu \bnabla^2 \uu_{in} = \00, 
\\
	 \uu_{in}=\ff_{in}
\quad \mbox{ on } \Gamma_{in} ,
\\
	 \uu_{in}=\00 
\quad \mbox{ on }  \Gamma_w
 \label{eq:LNharm_in}
 \end{gathered}
\end{equation}
for inlet forcing on $\Gamma_{in}$ \citep{Gar13}.
We write formally (\ref{eq:LNharm_vol}) as
$\uu_{vol} = \mathcal{R}_{vol}(\omega) \ff_{vol}$
and (\ref{eq:LNharm_in})  as 
$\uu_{in} = \mathcal{R}_{in}(\omega) \ff_{in}$,
where in both cases $\mathcal{R}(\omega)$ is the resolvent operator.
For a given forcing, 
one simply needs to invert a linear system to obtain the response.
We introduce the usual Hermitian scalar product
$(\aa \,|\, \bb)
=\int_\Omega \overline\aa \bcdot \bb \,\mathrm{d}\Omega
=\int_\Omega          \aa^H      \bb \,\mathrm{d}\Omega$ 
or
$(\aa \,|\, \bb)
=\int_{\partial\Omega} \overline\aa \bcdot \bb \,\mathrm{d}\Gamma
=\int_{\partial\Omega}          \aa^H      \bb \,\mathrm{d}\Gamma$
for complex fields defined respectively 
in the domain or 
on (part or all of) the boundary, where 
$\overline{\bcdot}$ and $\bcdot^H$ stand for  conjugate and conjugate transpose.
The norm induced by this scalar product is used to measure 
amplification in the flow, or  harmonic gain $G(\omega) = ||\uu||/||\ff||$.
A natural quantity to look at is the largest value the gain may take, or  optimal gain, together with  the associated ``most dangerous'' forcing, or optimal forcing. 
This worst-case scenario is classically investigated by introducing the adjoint operator of the resolvent, and recasting the harmonic gain as a Rayleigh quotient
\be 
G^2(\omega)
= \frac{||\uu||^2}{||\ff||^2}
= \frac{(\mathcal{R}\ff \,|\, \mathcal{R}\ff)}{(\ff\,|\,\ff)}
= \frac{(\mathcal{R}^\dag\mathcal{R}\ff \,|\, \ff)}{(\ff\,|\,\ff)}.
\label{eq:rayleigh}
\ee
The largest value of $G$ is by definition the induced norm of the resolvent 
$||\mathcal{R}||$, which can be calculated as the largest singular value of $\mathcal{R}$.
Alternatively, solving the symmetric eigenvalue problem
$\mathcal{R}^\dag \mathcal{R} \ff_k
=G_k^2 \ff_k$ 
yields a set of real positive eigenvalues $G_1^2\geq G_2^2\geq G_3^2\ldots$ and a set of orthogonal eigenvectors $\ff_k$,
 from which one deduces the optimal gain:
 \be
G_{1}(\omega)
= \max_\ff \frac{||\uu||}{||\ff||} 
= \frac{||\uu_1||}{||\ff_1||}.
\ee
The  response of the flow to the optimal forcing 
is the optimal response $\uu_1=\mathcal{R}\ff_1$.
One can  similarly define
 sub-optimal gains, forcings and responses as
 $G_{k}  = ||\uu_k|| / ||\ff_k|| $,
 $\uu_k=\mathcal{R}\ff_k$ \citep{Gar13,Der13}.

We now turn our attention to stochastic forcing.
We assume that the flow is continuously forced by componentwise uncorrelated, white noise of unit variance,  
$\ff'(\xx,t)=\int_{-\infty}^{\infty}  \ff(\xx,\omega) \e^{i \omega t} \,\mathrm{d}\omega$, 
such that 
$
\mathcal{E}\left( f_j(\xx,\omega_1) f_k(\xx,\omega_2) \right)
=
\delta_{jk} \delta(\omega_1-\omega_2)/2\pi
$
where $\mathcal{E}(\cdot)$ denotes the mean or expected value of a random variable. 
Unless one has specific knowledge about temporal and spatial characteristics of  incoming perturbations, this assumption has the advantage of being both reasonable and simple.
The stochastic response is then characterised by the stationary ensemble variance  \citet{Far96a, Zhou96}
\be 
E=\mathcal{E}\left( ||\uu||_2^2 \right)
=
\frac{1}{2\pi} \int_{-\infty}^\infty 
\Tr\left( \mathcal{R}(\omega)^\dag \mathcal{R}(\omega) \right)  \,\mathrm{d}\omega,
\label{eq:E_def}
\ee
which can be expressed in terms of eigenvalues of 
$\mathcal{R}^\dag \mathcal{R}$:
\be
E 
= \frac{1}{2\pi} \int_{-\infty}^\infty \sum_k G_k^2(\omega) 
\,\mathrm{d}\omega
= \frac{1}{\pi} \int_{0}^\infty \sum_k G_k^2(\omega)  
\,\mathrm{d}\omega
=  \sum_k I_k.
\label{eq:E_G}
\ee
For  convenience we call $E$ the \textit{stochastic gain}, in contrast with the \textit{harmonic gain} $G(\omega)$.

\subsection{Sensitivity of harmonic and stochastic gain}
\label{sec:method_sensitivity}

Since non-normal flows have the potential for large amplification of perturbations, 
one way to delay transition to unsteadiness and turbulence is to apply a control, for instance steady control in the volume or at the wall, with the aim of reducing  harmonic and stochastic gains.
In this section we give the expressions of  sensitivities of harmonic/stochastic gains with respect to steady flow modification and steady control. 
These sensitivities are gradients which predict the effect of small-amplitude flow modification and control on the asymptotic amplification of harmonic/stochastic perturbations, and allow one to identify regions of the domain
$\Omega$ and of the wall $\Gamma_w$ where gains are the most sensitive, i.e. can be modified most easily.
It is worth stressing that the time-dependent \textit{forcing} $\ff$ is an external, unwanted perturbation undergoing amplification as described in section~\ref{sec:resp_forc}, while the steady \textit{control} $\CC$ or $\UU_c$ introduced in this section is applied intentionally in the flow or at the wall with the aim of reducing  amplification.

Starting with harmonic gain, 
we look for the two-dimensional sensitivity field 
$\bnabla_\UU G_1^2$ defined in $\Omega$
such that a modification $\bdelta\UU$ of the base flow induces a variation of the (squared) optimal gain 
\be 
\delta G_1^2 = (\bnabla_\UU G_1^2 \,|\, \bdelta\UU). 
\label{eq:def_D_U_G2}
\ee
Using a Lagrangian-based variational technique,  
\cite{Bra11} derived an expression for the sensitivity of the optimal harmonic gain for the case of volume forcing.
This expression is straightforwardly generalised  to any sub-optimal gain $G_{vol,k}$, $k>1$, replacing the optimal forcing 
$\ff_{vol,1}$ and optimal response $\uu_{vol,1}$ 
by the $k$th sub-optimal forcing and response:
\be
\bnabla_\UU G_{vol,k}^2 
=         2 G_{vol,k}^2 \, \Real\{
- \bnabla \uu_{vol,k}^H \,           \ff_{vol,k}
+ \bnabla \ff_{vol,k}   \, \overline{\uu}_{vol,k} \}.
\label{eq:sens_Gharm_BF_vol}
\ee
Using the same technique, we derived an expression for the case of inlet forcing:
\be
\bnabla_\UU G_{in,k}^2 
= 2 \,\Real\{ 
- \bnabla \uu_{in,k}^H \,           \uua_{in,k}
+ \bnabla \uua_{in,k}  \, \overline{\uu}_{in,k} \}
\label{eq:sens_Gharm_BF_in}
\ee
where the adjoint perturbation 
$(\uua_{in,k},\pa_{in,k})$ 
is a solution of the linear system 
$\uua_{in,k} = \mathcal{R}_{vol}^\dag  \uu_{in,k} $:
\begin{equation}
\begin{gathered}
	\bnabla \bcdot \uua_{in,k} = 0,
\\
-i \omega \uua_{in,k} 
+ \bnabla \uua_{in,k} \, \UU_b 
- \bnabla \UU_b^T     \, \uua_{in,k} 
+ \bnabla \pa_{in,k} 
+ \nnu \bnabla^2 \uua_{in,k} = \uu_{in,k}, 
\\
	 \uua_{in,k}=\00 
\quad \mbox{ on } \Gamma_{in} \cup \Gamma_w.
 \label{eq:adj_in}
\end{gathered}
\end{equation}

\begin{figure}
\vspace{1cm}
  \centerline{
   	\begin{overpic}[height=8.5 cm,tics=10]{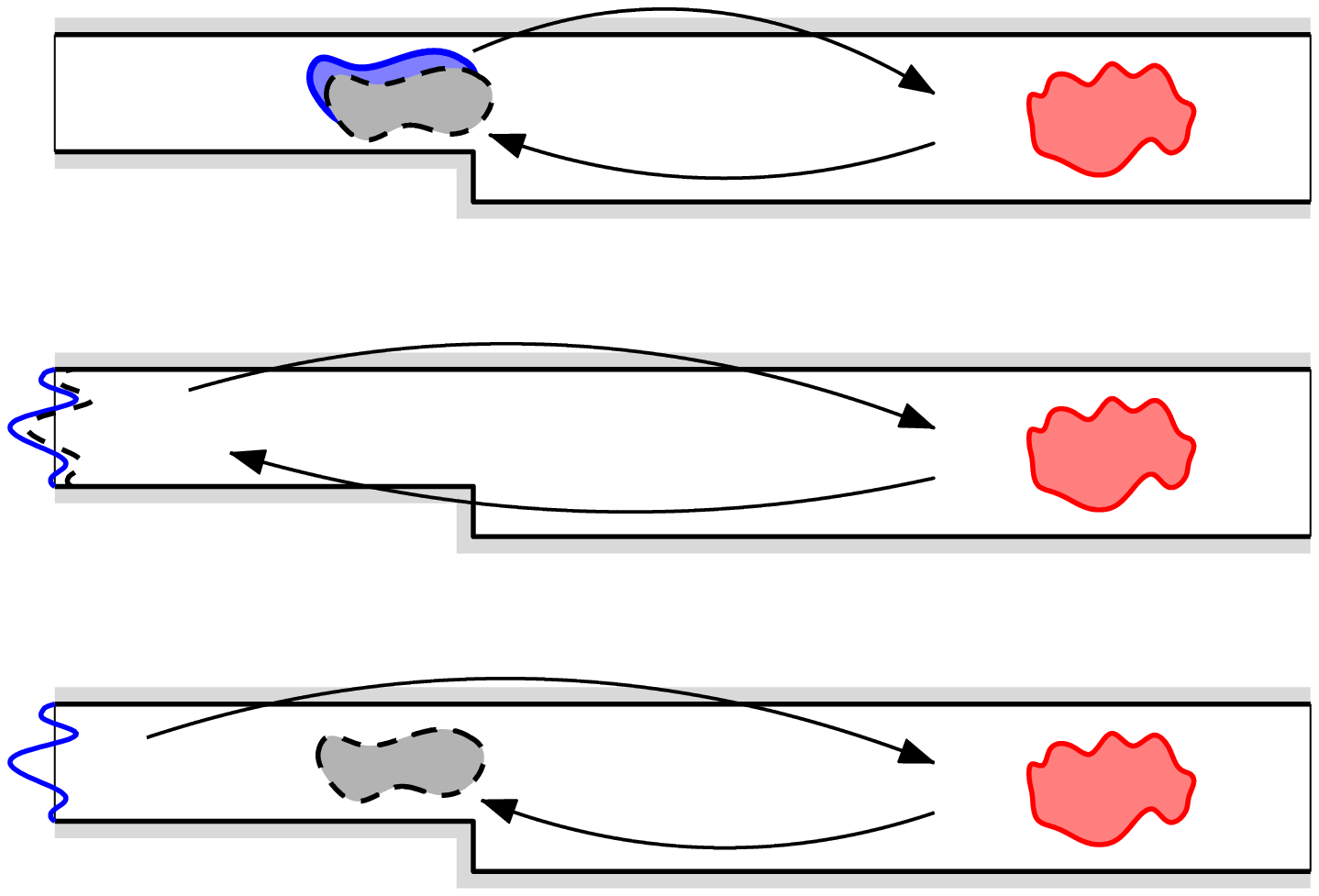}
   	\put(0,68.5) {$(a)$ Volume forcing: gain and sensitivity}
   	\put(58.5,  61.) {$\mathcal{R}_{vol}$}
   	\put(50,  55.7) {$\mathcal{R}_{vol}^{\dag}$}
   	\put(71.8,58)   {\textcolor{red}  {$\uu_{vol}$}}
   	\put(13,  61.5) {\textcolor{blue} {$\ff_{vol}$}}
   	\put(5.5, 57.7)   {\textcolor{black}{$G_{vol}^2 \ff_{vol}=\uu_{vol}^\dag$}}
   	\put(0,  43)   {$(b)$ Inlet forcing: gain }
   	\put(58.5,35.5) {$\mathcal{R}_{in}$}
   	\put(50,  30.5) {$\mathcal{R}_{in}^{\dag}$}
   	\put(71.8,32.4) {\textcolor{red}  {$\uu_{in}$}}   
	\put(10,  36.2) {\textcolor{blue} {$\ff_{in}$}}   	
	\put( 8,  32.4) {\textcolor{black}{$G_{in}^2 \ff_{in}$}}
   	\put(0,  17.8) {$(c)$ Inlet forcing: sensitivity}
   	\put(58.5,10.)  {$\mathcal{R}_{in}$}
   	\put(50,   5)   {$\mathcal{R}_{vol}^{\dag}$}
   	\put(71.8, 7.2) {\textcolor{red}  {$\uu_{in}$}}\\
   	\put(19,   8)   {\textcolor{black}{$\uu^\dag_{in}$}}
   	\put( 7,   10)   {\textcolor{blue} {$\ff_{in}$}}   	
   	\end{overpic}
  }
 \caption{
 Quantities and operators involved in the computation of the harmonic gain and of its sensitivity.
 $(a)$ In the case of volume forcing, the gain is given by 
$ G_{vol}^2 \ff_{vol}
= \mathcal{R}_{vol}^\dag \mathcal{R}_{vol} \ff_{vol}
= \mathcal{R}_{vol}^\dag \uu_{vol}$,
the operator $\mathcal{R}_{vol}^\dag \mathcal{R}_{vol}$ is self-adjoint,
and the adjoint perturbation $\uu_{vol}^\dag=G_{vol}^2 \ff_{vol}$ does not need to be computed to evaluate the gain sensitivity~(\ref{eq:sens_Gharm_BF_vol}).
$(b)$ In the case of inlet forcing, the gain is given by
 $ G_{in}^2 \ff_{in}
= \mathcal{R}_{in}^\dag \mathcal{R}_{in} \ff_{in}
= \mathcal{R}_{in}^\dag \uu_{in}$, where the operator 
$\mathcal{R}_{in}^\dag \mathcal{R}_{in}$ is self-adjoint;
$(c)$ however,  in order to evaluate the gain sensitivity~(\ref{eq:sens_Gharm_BF_in}) the adjoint perturbation 
$\uu_{in}^\dag=\mathcal{R}_{vol}^\dag \uu_{in}$  must be computed explicitly because the operator $\mathcal{R}_{vol}^\dag \mathcal{R}_{in}$ is not self-adjoint.
}
   \label{fig:sketch1}
\end{figure}

In the case of  sensitivity to \textit{volume} forcing (\ref{eq:sens_Gharm_BF_vol}),  no adjoint variable needs to be computed \citep{Bra11}.
This comes from the fact that the operator 
$\mathcal{R}_{vol}^\dag \mathcal{R}_{vol}$ involved in the \textit{volume} forcing  problem and associated gain sensitivity is self-adjoint (see figure~\ref{fig:sketch1}$(a)$).
In other words, even though an adjoint perturbation has to be included in the Lagrangian \textit{a priori}, calculations show that it 
can be replaced by $G_{vol}^2 \ff_{vol}$: 
indeed, equation (\ref{eq:LNharm_in})  is
$\uu_{vol} = \mathcal{R}_{vol} \ff_{vol}$
and implies
$ \mathcal{R}_{vol}^\dag \uu_{vol} 
= \mathcal{R}_{vol}^\dag \mathcal{R}_{vol} \ff_{vol} 
= G_{vol}^2 \ff_{vol}$;
at the same time, detailed calculations lead to the  adjoint perturbation equation
$\uua_{vol} = \mathcal{R}_{vol}^\dag \uu_{vol}$,
and therefore
$\uua_{vol} = G_{vol,k}^2 \ff_{vol}$.
The situation is quite different for \textit{inlet} forcing.
The operator 
$\mathcal{R}_{in}^\dag \mathcal{R}_{in}$
 involved in the \textit{inlet} forcing problem is self-adjoint too (fig.~\ref{fig:sketch1}$(b)$), 
 but the operator $\mathcal{R}_{vol}^\dag \mathcal{R}_{in}$
 needed to obtain the associated sensitivity is not (fig.~\ref{fig:sketch1}$(c)$).
 Consequently, the adjoint perturbation 
 $(\uua_{in},\pa_{in})$ which appears in the expression  of  sensitivity to 
 \textit{inlet} forcing  (\ref{eq:sens_Gharm_BF_in}) has to be computed on its own.
 Interestingly, note that although we are dealing with  \textit{inlet} forcing,
  the adjoint perturbation is a solution of an equation forced in the   
 \textit{volume} by the  response $\uu_{in}$.

Next, we turn to the sensitivity of harmonic gain to steady  volume control $\CC$ in  $\Omega$ and steady wall control $\UU_c$ on $\Gamma_w$. 
The former sensitivity is a two-dimensional field such that a small-amplitude volume control produces the variation   
\be 
\delta G_k^2 = (\bnabla_{\CC} G_k^2 \,|\, \bdelta\CC), 
\label{eq:def_D_C_G2}
\ee
while the latter is a one-dimensional field defined on $\Gamma_w$ such that 
\be 
\delta G_k^2 = (\bnabla_{\UU_c} G_k^2 \,|\, \bdelta\UU_c). 
\label{eq:def_D_Uc_G2}
\ee
Again, one can generalise the expression  of \cite{Bra11} for the optimal gain $G_1$ to any sub-optimal gain $G_k$, $k>1$:
\begin{equation} 
\begin{gathered} 
  \bnabla_\CC G_k^2 = \UUa_k,
	\\
\bnabla_{\UU_c} G_k^2 =  \Pa_k \nn + \nnu  \nabla \UUa_k \,\nn,
\label{eq:sens_Gharm_ctrl}
\end{gathered} 
\end{equation} 
where $\nn$ is the outward unit normal  vector,
and the adjoint base flow $(\UUa_k,\Pa_k)$ is solution of the following  linear 
system forced by the sensitivity to base-flow modification
$\bnabla_\UU G_k^2$
defined in (\ref{eq:sens_Gharm_BF_vol})-(\ref{eq:sens_Gharm_BF_in}):
\begin{equation} 
\begin{gathered} 
\bnabla \bcdot \UUa_k = 0,
\\
- \bnabla \UUa_k  \, \UU_b 
+ \bnabla \UU_b^T \, \UUa_k 
- \bnabla \Pa_k - \nnu \bnabla^2 \UUa_k 
= \bnabla_\UU G_k^2,
\\  
\UUa_k = \00  \quad \mbox{ on }  \Gamma_{in} \cup \Gamma_w.
\label{eq:adjBFharm}
\end{gathered} 
\end{equation} 
This time, the same method holds for both inlet and volume forcing, so we omitted subscripts $in$ and $vol$ in (\ref{eq:def_D_C_G2}) to (\ref{eq:adjBFharm}).

Finally, the sensitivity of the stochastic gain (\ref{eq:E_G}) is defined by
\be 
\delta E = (\bnabla_{\boldsymbol{*}} E \,|\, \bdelta \boldsymbol{*})
\label{eq:def_D_E}
\ee
and can be expressed by linearity in terms of the sensitivity of harmonic gains 
\be
\bnabla_{\boldsymbol{*}} E
=
\frac{1}{\pi} \int_{0}^\infty \sum_k
\bnabla_{\boldsymbol{*}} G_k^2(\omega)
 \,\mathrm{d}\omega 
=
 \sum_k \bnabla_{\boldsymbol{*}} I_k,
\label{eq:sens_E_BF}
\ee
where $\boldsymbol{*}$  stands for either
base-flow modification $\UU$, 
volume control         $\CC$ or 
wall   control         $\UU_c$. 
Again, expression (\ref{eq:sens_E_BF}) is valid for both inlet and volume forcing.

\section{Numerical method and validation}
\label{sec:num_valid}

All calculations  are performed using methods described in \cite{Bou13}.
The finite element software \textit{FreeFem++} is used to 
generate a two-dimensional triangulation of the domain $\Omega$ and, based on P2 and P1 Taylor-Hood elements for velocity and presure respectively, to
build all  the discrete operators involved in calculations of base flow, eigenvalue and sensitivity, from  their corresponding continuous expression in variational form.
Steady-state base flows are obtained with an iterative Newton method,
while eigenvalue calculations are conducted with an implicitly restarted Arnoldi method.  
Careful validation and convergence study (described below) led us to set the outlet length to $L_{out}=50$ for $\Gamma=0.5$ and $L_{out}=250$ for $\Gamma=0.3$,
 the entrance length to  $L_{in}=5$ for both geometries, and a mesh density distribution yielding 
 216340 and 298484 elements 
 (0.98 and 1.36 million degrees of freedom)
 for $\Gamma=0.5$ and $\Gamma=0.3$ respectively.

Our choice of outlet length $L_{out}$ is such that the outlet velocity profile  is well developed for all conditions: specifically, it ensures that the difference between the base flow and the fully developed parabolic Poiseuille profile
$\Delta U(y)=U_b(L_{out},y)-U_P(y)$  
is less than 1\% for $\Gamma=0.5$ and 3\% for  $\Gamma=0.3$,   
both in $L^2$ norm $||\Delta U||_2$ 
(relative to $||U_P||_2$) 
and 
$L^\infty$ norm 
$||\Delta U||_\infty$ 
(relative to $U_P(y^*)$ at the height $y^*$ of largest $|\Delta U|$). 
Validation included a three-dimensional stability analysis: using global modes 
$\uu'(x,y,z,t)=\uu(x,y) \e^{i\beta z+\sigma t}$ we calculated the  
critical Reynolds number $\Rey_c$ (i.e. the smallest  $\Rey$ for which one global mode becomes unstable, $\Real\{\sigma\}\geq0$) and corresponding spanwise wavenumber $\beta_c$. Results are given in
 table~\ref{tab:valid1} and show an 
excellent agreement with those of \cite{bar02} 
and \cite{Lan12-BF}, with differences smaller than 0.5\%.
We also looked at the positions of reattachment and separation points $(x_{lr},x_{us},x_{ur})$ (characterised by zero wall shear stress, see fig.~\ref{fig:sketch_step}) for $\Gamma=0.5$.
At all Reynolds numbers up to $\Rey\leq1000$, our values were indistinguishable  with data extracted from  figures in \cite{bar02} and \cite{bla08}.
At $\Rey=600$ we find the values  given in table~\ref{tab:valid2}, in excellent agreement
with those reported by \cite{bar02}.
The secondary recirculation zone appears at the upper wall at $\Rey=272$, $x_u=8.2$, consistent with the values $\Rey\simeq275$, $x_u\simeq8.1$ of \cite{bla08}.
Tables~\ref{tab:valid1} and \ref{tab:valid2} show that the choice $L_{in}=5$ is justified since all values are well converged.
Mesh independence was checked by increasing the number of  elements by 20\% with a global and uniform refinement, which led to less than 0.05\%  variation for critical conditions (Reynolds number and wavenumber) and the locations of stagnation points.

\begin{table} \small
  \begin{center}
  \begin{tabular}[]{l c ccc c ccc}
\multicolumn{1}{c}{$\Gamma$}	&& \multicolumn{3}{c}{$0.5$} 				&& \multicolumn{3}{c}{$0.3$} \\
\multicolumn{1}{c}{$L_{in}$}		&& $1$ 		  & $5$ 		 & $10$ 		&& $1$ 		    & $5$ 		   & $10$ \\
\cline{1-1} \cline{3-5} \cline{7-9}
\\
$(a)$ BGH02   						&& (748, 0.91) & - 			 & -			&& - 			& -		 	   & -\\
$(b)$ LK12							&& (748, 0.92) & (714, 0.88) & (714, 0.88)	&& - 		    & -			   & (2948, 1.02) \\
$(c)$ Present  						&& (750, 0.92) & (715, 0.88) & (715, 0.88) 	&& (3206, 1.13) & (2966, 1.01) & (2964, 1.01)
  \end{tabular}
  \caption{\small Critical Reynolds number and spanwise wavenumber $(\Rey_c,\beta_c)$ for different expansion ratios $\Gamma$ and  entrance lengths $L_{in}$: 
  $(a)$ \cite{bar02}, $(b)$ \cite{Lan12-BF}, $(c)$ present study.}
  \label{tab:valid1}
  \end{center}
\end{table}

\begin{table} \small
  \begin{center}
  \begin{tabular}[]{l c ccc }
\multicolumn{1}{c}{$L_{in}$}	&& $1$ 		  		    & $5$ 		 	       & $10$  \\
\cline{1-1} \cline{3-5}
\\
$(a)$ BGH02   					&& (11.91, 9.5, 20.6)   & - 			       & -			\\
$(b)$ Present  					&& (11.93, 9.45, 20.60) & (11.82, 9.34, 20.59) & (11.82, 9.34, 20.59)
  \end{tabular}
  \caption{\small Locations $(x_{lr},x_{us},x_{ur})$ of lower reattachment point, upper separation point and upper reattachment point  at $\Rey=600$, for $\Gamma=0.5$ and different entrance lengths $L_{in}$: 
  $(a)$ \cite{bar02}, $(b)$ present study.}
  \label{tab:valid2}
  \end{center}
\end{table}

In the case of inlet forcing,
the stochastic response (\ref{eq:E_G}) and its sensitivities (\ref{eq:sens_E_BF}) are evaluated as follows.
Integrals $I_k$ and $\bnabla_{\boldsymbol{*}} I_k$  are calculated
using a trapezoidal rule with $n_\omega=41$ points regularly  distributed over the range of frequencies  $\omega\in[0;\omega_c]$.
Halving or doubling $n_\omega$   modifies the value of $E$ 
by less than 1\%.
The cut-off frequency is set to $\omega_c=2$ and kept fixed throughout the study. This value is well above the main peak of $G_{1}(\omega)$ at $\omega_0=0.5$, so as to include the contribution of amplification mechanisms, while optimal and sub-optimal forcings and responses at higher frequencies correspond only to advection and diffusion; 
therefore the exact value of $E$  does depend on $\omega_c$ but qualitative results are unaffected
(see appendix~\ref{sec:appB}).
Sums over $k$ are computed with the full set of optimal and sub-optimals, which is computationally tractable in the case of inlet forcing since their number is equal to the number of  degrees of freedom at the inlet.
The effect of taking a limited number of sub-optimals is reported in appendix~\ref{sec:appC}.
Note that although evaluating $E$ and $\bnabla_{\boldsymbol{*}} E$ is relatively costly, computations are largely parallelisable since different frequencies can be treated independently.
%

\section{Harmonic response}
\label{sec:harm_resp}

In this section we present results about the response of the flow to small-amplitude harmonic forcing.
Although the focus of this paper is on \textit{stochastic} forcing at the \textit{inlet} of the domain as a realistic model of random perturbations advected by the  flow, we  mention here \textit{harmonic} forcing  since it is a building block of the stochastic problem, and forcing in the \textit{volume} for comparison purposes.
The configuration $\Gamma=0.5$, $\Rey=500$, is considered unless otherwise stated.

\begin{figure}
  \psfrag{om}[t][]{$\omega$}
  \psfrag{re}[t][]{$\Rey$}
  \psfrag{Gi}[r][][1][-90]{} 	
  \psfrag{maxG1}[r][][1][-90]{} 
  \centerline{
   	\begin{overpic}[width=6.5 cm,tics=10]{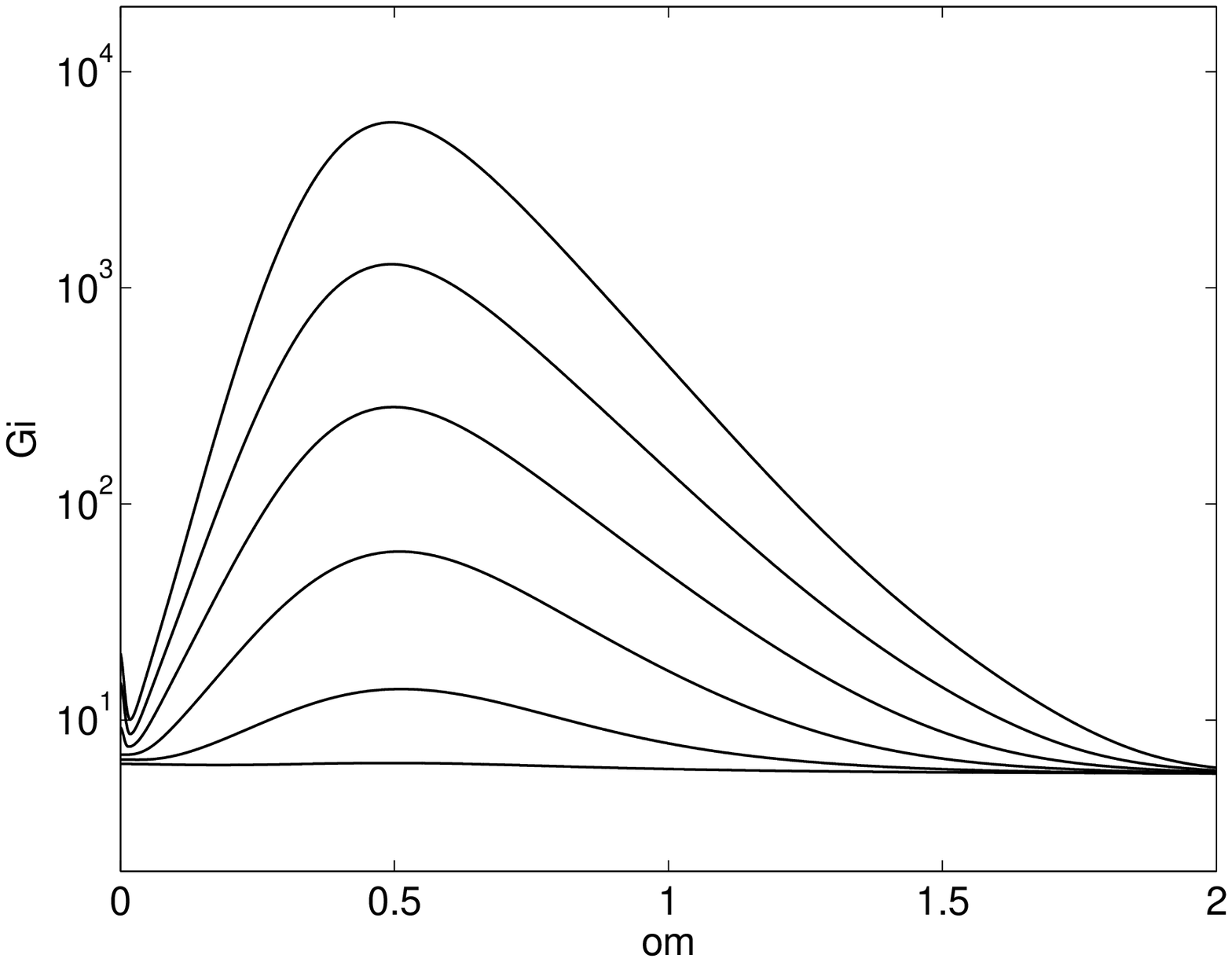}
		\put(12, 72){$(a)$ $G_{in,1}$}
 		\put(75, 30){$\Rey=600$}
  		\put(69, 10.5){$\Rey=100$}
   	\end{overpic}
   	\hspace{0.01 cm}
   	\begin{overpic}[width=6.5 cm,tics=10]{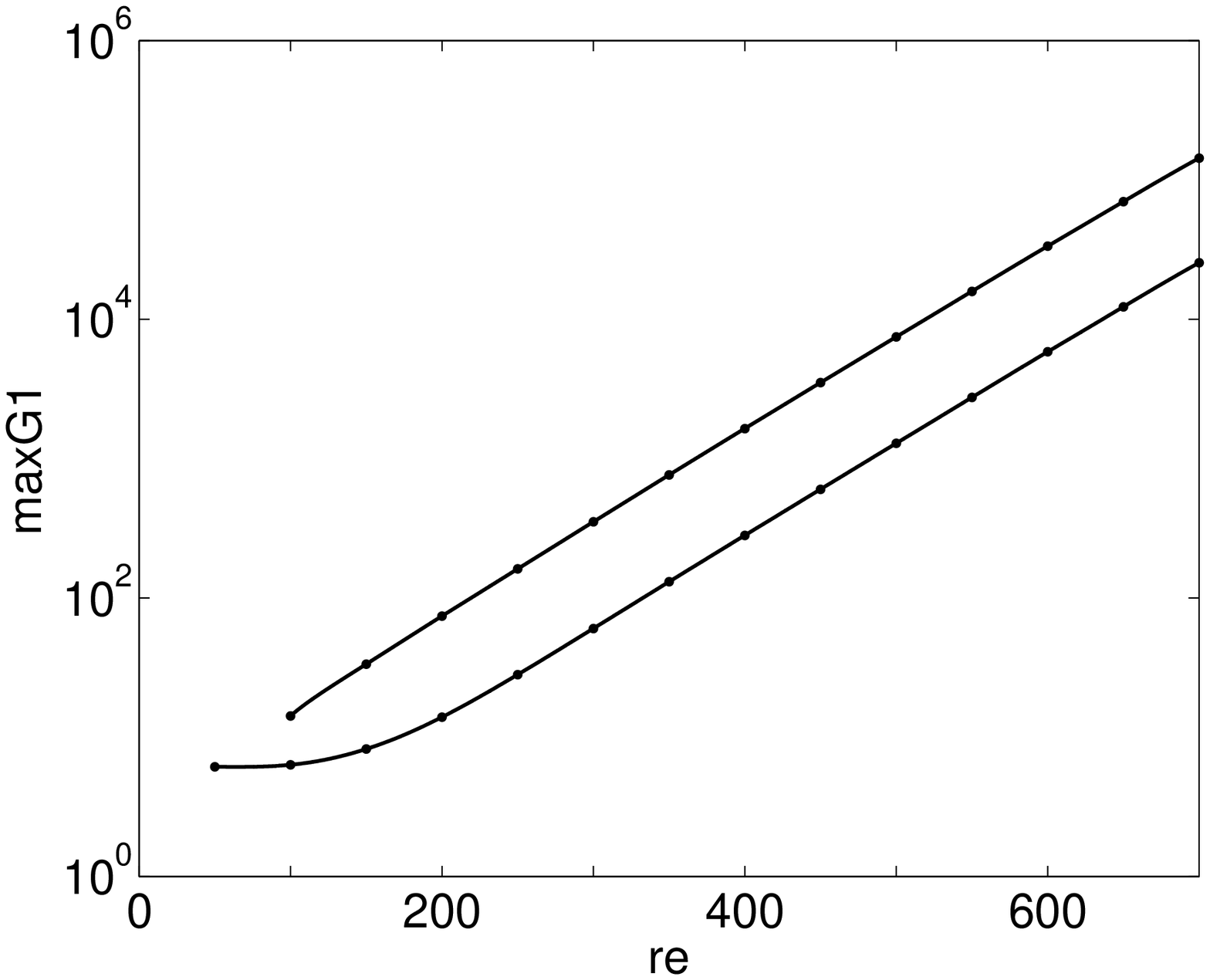}
		\put(14, 72){$(b)$}
		\put(40, 50){$\displaystyle \max_\omega G_{vol,1}$}
		\put(60, 30){$\displaystyle \max_\omega G_{in,1}$}				
   	\end{overpic}
  }
 \caption{
 $(a)$~Optimal harmonic gain for inlet  forcing. $\Gamma=0.5$, $\Rey=100,$ 200$\ldots$, 600.
$(b)$~Maximum harmonic gain for volume and inlet forcing. $\Gamma=0.5$.
}
   \label{fig:Gin_gamma05_re100to600}
\end{figure}

Figure~\ref{fig:Gin_gamma05_re100to600}$(a)$ shows the optimal harmonic gain for inlet forcing $G_{in,1}$ for $\Gamma=0.5$. 
The maximal optimal gain increases from 
$6.33$ at $\Rey=100$
to 
$5.83\times10^3$ at $\Rey=600$, the optimal frequency being close to $\omega_0=0.5$ for all investigated Reynolds numbers.
The dependence of the maximal optimal gain on Reynolds number is  exponential beyond some value of $\Rey$, as illustrated in Figure~\ref{fig:Gin_gamma05_re100to600}$(b)$, with 
$\log(\max_\omega G_{in,1})$ and
$\log(\max_\omega G_{vol,1})$ scaling like 
$0.66\times10^{-2}\Rey$.
The exponent for volume forcing was 
$\simeq0.6\times10^{-2}\Rey$ in the flow past a wall-mounted bump with $\Rey$ based on the bump height~\citep{Bou13}.
This exponential dependence contrasts with parallel flows, where the maximal optimal harmonic gain only increases like $\Rey^2$~\citep{Sch01}.

\cite{Cantwell10Exp} observed the same phenomenon for the maximal transient growth in several wall-bounded separated flows: 
they reported exponential dependence on $\Rey$ with exponent of order  $10^{-2}$
($0.45\times10^{-2}$ for a sudden axisymmetric expansion,
$0.61\times10^{-2}$ for a stenotic flow, and
$1.18\times10^{-2}$ for the $\Gamma=0.5$ backward-facing step with $\Rey$ based on the centreline velocity),
while in parallel flows the maximal transient growth only increases like $\Rey^2$~\citep{Sch01}.

In these two analyses (harmonic response and transient growth), the spatial growth of perturbations is involved. In spatially developing flows, convective non-normality allows for an exponential growth  over the entire shear layer length $l$. Since the latter scales like $l=\alpha\Rey$, perturbations grow like $\exp{(\alpha\Rey)}$, with $\alpha$ a constant. In contrast, in parallel flows, component-type non-normality (e.g. lift-up and Orr mechanisms) results in algebraic growth, scaling like $\Rey^2$ as can be shown by considering the Orr-Sommerfeld-Squire equations~\citep{Sch01} or a simple model thereof~\citep{Cossu14}.

\begin{figure}
  \centerline{
  \psfrag{om}[t][]{$\omega$}
  \psfrag{Gv}[r][]{} 
  \psfrag{Gi}[r][]{} 	
	\begin{overpic}[width=6.5cm,tics=10]{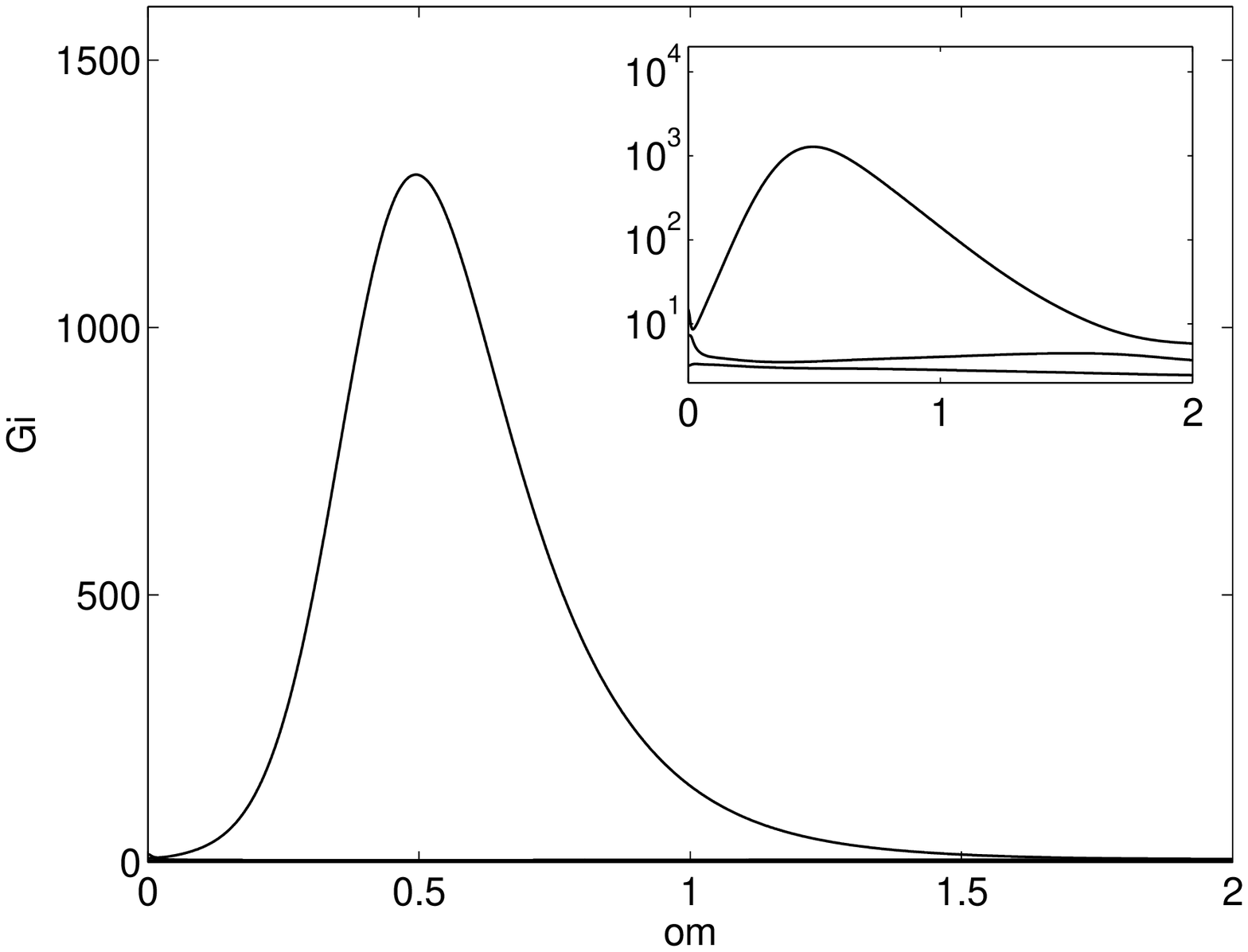}
  		\put(14,70){$(a)$ $G_{in,k}$}
   	\end{overpic} 
   	\begin{overpic}[width=6.5cm,tics=10]{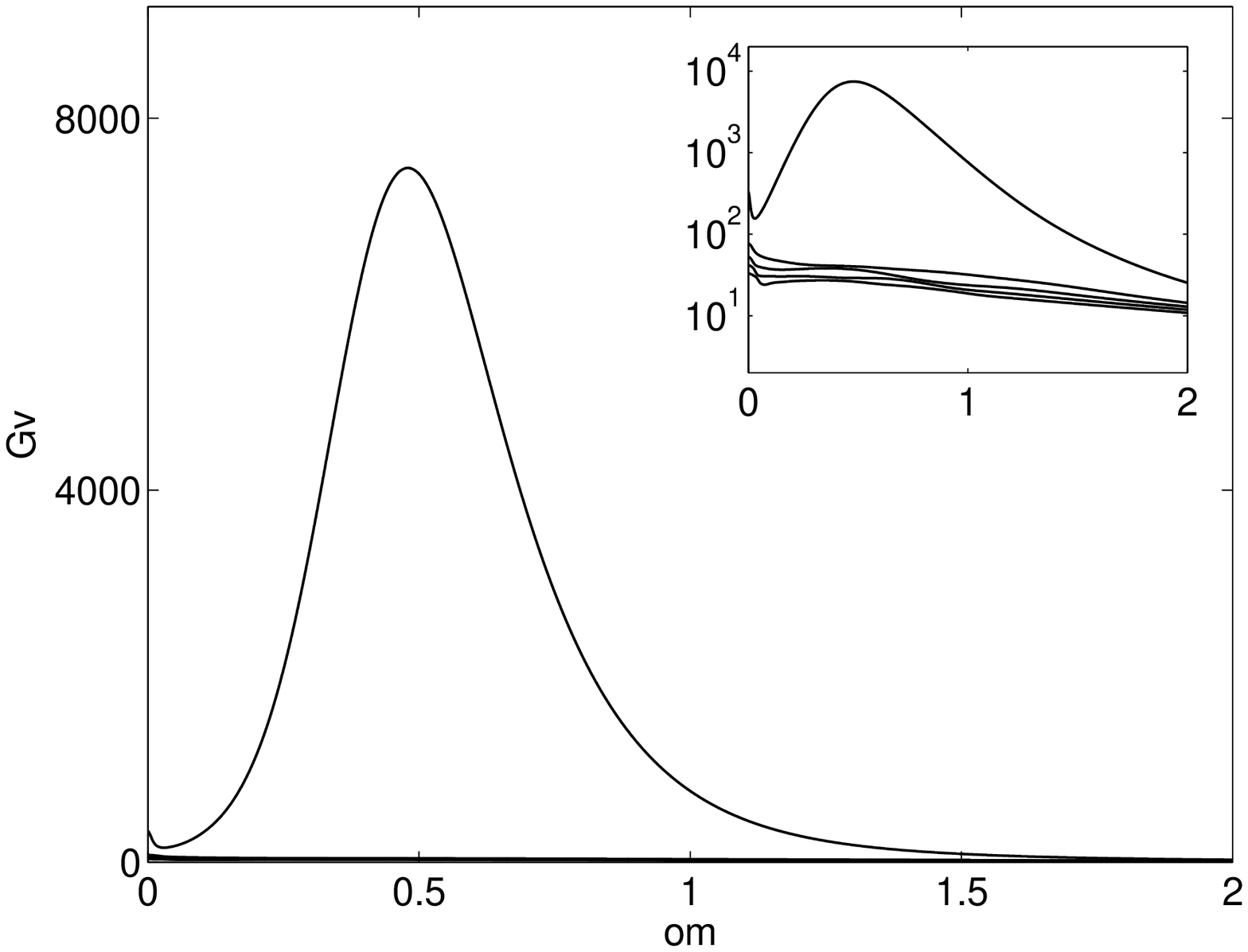}
   		\put(14,70){$(b)$ $G_{vol,k}$}
   	\end{overpic}  
  }
 \caption{
 Optimal harmonic gain for $(a)$ inlet and $(b)$ volume forcing. 
  $\Gamma=0.5$, $\Rey=500$.
Insets show the optimal gain and a few sub-optimal gains in logarithmic scale.
 }
   \label{fig:Gin_Gvol_gamma05_re500}
\end{figure}

Focusing on $\Rey=500$ from now on, we compare the 
optimal harmonic gain for inlet forcing and volume forcing in figure~\ref{fig:Gin_Gvol_gamma05_re500}.
The gain is larger in the case of volume forcing, which is a consequence (i) of the choice of the norm used to measure forcing amplitude (two-dimensional versus one-dimensional), and 
(ii) of the greater efficiency with which two-dimensional forcing structures (allowed to occupy the whole domain) excite the flow, compared to one-dimensional forcing structures (restricted to the inlet).
The maximum gain is 
$\max_\omega G_{in,1}=1.29\times10^3$ at  
$\omega_0 = 0.49$ for inlet forcing,
and 
$\max_\omega G_{vol,1}=7.46\times 10^3$ at
$\omega_0 = 0.48$ for volume forcing.
 The latter values are in excellent agreement with those reported by \cite{Marquet10-BF1}: 
$\max_\omega G_{vol,1}=7.5\times 10^3$ at $\omega_0=0.47$. 
Sub-optimal branches in figure~\ref{fig:Gin_Gvol_gamma05_re500} have a much lower gain and mainly correspond to advection and diffusion.
As pointed out by \cite{Marquet10-BF1}, the flow response should therefore be dominated by the optimal response.  
\cite{Bou13}  observed this behaviour in a direct numerical simulation of a different geometry, with a predominance of the optimal response at optimal frequency. 
As discussed later in section \ref{sec:sensit}, the sensitivity of the stochastic gain is also dominated by the sensitivity of the optimal harmonic gain at the optimal frequency.

\begin{figure}
  \psfrag{x}[t][]{$x$}
  \psfrag{y}[][][1][-90]{$y$}	
  \centerline{
   	\begin{overpic}[width=13.cm,tics=10]{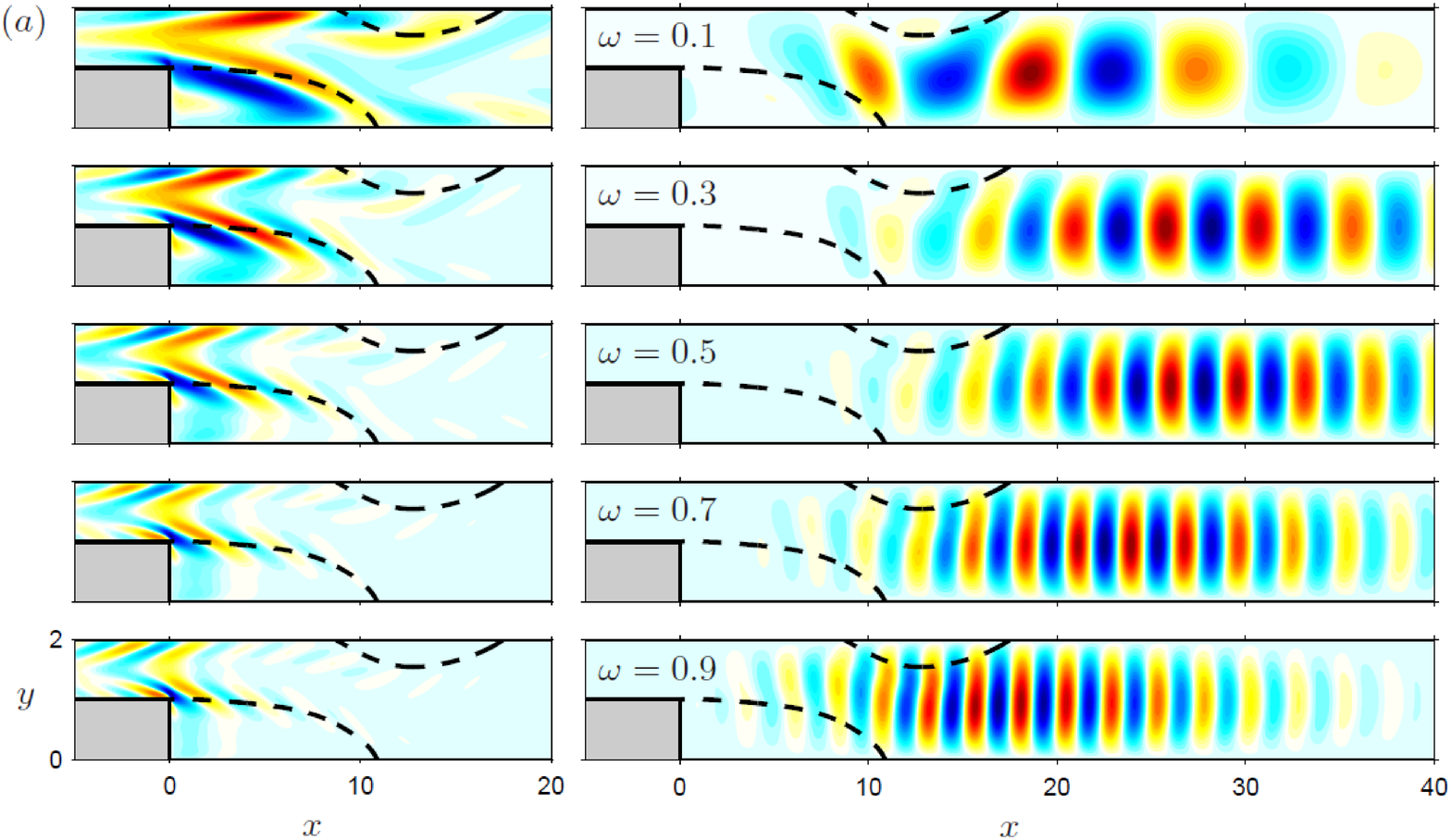}
   	\end{overpic}
  }
  \vspace{0.3cm}
  \centerline{
   	\begin{overpic}[width=13.1cm,tics=10]{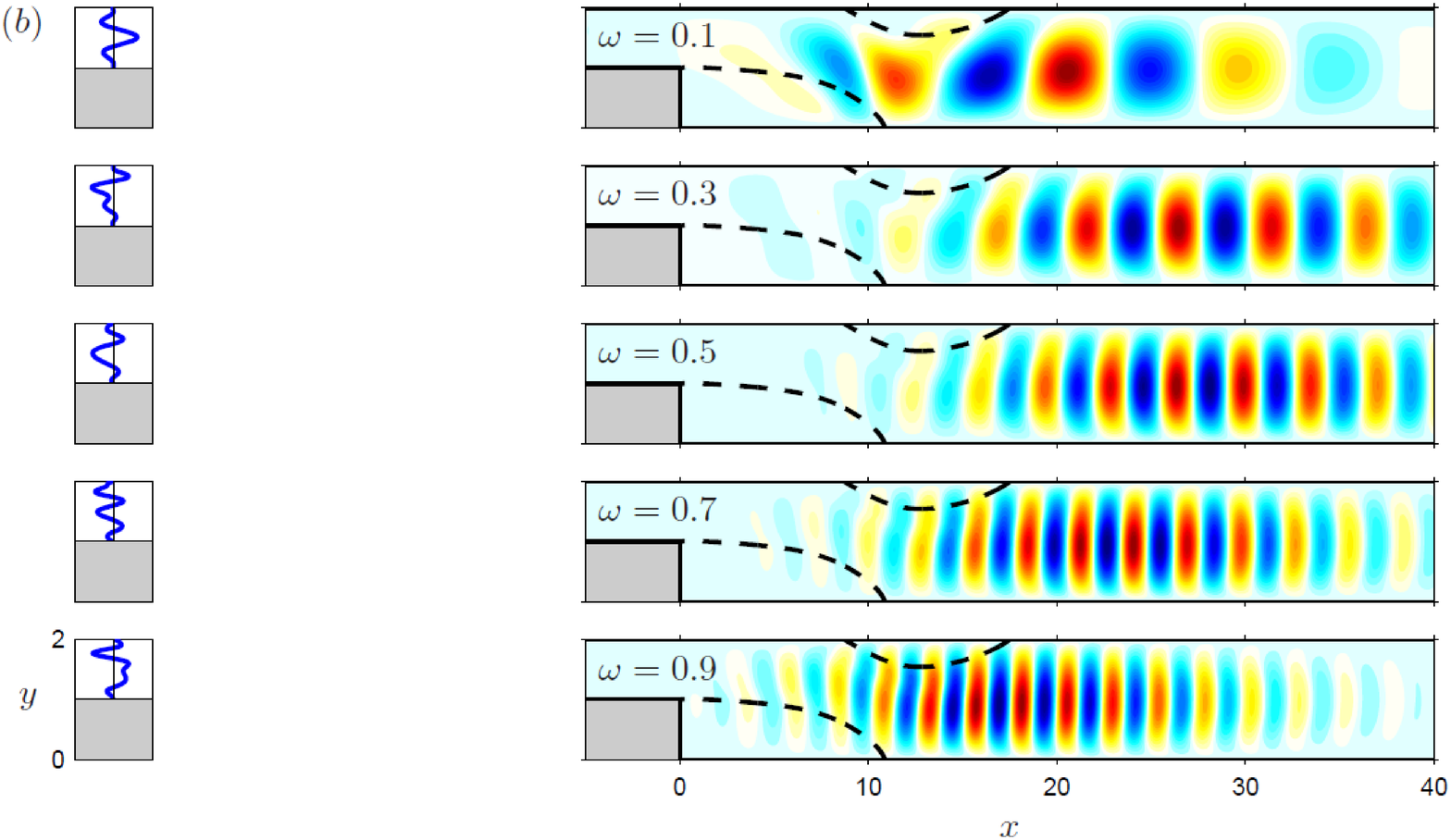}
   	\end{overpic}
  }
 \caption{
 Optimal harmonic forcing (left; real part of streamwise component $\ff_1\bcdot\ex$) and 
 optimal harmonic response (right; real part of cross-stream component $v_1$):
$(a)$ volume forcing and corresponding response,
$(b)$ inlet forcing and corresponding response. 
     $\Gamma=0.5$, $\Rey=500$.
     }
   \label{fig:opt_forc_resp_in_vol_gamma05_re500}
\end{figure}

\begin{figure}
  \psfrag{x}[t][]{$x$}
  \psfrag{y}[][][1][-90]{$y$}	
  \centerline{
   	\begin{overpic}[width=13.cm,tics=10]{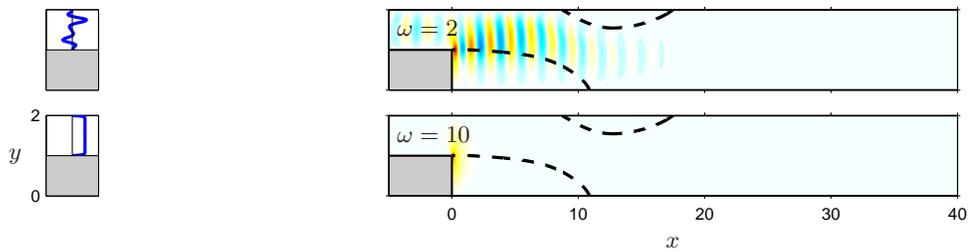}
  	\put(41, 22.0)  {\footnotesize$\omega=2$}
 	\put(41, 11.1)  {\footnotesize$\omega=10$}
   	\end{overpic}
  }
 \caption{
Same as figure \ref{fig:opt_forc_resp_in_vol_gamma05_re500}$(b)$ at higher frequencies.
 }
 \label{fig:opt_forc_resp_in_vol_gamma05_re500_higher_freq}
\end{figure}

The optimal forcing and response are shown in figure~\ref{fig:opt_forc_resp_in_vol_gamma05_re500} for different frequencies.
They have a shorter wavelength as  $\omega$ increases, consistent with most studies of harmonic optimal gain in convective flows \citep{ali09,Gar13,Der13,Bou13}.
The volume optimal forcing is maximal close to the step corner for all frequencies.
The optimal response is maximal  downstream of the step corner: close to the upper reattachment point at $\omega=0.1$, farther downstream as $\omega$ increases towards the optimal frequency $\omega_0\simeq0.5$,  then farther upstream as frequency continues to increase.
Interestingly, the optimal inlet  forcing 
$\ff_{in,1}(y)$ in figure~\ref{fig:opt_forc_resp_in_vol_gamma05_re500}$(b)$ is very similar  to the profile of optimal volume forcing in figure~\ref{fig:opt_forc_resp_in_vol_gamma05_re500}$(a)$) close to the inlet  $\ff_{vol,1}(x\rightarrow-L_{in}^+,y)$. 
Furthermore, these optimal inlet and volume  forcings lead to very similar optimal responses (except for slight differences best seen at low frequency near the step corner and the lower recirculation region).
\cite{Gar13} observed the same phenomenon over a broad range of frequencies.
At higher frequencies, where amplification is small, the optimal inlet forcing tends to a plug profile and the response is concentrated at the step corner 
(figure \ref{fig:opt_forc_resp_in_vol_gamma05_re500_higher_freq}).


\begin{figure}
  \psfrag{x}  [][]{$x$}
  \psfrag{y}[r][][1][-90]{$y\,\,$}
  \psfrag{k}  [][]{$k$}
  \psfrag{oom}[][]{$\omega$}
  \psfrag{-ki}[][][1][-90]{$-k_i^{(S)}$}	
  \psfrag{wi} [][][1][-90]{$\omega_i^{(T)}$}	
  \centerline{
     \hspace{0.3cm}
     \begin{overpic}[width=13.cm,tics=10]{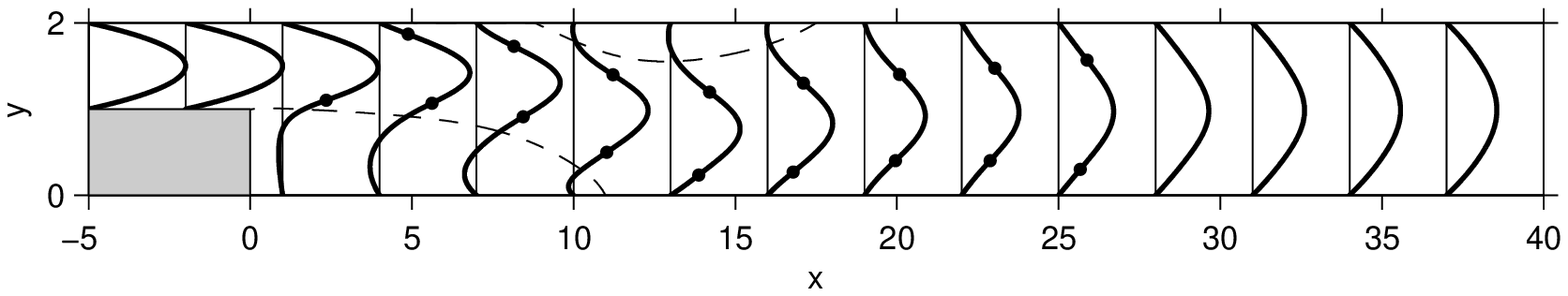}
     \put(-2, 16)  {$(a)$}   
     \end{overpic}
  }
  \vspace{0.2cm}
  \centerline{
   	\begin{overpic}[width=13.cm,tics=10]{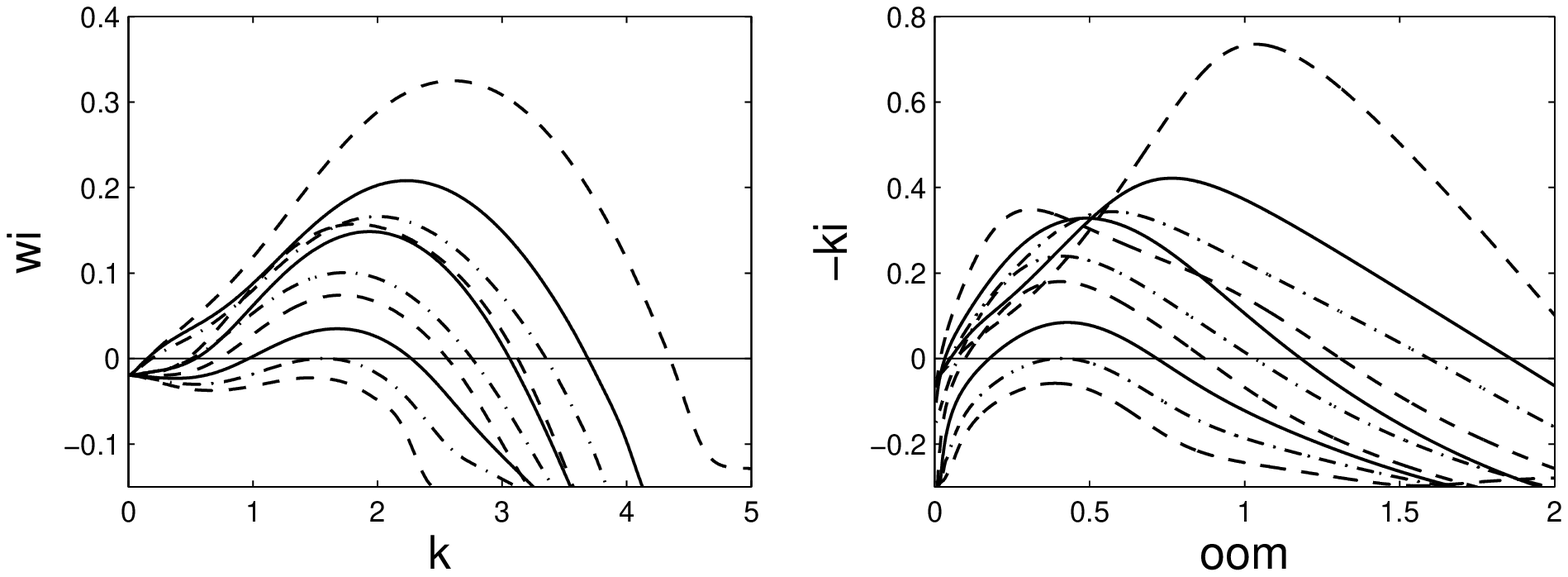}
   	\put(0, 34)  {$(b)$}
   	\put(32.0, 26.0)  {\tiny  $x=1$}   
	\put(33.0, 22.0)  {\tiny    $3$}
	\put(31.5, 19.7)  {\tiny    $5$}
	\put(30.3, 18.5)  {\tiny    $8$}	
	\put(28.3, 17.8)  {\tiny   $11$}	
	\put(26.3, 17.0)  {\tiny   $14$}	
	\put(24.1, 16.0)  {\tiny   $17$}	
	\put(22.3, 15.0)  {\tiny   $22$}		
	\put(20.5, 13.3)  {\tiny   $27$}
	\put(16.0, 11.5)  {\tiny $x=32$}	
	\put(50.5, 34)  {$(c)$}
   	\put(87,   24.0)  {\tiny  $x=1$}   
	\put(87,   20.0)  {\tiny    $3$}
	\put(82.5, 18.5)  {\tiny    $5$}
	\put(80.2, 17.7)  {\tiny    $8$}	
	\put(76.5, 17.0)  {\tiny   $11$}	
	\put(73.5, 16.5)  {\tiny   $14$}	
	\put(71,   16.0)  {\tiny   $17$}	
	\put(69.5, 14.8)  {\tiny   $22$}		
	\put(68.,  13.0)  {\tiny   $27$}
	\put(64.5, 11.0)  {\tiny $x=32$}	
   	\end{overpic}
  }    
  \caption{$(a)$ Profiles of streamwise velocity for $\Gamma=0.5$, $\Rey=500$, with inflexion points shown as dots.
  $(b)$  Temporal and $(c)$ spatial growth rates obtained from local stability analysis.
  }
  \label{fig:localstab0}
\end{figure}

\begin{figure}
  \psfrag{oom}[t][]{$\omega$}
  \psfrag{x}  [t][]{$x$}
  \psfrag{xgrowth} [t][]{$x$}
  \psfrag{om}   [r][][1][-90]{$\omega$}	
  \psfrag{-kis} [][][1][-90]{$-k_i^{(S)}\,\,$}	
  \psfrag{-ki}  [][][1][-90]{$-k_i^{(S)}\,\,$}
  \psfrag{Gi}   [][][1][-90]{}	
  \psfrag{exp(-int(ki(x)dx))} [][]{} 
  \centerline{
    \hspace{0.1cm}
   	\begin{overpic}[width=13.cm,tics=10]{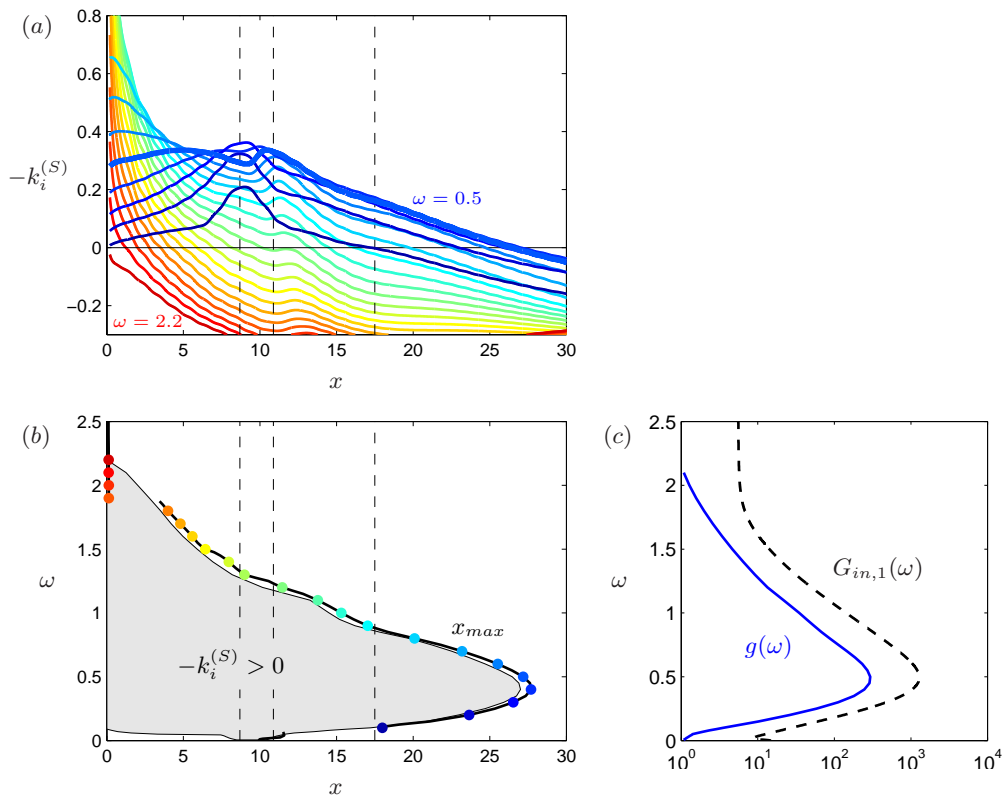}
   	\put(-1, 75.5)  {$(a)$}
   	\put(39,   58.0)  {\scriptsize\textcolor{blue}{$\omega=0.5$}}
   	\put( 8.4, 45.4)  {\scriptsize\textcolor{red} {$\omega=2.2$}}
	\put(-1, 33.5)  {$(b)$}
   	\put(43, 14)  {\textcolor{black}{$x_{max}$}}
	\put(15, 10.){\textcolor{black}{$-k_i^{(S)}>0$}}
	\put(58.5, 33.5)  {$(c)$}
   	\put(82, 20)  {$G_{in,1}(\omega)$}
	\put(73, 12)  {\textcolor{blue}{$g(\omega)$}}
   	\end{overpic}
  }    
  \vspace{0.3cm}
  \caption{Local stability analysis for $\Gamma=0.5$, $\Rey=500$.
$(a)$~Streamwise evolution of the spatial growth rate 
$-k_i^{(S)}$ 
at different frequencies.
$(b)$~Region of local instability $-k_i^{(S)}>0$ (shaded area) 
and location of the maximum of the optimal response energy density $x_{max}$ (solid line and symbols).
$(c)$~Integral amplification factor $g$ (solid line) 
and optimal harmonic gain $G_{in,1}$ (dashed line, reproduced from fig.~\ref{fig:Gin_Gvol_gamma05_re500}$(a)$).
The thick line in $(a)$ corresponds to the most amplified frequency $\omega=0.5$.
  }
  \label{fig:localstab}
\end{figure}

Interesting complementary information can be obtained from a local linear stability analysis, assuming the flow is parallel. 
At each streamwise location, the Orr-Sommerfeld equation was first solved for the \textit{temporal} problem, the eigenvalue problem of the complex frequency $\omega^{(T)}$ for a given real streamwise wavenumber $k$. 
We found at most one unstable eigenmode for all values of $x$ and $k$. 
Next, the same Orr-Sommerfeld equation was solved for the \textit{spatial} problem,  the eigenvalue problem of the complex wavenumber $k^{(S)}$ for a given real frequency $\omega$.
The identification of the correct eigenvalue in the spatial problem is difficult in general. 
In order to make this process easier,
 Gaster's relation $-k_i^{(S)}=\omega_i^{(T)}/c_g$ \citep{Gas62} was used to estimate the spatial growth rate $-k_i^{(S)}$
from the available temporal growth rate $\omega_i^{(T)}$ and group velocity $c_g=\partial \omega_r^{(T)}/\partial k_r^{(T)}$, 
 close to the neutral curve where $k_i^{(S)}=\omega_i^{(T)}=0$. 
The eigenvalue of interest was then identified and  followed while varying $\omega$.
Figure~\ref{fig:localstab0} shows the temporal and spatial growth rates at several locations. 
The flow is unstable between 
$x=0$ and $x=27$, which corresponds to the region where streamwise velocity profiles  contain inflection points (fig.~\ref{fig:localstab0}$(a)$). 
The temporal growth rate (fig.~\ref{fig:localstab0}$(b)$) is decreasing  monotonously with $x$, as shear gradually weakens downstream;
in contrast, the spatial growth rate (fig.~\ref{fig:localstab0}$(c)$) is also globally decreasing with $x$ but not monotonously at all frequencies, consistent with results from 
\cite{Kai96} for similar geometry and flow conditions ($\Gamma\simeq0.49$, $\Rey\simeq510$).
This is best seen in figure~\ref{fig:localstab}$(a)$, showing that 
$-k_i^{(S)}(x)$ has a local maximum around 
$x \simeq 9-11$, i.e. between the upper separation and the lower reattachment. 
The spatially unstable domain in the $x-\omega$ plane is summarised in figure~\ref{fig:localstab}$(b)$.
As frequency increases, the unstable region  quickly widens, from  $x\simeq9$-11 at $\omega=0$, 
to a long region extending downstream up to $x=27$ at $\omega=0.5$,
before shrinking back towards the step corner $x=0$, until the flow finally becomes stable everywhere for high frequencies $\omega\geq2.2$.
The downstream neutral curve $-k_i^{(S)}=0$ is  followed closely by 
the location $x_{max}$ where  the  energy density 
of the global optimal harmonic response 
is maximal,  
\be 
x_{max} = \arg \max_x \int ||\uu_1(x,y)||_2^2 \,\mathrm{d}y,
\ee 
consistent with the idea that perturbations grow spatially as long as 
$-k_i^{(S)}>0$, and then decay.
This is also reminiscent of the observation from \cite{Sipp13} in a flat-plate boundary layer, where the optimal response shows a peak at $x_{max}$ close to the downstream neutral curve obtained from local stability analysis.

One can quantify more precisely how much perturbations are amplified as they propagate downstream. To this aim, we compute for each frequency the integral amplification factor 
\be
g(\omega) = \exp{\left( \int -k_i^{(S)}(\omega,x) \,\mathrm{d}x \right)}
\label{eq:IAF}
\ee
over the  unstable region where $-k_i^{(S)}(\omega)>0$.
The integral amplification factor shown in figure~\ref{fig:localstab}$(c)$ (solid line) is maximum close to $\omega=0.5$.
This is in very good agreement with the global optimal harmonic gain $G_{in,1}(\omega)$ (dashed line), also reported  for qualitative comparison. 
Note that in this local analysis, $\omega=0.5$ is not the most unstable frequency at all locations: high frequencies display a much larger growth rate close to the step, as seen in fig.~\ref{fig:localstab}$(a)$; 
however, perturbations at $\omega=0.5$ do show the largest growth rate further downstream and  take advantage of the longest possible unstable region, resulting in the largest integral amplification factor.
The  shapes of  $g(\omega)$
and $G_{in,1}(\omega)$ are similar, too, except for $\omega \gtrsim 2$; 
in this frequency range the parallel flow is locally stable, 
whereas in the global flow non-parallel effects and component-type non-normality (\eg  Orr mechanism) are at work.
Overall, figures~\ref{fig:localstab}$(b)$-$(c)$ indicate that the  agreement between local and global stability analyses is remarkable,
both in terms of  instability/amplification region and in terms of most amplified frequency.
Note also that \cite{bla08} obtained a maximum mean perturbation energy at $x=26-27$ in two and three-dimensional calculations, 
both in the optimal perturbations obtained from transient growth analysis and
in a direct numerical simulation forced with small-amplitude perturbations (Gaussian white noise) at the inlet; 
this is consistent with \cite{Kai96} and in very good agreement with our spatial and global analyses.

It is worth comparing further the results of the present harmonic response analysis with those of the transient growth analysis of \cite{bla08}.
These authors reported a centre frequency $\omega=0.55$ in the spectral density of their forced three-dimensional direct numerical simulation, close to the most amplified frequency obtained in our harmonic response analysis.
This frequency was identical to that of wave packets of streamwise wavelength 3.73 (measured in optimal perturbations from transient growth analysis) advected at the mean downstream channel velocity $U_\infty/3$.
Finally, the optimal initial perturbation 
in their transient growth analysis is strikingly similar to our optimal harmonic volume forcing at $\omega=0.5$ (fig.~\ref{fig:opt_forc_resp_in_vol_gamma05_re500}$(a)$).
Although responses to initial perturbations, to sustained harmonic forcing and to sustained white noise
are three different processes,  the above-mentioned similarities point to the close relationship between the physical amplification phenomena involved in these processes.

\section{Sensitivity and control of harmonic and stochastic amplification}
\label{sec:sensit}

As shown in section~\ref{sec:harm_resp}, the backward-facing step flow is a strong amplifier, with a potential for large amplification of harmonic and stochastic forcing.
The objective of this section is to investigate how this amplification can be reduced by steady control. 
We stress  that \textit{forcing} is considered as an external perturbation, generally unwanted, whereas \textit{control} is applied on purpose with the aim of reducing the amplification of the forcing. Hereafter, forcing at the inlet is considered.

We use the method presented in section~\ref{sec:method_sensitivity} to compute the sensitivity of the optimal harmonic/stochastic gain  
to steady flow modification and steady control.
Sensitivity maps provide quantitative and qualitative  information (sensitivity values, regions of largest sensitivity) that are useful to design efficient control configurations in the volume (section~\ref{sec:volume}) or at the wall (section~\ref{sec:wall}).

\subsection{Volume control}
\label{sec:volume}

\subsubsection{Flow modification}
\label{sec:flowmodif}

\begin{figure}
  \psfrag{x}[t][]{$x$}
  \psfrag{y}[r][][1][-90]{$y$}	
  \centerline{
   	\begin{overpic}[width=13.cm,tics=10]{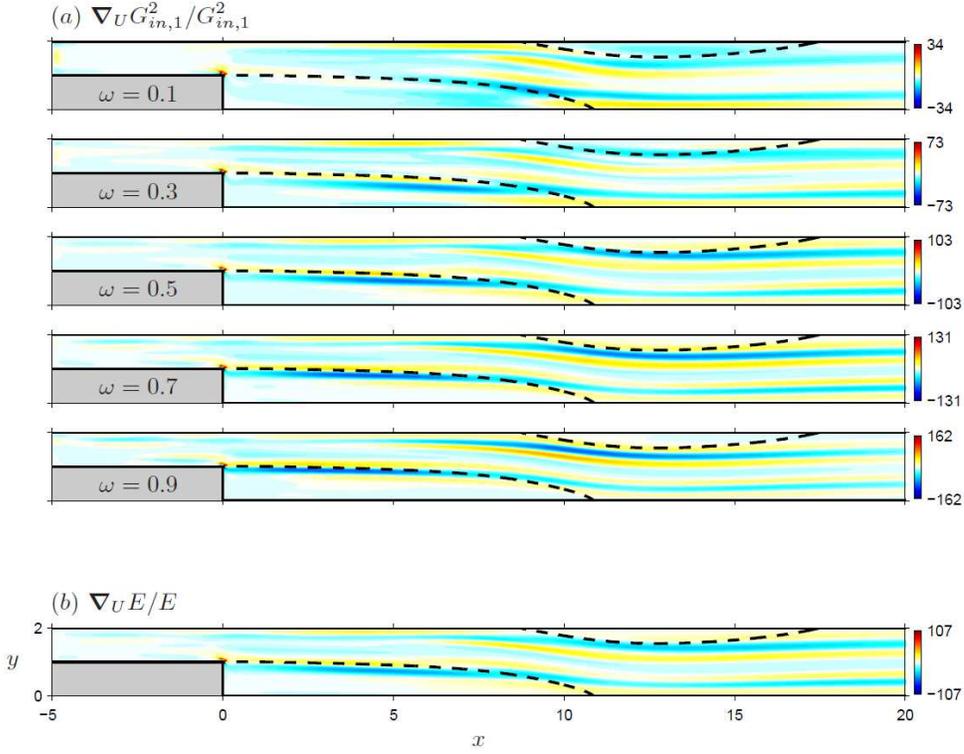}
   	\end{overpic}
  }    
 \caption{Sensitivity of  $(a)$ optimal harmonic gain and  $(b)$  stochastic gain
 to base-flow modification (streamwise component $\bnabla_U  = \ex\bcdot\bnabla_\UU $).  $\Gamma=0.5$, $\Rey=500$.
}
   \label{fig:DU_gamma05_re500}
\end{figure}

Figure~\ref{fig:DU_gamma05_re500} shows sensitivities to base-flow modification in the streamwise direction: regions of positive (resp. negative) sensitivity indicate where  a small-amplitude increase of streamwise velocity would increase (resp. decrease) the gain.
The optimal harmonic gain $G_{in,1}$ is most sensitive at the step corner, and in elongated regions parallel to  the direction of the main flow (fig.~\ref{fig:DU_gamma05_re500}($a$)). 
The structure of these positive/negative sensitivity regions indicates that, for most frequencies, increasing shear in the shear layers near the boundaries of the upper/lower recirculation regions would result in a larger gain. Further downstream, increasing shear at the $y$ location of the base flow inflection points would also result in a larger gain.
This is consistent with the global amplification process being linked to the local shear instability mechanism.
In contrast, some regions display a sensitivity which changes sign with $\omega$, meaning that a  modification of the base flow in such a region  would increase  $G_{in,1}$ at some frequencies and decrease it at others.
 \cite{Bou13} made the same observation for the flow past a wall-mounted bump, and  empirically proposed to design their control based on the optimal frequency alone.
This approach appears justified for the present $\Gamma-\Rey$ conditions since the sensitivity of the stochastic gain $E$ (fig.~\ref{fig:DU_gamma05_re500}($b$)) is essentially similar to that of $G_{in,1}$ at the optimal frequency $\omega_0=0.5$. Therefore, reducing the response to stochastic forcing is tantamount to reducing the optimal harmonic gain at the optimal frequency, which brings substantial benefits in terms of simplicity and computational cost.

\subsubsection{Volume control}
\label{sec:volumeforce}

\begin{figure}  
  \psfrag{x}[t][]{$x$}
  \psfrag{y}[r][][1][-90]{$y$}	
  \centerline{
   	\begin{overpic}[width=13.cm,tics=10]{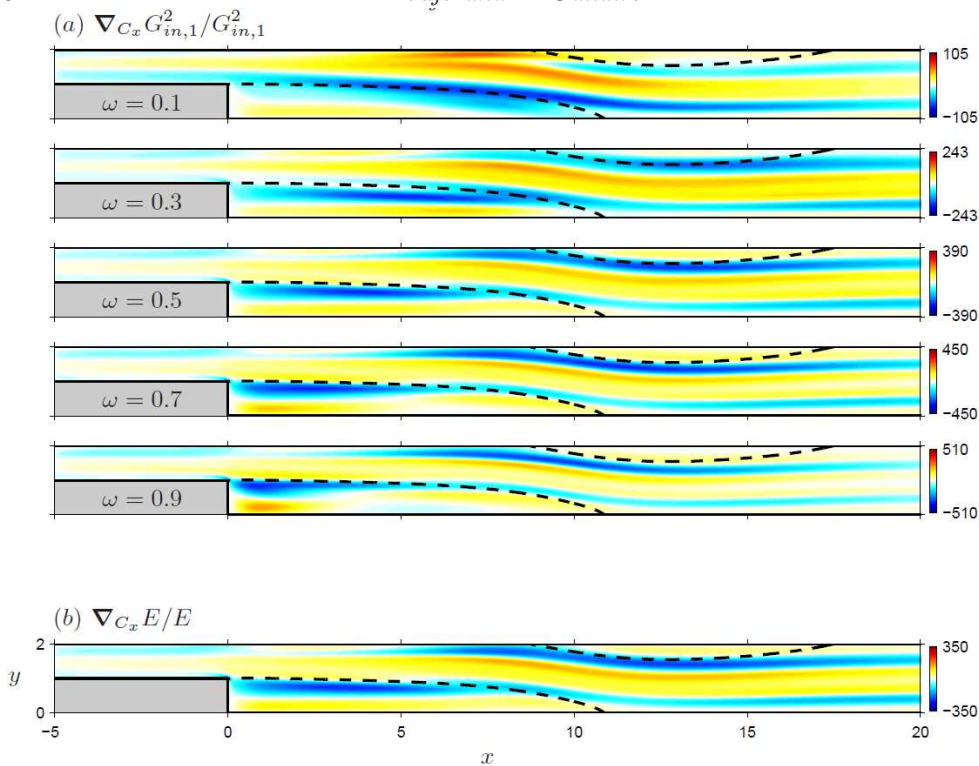}
   	\end{overpic}
  }    
 \caption{Sensitivity of $(a)$ optimal harmonic gain and $(b)$ stochastic gain
 to volume control (streamwise component $\bnabla_{C_x}  = \ex\bcdot\bnabla_\CC  $).  $\Gamma=0.5$, $\Rey=500$.
}
   \label{fig:DF_gamma05_re500}
\end{figure}

Figure~\ref{fig:DF_gamma05_re500} shows in the same way
sensitivities  to streamwise volume control:
 regions of positive (resp. negative) sensitivity indicate where  a small-amplitude body force pointing in the $+x$  direction  would increase (resp. decrease) the gain.
Again, elongated regions of large sensitivity are found along the shear layers, but also in the lower recirculation zone, while the step corner is less sensitive.
For the optimal harmonic  gain (fig.~\ref{fig:DF_gamma05_re500}$(a)$), variations with $\omega$ are still present, for instance near the upper wall  upstream of the upper separation point.
In this case too, the stochastic gain  exhibits a sensitivity (fig.~\ref{fig:DF_gamma05_re500}$(b)$) dominated by that of the optimal harmonic  gain at the optimal frequency.

\subsubsection{Control cylinder}
\label{sec:cylinder}

In practice, it is difficult to generate arbitrary volume forces in the flow. It is more common, both experimentally  and numerically, to use open-loop devices to control, for instance, aerodynamic forces or vortex shedding  \citep{Stry90,Iga97,Mit01,Dal01,Cad09,Par09,Par12}.
Sensitivity analysis is well suited to estimate the effect of a wire, i.e. a small control cylinder of diameter $d$, using (i) the sensitivity to volume forcing already computed, and (ii) a simple model for the control force $\CC$ exerted by the cylinder on the flow, namely a force opposite to the drag felt by the control cylinder in a steady uniform flow at the local velocity $\UU_b(x,y)$ \citep{Hill92AIAA,Marquet08cyl,Mel10,Pralits10,Fani12}:
$\bdelta \CC(x,y) = -\frac{1}{2} d C_d(x,y) ||\UU_b(x,y)|| \UU_b(x,y) \delta(x-x_c,y-y_c)$,
where $\delta$ is the two-dimensional Dirac delta function, and the drag coefficient $C_d$  depends on the local Reynolds number $\Rey_d(x,y)=||\UU_b(x,y)|| d/\nu$.
Here we choose $d=0.05$, corresponding to Reynolds numbers $\Rey_d\leq25$ everywhere in the flow. We therefore use
the following composite expression for the drag coefficient:
\begin{align}
\displaystyle
0.5\leq \Rey_d \leq25: \quad
& C_d = a+b \Rey_d^c, 
a=0.85, \, b=10.6, \, c=-0.72, 
\label{eq:cd1}
\\
 \Rey_d\leq0.5: \quad&
C_d = \frac{8\pi}{\Rey_d S} \left( 1-\frac{\Rey_d^2}{32} \left(S-\frac{1}{2}+\frac{5}{16S}\right) \right), 
\label{eq:cd2}
\end{align}
where 
$S=\frac{1}{2}-\gamma-\log\left(\Rey_d/8\right)$ and  
$\gamma$ is Euler's constant.
Expression (\ref{eq:cd1}) is a fit of experimental data from \cite{Tri59} and in-house numerical results, while (\ref{eq:cd2})  is an extension of Oseen's formula 
$C_d = 8\pi/\Rey_d S$
for low Reynolds numbers \citep{Oseen10,Proudman57}  derived by \cite{Tomo51}.
In practice, the exact value at low Reynolds number is of little importance: although the drag coefficient goes to infinity 
like 
$ \sim 1/(\Rey_d \log\Rey_d)$ as $\Rey_d \rightarrow 0$ as an artificial consequence of the aerodynamic definition of $C_d$,
the actual force exerted on the cylinder goes to zero like 
$\sim \Rey_d / \log\Rey_d$.

\begin{figure}
  \psfrag{x}[t][]{$x$}
  \psfrag{y}[r][][1][-90]{$y$}	
  \centerline{
   	\begin{overpic}[width=13.cm,tics=10]{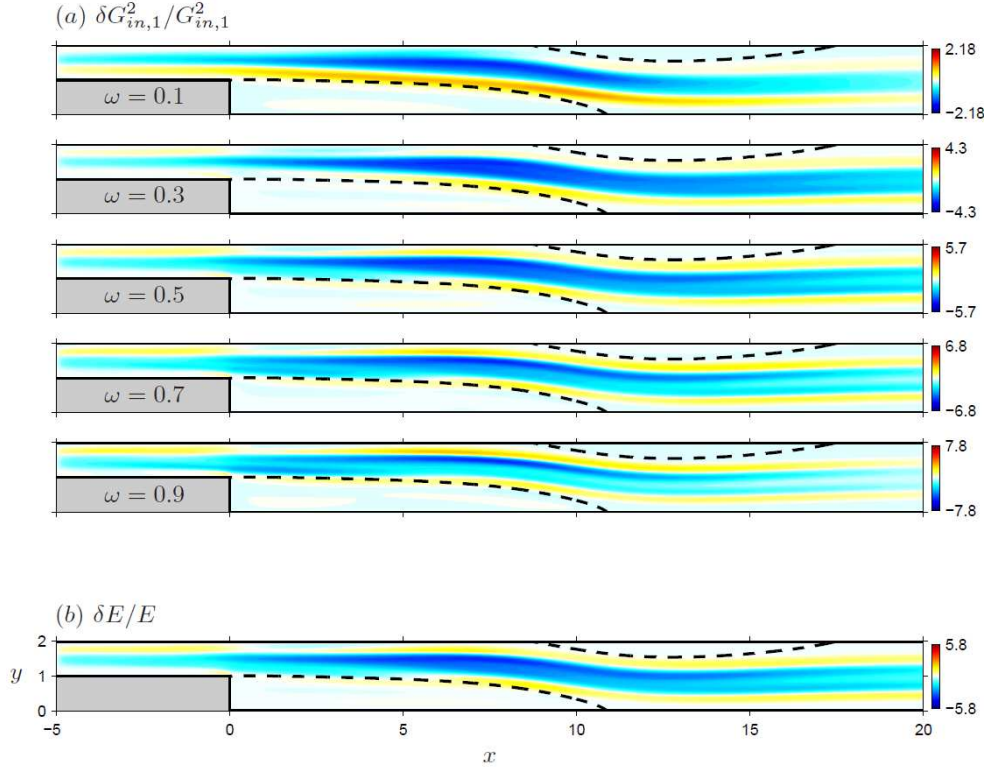}
   	\end{overpic}
  }    
 \caption{Effect of a small control cylinder of diameter $d=0.05$ on 
 $(a)$ optimal harmonic gain and  $(b)$ stochastic gain. $\Gamma=0.5$, $\Rey=500$.
}
   \label{fig:SmallCyl}
\end{figure}

The effect of a small control cylinder is shown in figure~\ref{fig:SmallCyl}. 
The amplification of stochastic noise is best reduced when inserting the control cylinder  in an elongated region extending from the step to the upper reattachment point, where the main stream velocity is large.
Conversely, noise amplification increases when the control cylinder is inserted in the outer vicinity of recirculation regions, where shear is large.
Again, the basic mechanism is shear strengthening (resp. shear weakening) due to the cylinder's wake in the main stream (resp. in the shear layers).
Recirculation regions themselves have no significant effect as a result of their low velocities.

\begin{figure}
\vspace{0.5cm}
  \centerline{
   	\begin{overpic}[width=13.cm,tics=10]{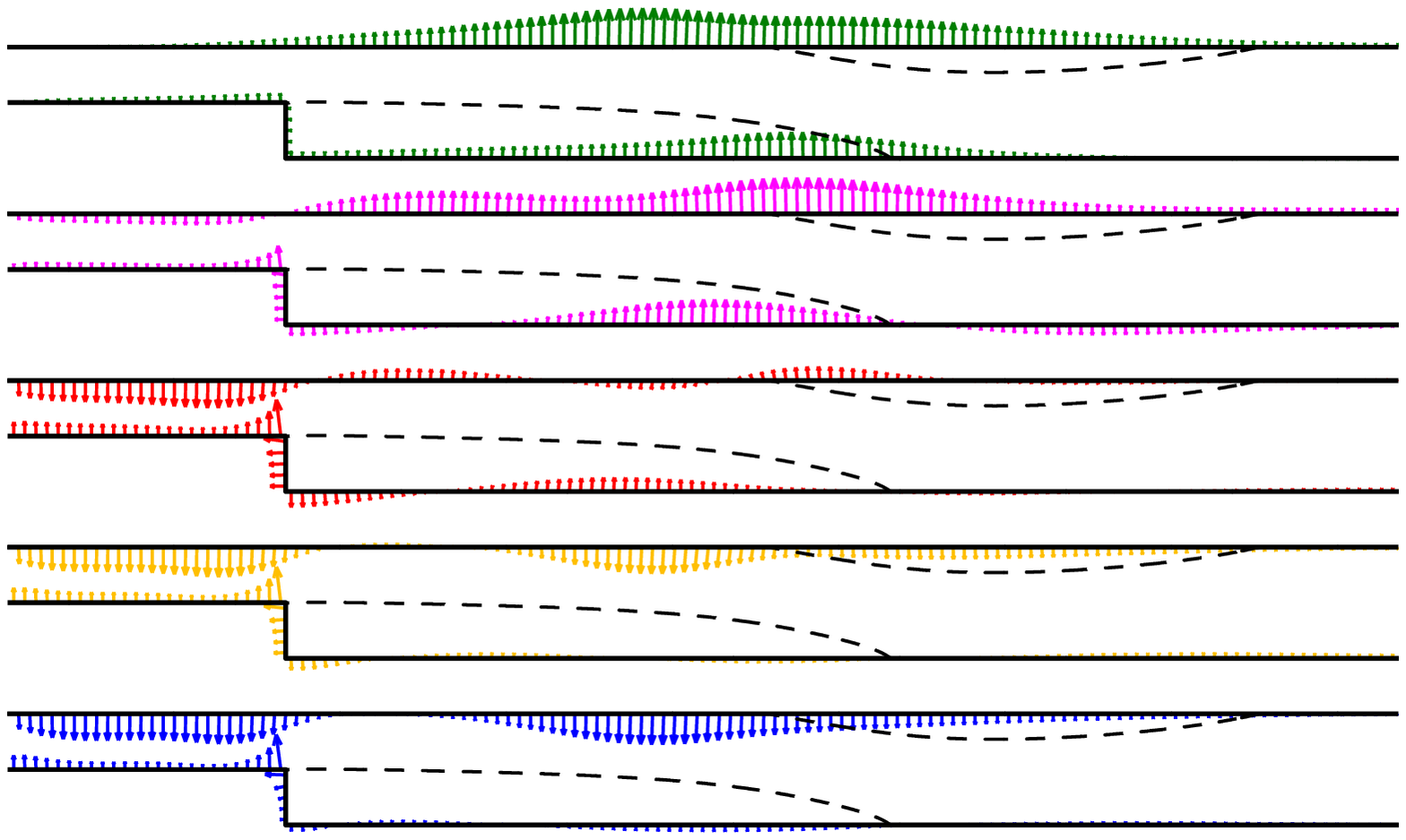}
   	\put( 0,59)  {$(a)$ $\bnabla_{\UU_c} G_{in,1}^2$}   	
   	\put(6,49.4)  {\footnotesize$\omega=0.1$}
   	\put(6,37.4)  {\footnotesize$\omega=0.3$}
   	\put(6,25.5)  {\footnotesize$\omega=0.5$}
   	\put(6,13.5)  {\footnotesize$\omega=0.7$}
   	\put(6, 1.5)  {\footnotesize$\omega=0.9$}
   	\put(75,49.30)  {\footnotesize$\max ||\bcdot||_2=3.2\times10^4$}
   	\put(75,37.45)  {\footnotesize$\max ||\bcdot||_2=1.1\times10^7$}
   	\put(75,25.50)  {\footnotesize$\max ||\bcdot||_2=1.3\times10^8$}
   	\put(75,13.55)  {\footnotesize$\max ||\bcdot||_2=5.0\times10^7$}
   	\put(75, 1.60)  {\footnotesize$\max ||\bcdot||_2=6.8\times10^6$}
   	\end{overpic}
  }
  \vspace{1.2cm}
  \centerline{
  	\begin{overpic}[width=13.cm,tics=10]{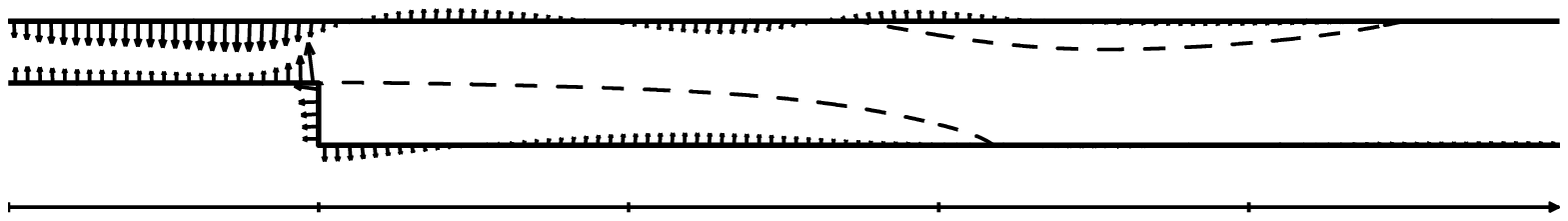}  	
   	\put( 0,15)  {$(b)$ $\bnabla_{\UU_c} E$}
   	\put(49,-4)  {$x$}
   	\put(75,5.5)  {\footnotesize$\max ||\bcdot||_2=1.3\times10^7$}
    \put(-1.3, -2.) {\footnotesize$-5$}
    \put(19.5, -2.) {\footnotesize$ 0$}
    \put(39.4, -2.) {\footnotesize$ 5$}
    \put(58.5, -2.) {\footnotesize$10$}
    \put(78.5, -2.) {\footnotesize$15$}
   	\end{overpic}
  }    
 \vspace{0.5cm}
 \caption{Sensitivity to wall control of  $(a)$ the optimal harmonic gain
 and  $(b)$ the stochastic gain.  $\Gamma=0.5$, $\Rey=500$. 
 Arrows point in the direction of positive sensitivity (i.e. increasing gain).
}
   \label{fig:DUw_gamma05_re500}
\end{figure}

\subsection{Wall control} 
\label{sec:wall}

\subsubsection{Sensitivity maps}
\label{sec:wallsensit}

Sensitivities to wall control  are shown in figure~\ref{fig:DUw_gamma05_re500}. 
Arrows point in the direction of positive sensitivity: at each point of the wall,  blowing or suction in the direction of the corresponding arrow  would increase the gain.
More generally, an actuation direction whose scalar product 
with the sensitivity is positive would increase the gain, while an actuation direction orthogonal to the sensitivity would have no first-order effect. 
The sensitivity of the harmonic optimal gain (fig.~\ref{fig:DUw_gamma05_re500}$(a)$) is essentially normal to the wall and, again, changes sign with frequency, except notably at the lower wall of the inlet channel.
The sensitivity of the stochastic gain (fig.~\ref{fig:DUw_gamma05_re500}$(b)$), once again mostly dominated by that of the optimal harmonic  gain at the optimal frequency, is maximum just  upstream of the step corner.
Given the sign of the sensitivity, wall suction at this location   should reduce noise amplification. 
Note that the step corner is often chosen for wall control in this flow and similar ones~\citep{Pas05,Bea06,Hen07,Her12}, but other locations have a sensitivity of comparable magnitude, like the vertical wall (where blowing should reduce $E$) and the  upper wall of the inlet channel (where suction should reduce $E$). These locations could offer interesting alternatives to the step corner depending on technical feasibility constraints.

\subsubsection{Validation}
\label{sec:valid}

\begin{figure}
\vspace{0.5cm}
  \centerline{
   	\begin{overpic}[width=10 cm,tics=10]{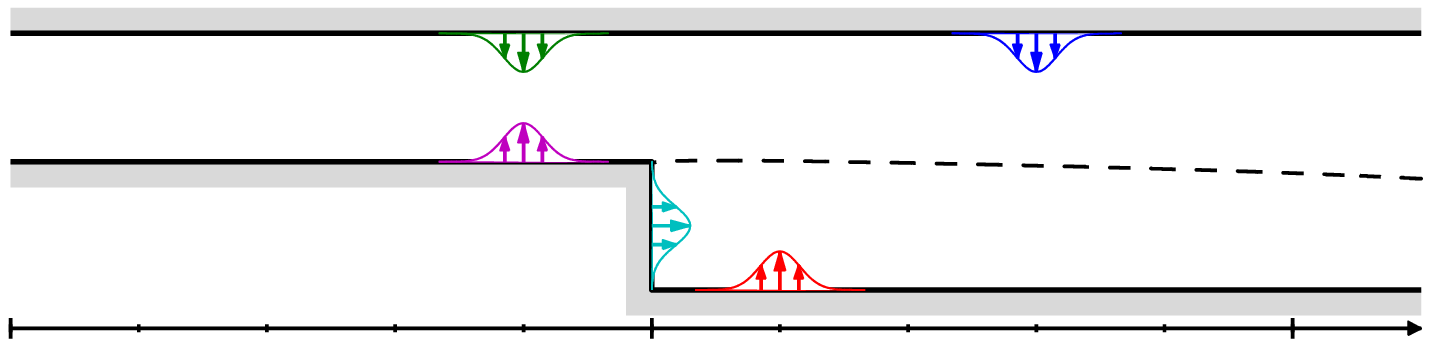}
    \put(1,   26)   {$(a)$}
   	\put(28., 14.)  {\textcolor[rgb]{0.75,0,   0.75}{$W_1$}}
   	\put(48.5, 8)   {\textcolor[rgb]{0,   0.75,0.75}{$W_2$}}
   	\put(57,   5.)  {\textcolor[rgb]{1,   0,   0}   {$W_3$}}
   	\put(39,  17.5) {\textcolor[rgb]{0,   0.5, 0}   {$W_4$}}
	\put(75,  17.5) {\textcolor[rgb]{0,   0,   1}   {$W_5$}}		 
    \put(-2,  -3)   {\footnotesize$ -5$}
    \put(44.5,-3)   {\footnotesize$ 0$}          
    \put(89,  -3)   {\footnotesize$ 5$}
 	\put(50,  -6)   {$x$}
   	\end{overpic}   	
  }
  \vspace{1.cm}
  \centerline{
    \psfrag{om}[t][]{$\omega$}
    \psfrag{W}[t][]{flow rate $W$}
    \psfrag{W4}[t][]{$W_4$}
    \psfrag{G1}[r][][1][-90]{$k=1$}	
    \psfrag{G2G3}[r][][1][-90]{$k=2,3$}	
    \psfrag{GW4}[r][][1][-90]{ }
    \psfrag{GW4**2}[r][][1][-90]{ }	
    \psfrag{k=1}[r][][1][-90]{}	
    \psfrag{k=2,3}[r][][1][-90]{}	
    \psfrag{E/E0}[r][][1][-90]{ }	
    \hspace{-0.6cm}
   	\begin{overpic}[height=4.9cm,tics=10]{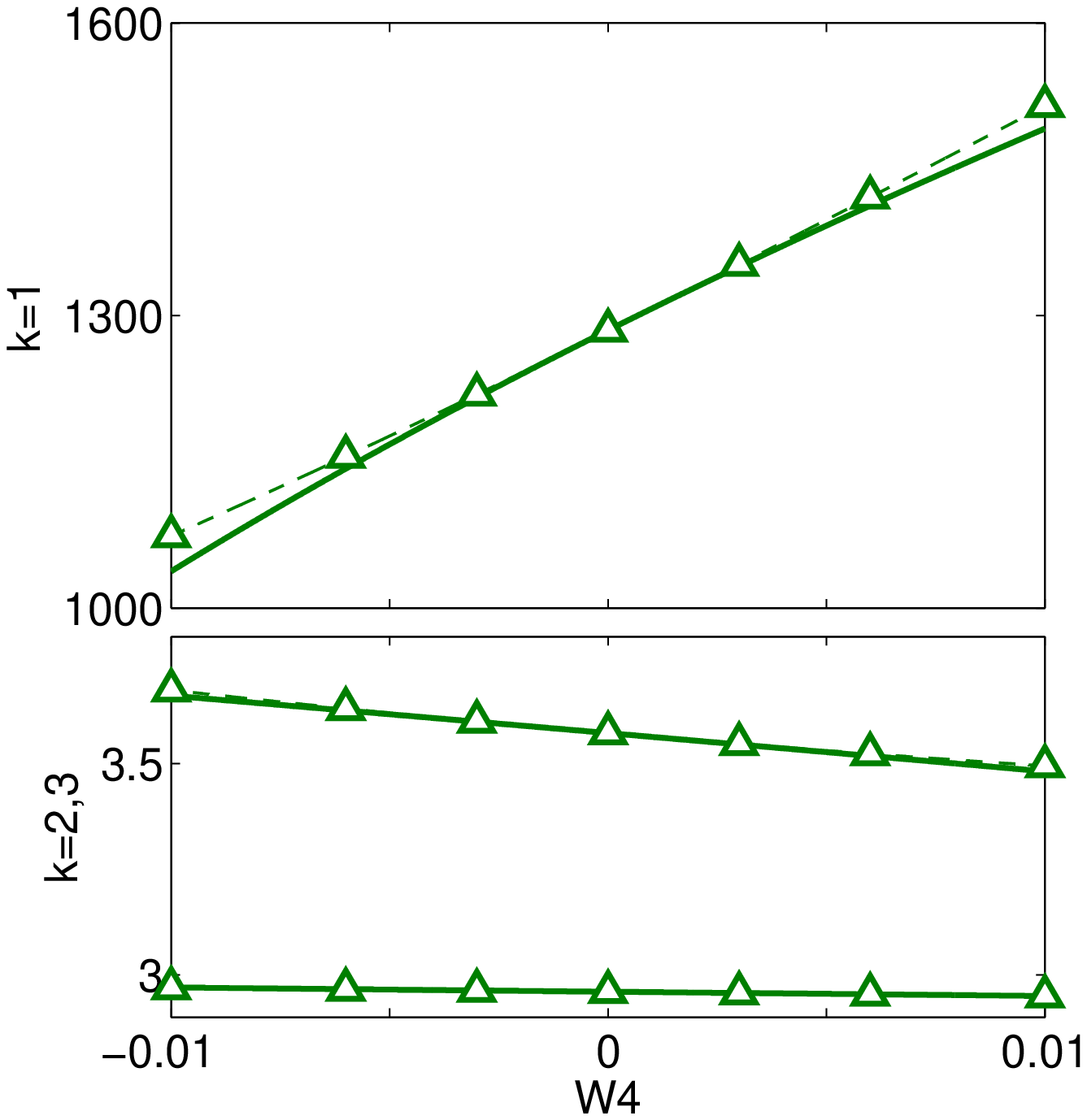}
   	\put(18,90) {$(b)$ $G_{in,k}(\omega=0.5)$}
   	\put(75,74) {\footnotesize $k=1$}
   	\put(75,36) {\footnotesize $k=2$}
   	\put(75,15) {\footnotesize $k=3$}
   	\end{overpic}
   	\hspace{-0.2cm}
   	\begin{overpic}[height=5.08cm,tics=10]{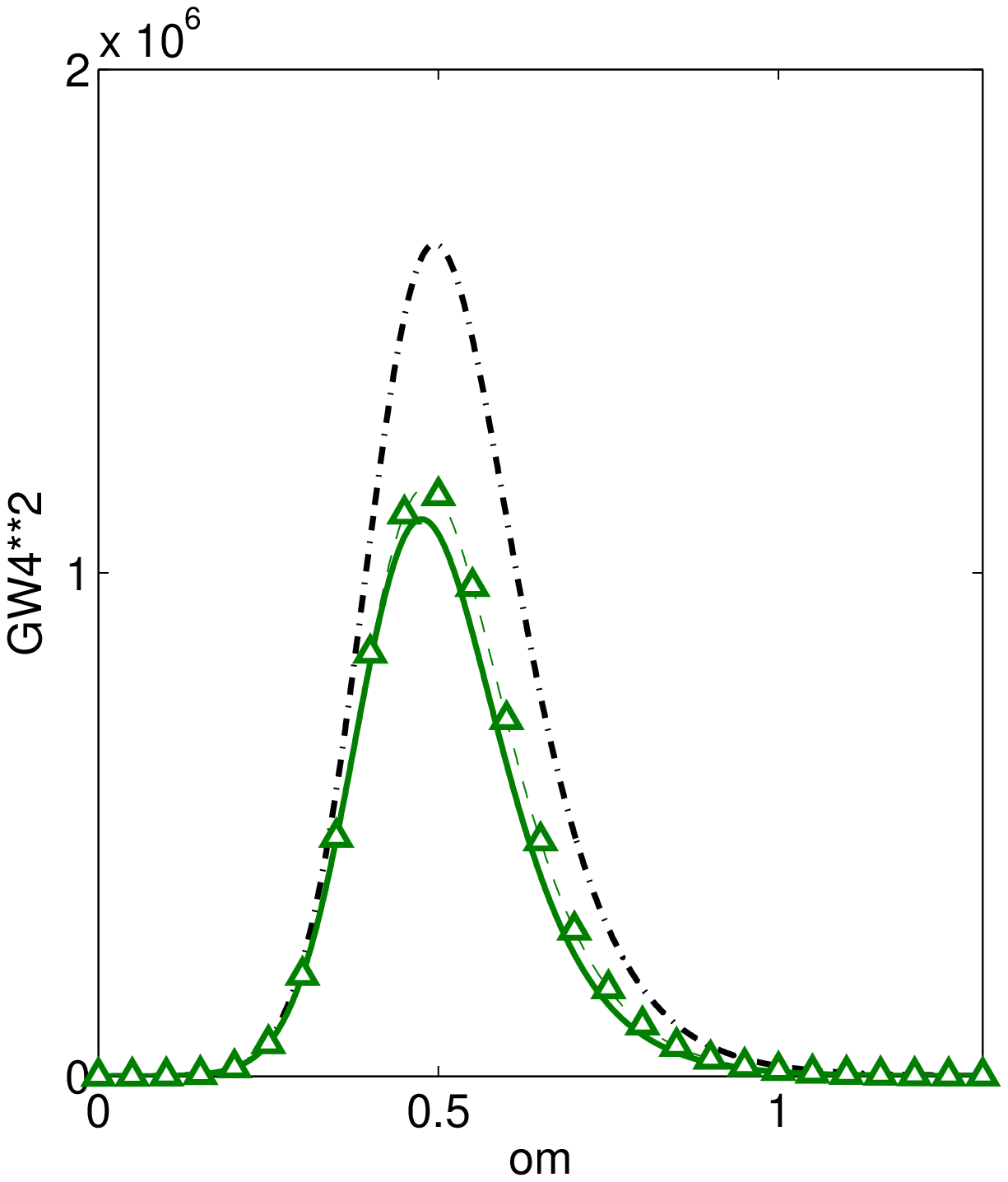}
   	\put(11, 86.5) {$(c)$ $G_{in,1}^2$}
   	\put(45, 62) {\footnotesize                          $W=0$}
  	\put(30, 20) {\footnotesize \textcolor[rgb]{0,0.5,0}{$W_4=$}}
  	\put(30, 14) {\footnotesize \textcolor[rgb]{0,0.5,0}{$-0.01$}}
   	\end{overpic}
   	\hspace{-0.2cm}
   	\begin{overpic}[height=4.9cm,tics=10]{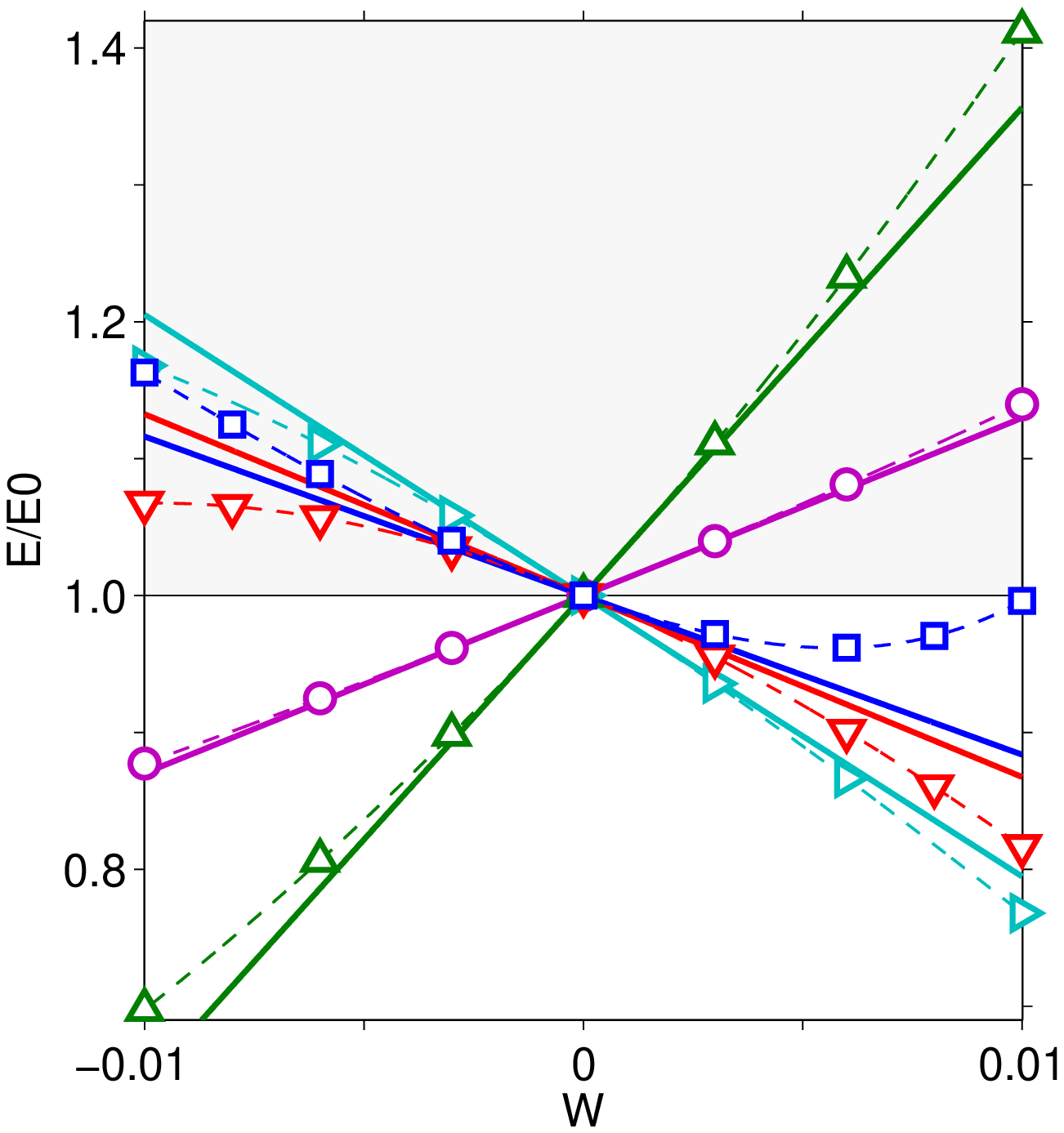}
   	\put(15, 90) {$(d)$ $E/E_0$}
   	\put(15, 36){\footnotesize \textcolor[rgb]{0.75,0,0.75}{1}}
   	\put(15, 19){\footnotesize \textcolor[rgb]{0,0.5,0}{4}}
   	\put(92, 18){\footnotesize \textcolor[rgb]{0,0.75,0.75}{2}}
   	\put(92, 24){\footnotesize \textcolor[rgb]{1,0,0}{3}}
   	\put(92, 45){\footnotesize \textcolor[rgb]{0,0,1}{5}}
   	\end{overpic}   	
  }
 \caption{
 Wall control with steady Gaussian 
 blowing/suction for $\Gamma=0.5$, $\Rey=500$.
 $(a)$~Sketch of control configurations 1 to 5. 
 Blowing corresponds to positive flow rates ($W>0$), 
 suction             to negative flow rates ($W<0$).
 $(b)$~Variation of harmonic gain at $\omega=0.5$ with flow rate in configuration 4 
 ($x_c=-1$, $y_c=2$).
 $(c)$~Overall optimal harmonic gain reduction in control configuration 4 with suction flow rate $W_4=-0.01$.
 $(d)$~Variation of stochastic gain (normalised by its uncontrolled value) with flow rate.
Solid lines show predictions from sensitivity analysis, symbols are actual results from controlled flows.
}
   \label{fig:wall_actuation_all}
\end{figure}

We illustrate the effect of steady wall control on the harmonic/stochastic gain, and use this opportunity to validate the sensitivity analysis.
We consider several locations upstream and downstream of the step corner, both at the lower and upper walls, as represented in 
figure~\ref{fig:wall_actuation_all}$(a)$.
For the sake of simplicity, and since we observed that sensitivity to wall-normal control was much larger than that of tangential control, we  use blowing and suction in the normal direction only. 
We choose Gaussian actuation profiles
$ V_c(x)=(-\nn\bcdot\ey)\,W \exp(-(x-x_c)^2/\sigma_c^2)/(\sigma_c \sqrt{\pi})$ 
for control on horizontal walls,
and
$ U_c(y)=(-\nn\bcdot\ex)\,W \exp(-(y-y_c)^2/\sigma_c^2)/(\sigma_c \sqrt{\pi})$ 
for control on the vertical wall, with  characteristic width $\sigma_c=0.1$
 and flow rate $W$,   positive for blowing and negative for suction (recall $\nn$ points outward).
We compare  results from full gain calculations for flows actually controlled with the Gaussian actuation profile applied as a boundary condition, 
and predictions from sensitivity analysis  according to equations
(\ref{eq:def_D_Uc_G2})-(\ref{eq:def_D_E}), 
i.e.
$\delta G_{in,k}^2 = (\bnabla_{\UU_c} G_{in,k}^2 \,|\, \bdelta\UU_c)$
and  
$\delta E          = (\bnabla_{\UU_c} E          \,|\, \bdelta\UU_c)$.

Figure~\ref{fig:wall_actuation_all}$(b)$
shows the effect of steady blowing and suction at the upper wall upstream of the step (configuration 4) on the optimal harmonic  gain and the first two sub-optimal gains at the optimal frequency $\omega_0=0.5$.
Predictions from sensitivity analysis (solid lines) are in good agreement with full calculations (symbols), as could be expected for  small control amplitudes ($W$ less than 0.01, to be compared to the inlet flow rate 
$2 U_\infty h_{in}/3=2/3$). 
The effect of wall suction on the optimal gain at other frequencies is shown in 
figure~\ref{fig:wall_actuation_all}$(c)$.
As predicted by sensitivity (figure~\ref{fig:DUw_gamma05_re500}), the gain is reduced over the whole range of most amplified frequencies.
Moving to the stochastic gain in figure~\ref{fig:wall_actuation_all}$(d)$, we observe  a good agreement too: sensitivity analysis (solid line) captures well the reduction in $E$ from its uncontrolled value $E_0$ compared to actual results from full calculations (symbols).
In some control configurations, the actual variation of $E$ departs from sensitivity predictions as the flow rate becomes larger, due to non-linear effects not taken into account in the first-order linear sensitivity analysis based on the assumption of small-amplitude control. These non-linear effects are larger in configurations 3 and 5, i.e. for wall actuation downstream of the step.

\subsubsection{Towards passive wall actuation}
\label{sec:passive_wall}

Wall control allows us to choose the actuation direction (blowing or suction) and orientation (angle) freely. This is an advantage over volume control: typically, a small cylinder can only produce a force in the direction of the flow. 
The price to pay is a more complex actuation system  and a potentially higher power requirement to drive the control, 
 unless one takes advantage of pressure differences: connecting wall regions of  relative higher and lower pressure with a small channel could induce natural  suction and blowing at the inlet and at the outlet, respectively.
This configuration would not require any mechanical device and would therefore constitute a means of passive control.

\begin{figure}
 \hspace{-0.2cm}
 \centerline{
    \psfrag{p,DUw}[t][]{} 
    \psfrag{x}[t][]{$x$} 
    \psfrag{yy}[t][]{$y$} 
    \psfrag{y} [r][][1][-90]{$y\quad$}       
    \psfrag{om}[t][]{$\omega$} 
    \psfrag{G2} [r][][1][-90]{}    	 	
   	\begin{overpic}[height=5.6 cm,tics=10]{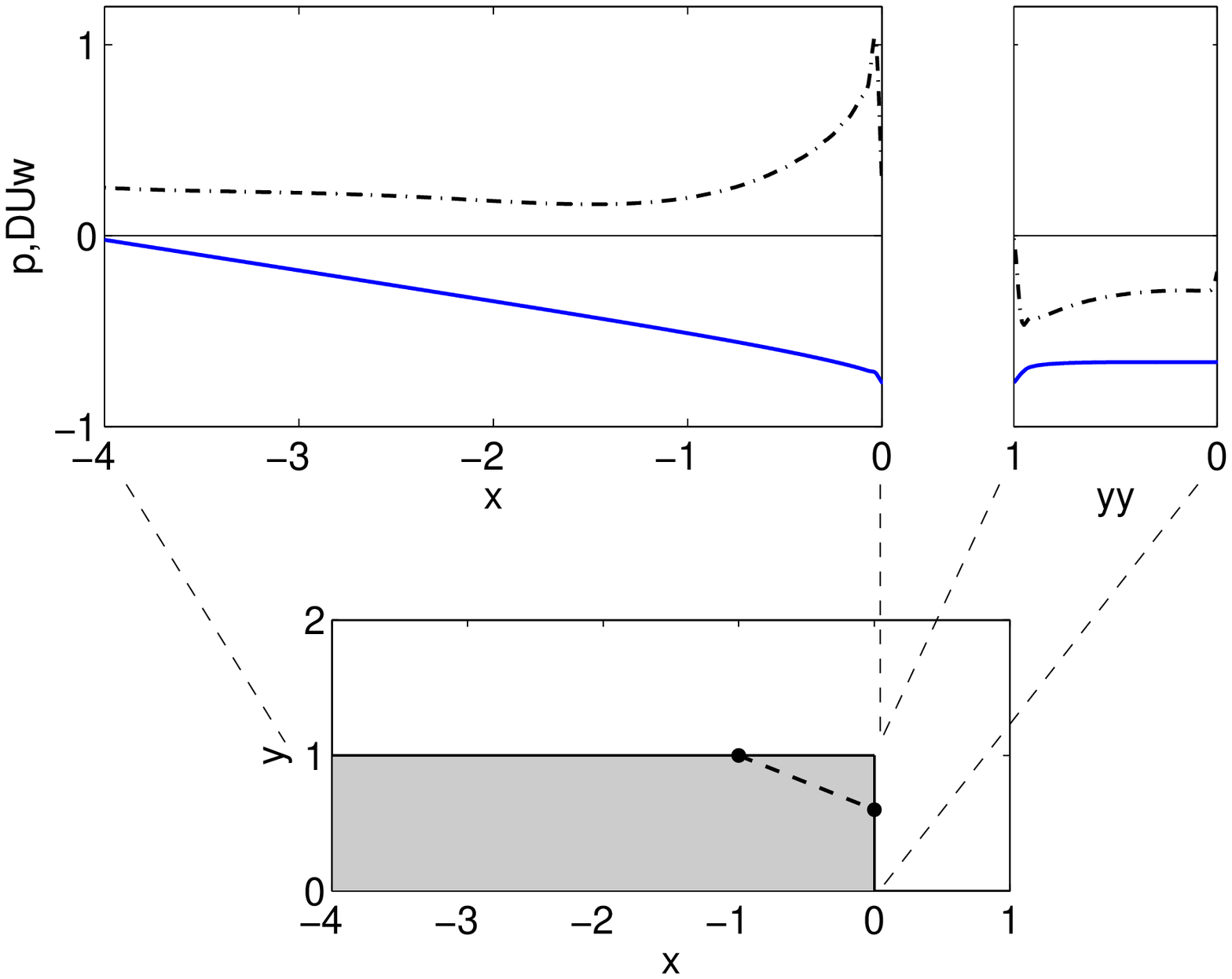}
   		\put(14, 71) {\footnotesize $(a)$}
   		\put(35, 65) {\footnotesize $\bnabla_{\UU_c}E \bcdot \nn/10^7$}
   		\put(30, 50) {\footnotesize \textcolor[rgb]{0,0,1}{$P$}}
   		\put(60, 20) {\footnotesize $A$}
   		\put(73.5, 14) {\footnotesize $B$}
   	\end{overpic}
   	\hspace{1cm}
   	\begin{overpic}[height=6 cm,tics=10]{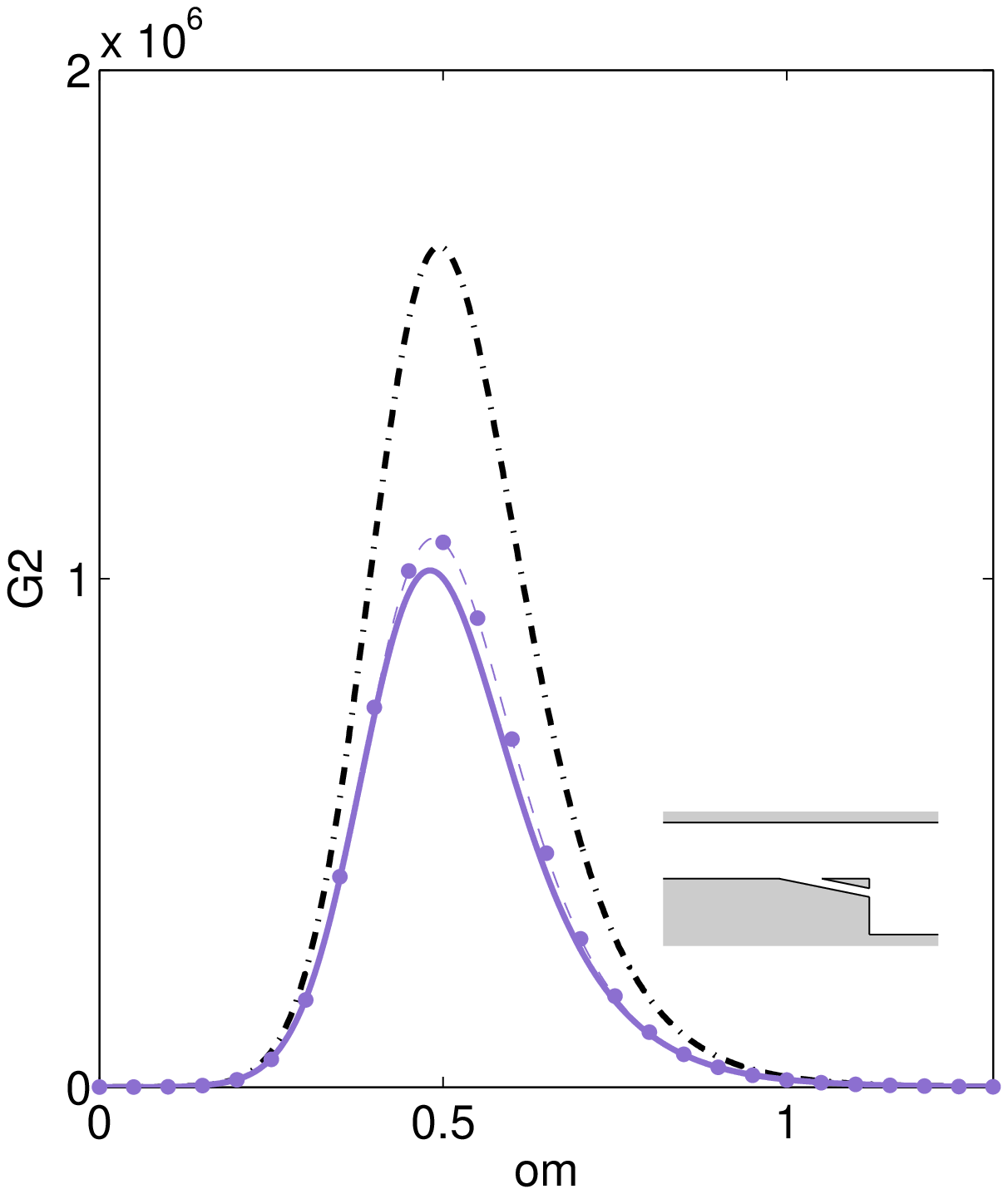}
   		\put(11, 87) {\footnotesize $(b)$ $G_{in,1}^2$}
   		\put(45, 60) {\footnotesize uncontrolled}
   		\put(33, 22) {\footnotesize \textcolor[rgb]{0.55,0.44,0.82}{$AB$}}
   		\put(29, 16) {\footnotesize \textcolor[rgb]{0.55,0.44,0.82}{channel}}
   	\end{overpic}
 } 
\caption{$(a)$ Passive  control  by means of a channel connecting regions of high and low pressure (solid line). The  induced flow results in wall suction in  $A$ and wall blowing in  $B$,  and reduces the  stochastic response $E$ as predicted by the sensitivity to  wall normal actuation (the dash-dotted line shows the normal component, much larger than the tangential component as observed in figure  \ref{fig:DUw_gamma05_re500}).
 $(b)$ Reduction in optimal harmonic gain for $x_A=-1$, $y_B=0.75$ and channel height $h_c=0.15$.
 Dash-dotted line: uncontrolled flow,
 solid line: estimation from (\ref{eq:dE_smallchannel})  (Poiseuille flow concentrated in $A$ and $B$),
symbols: actual optimal harmonic gain for the flow with channel.
 }
   \label{fig:smallchannel}
\end{figure}

This could be implemented in the backward-facing step flow between the lower wall 
upstream of the corner (suction at higher pressure) and the vertical wall (blowing at lower pressure):  
figure~\ref{fig:smallchannel}$(a)$ shows that pressure (solid line)  along the horizontal wall is larger than on the vertical wall, and that the sign of the sensitivity of $E$  (dash-dotted line) is such that the  stochastic gain can be reduced precisely with suction on the horizontal wall and blowing on the vertical wall.

A crude estimate of the expected reduction in $E$ can be obtained by assuming that connecting points $A$ and $B$ of coordinates $\xx_A=(x_A,h_s)$ and $\xx_B=(0,y_B)$ with a straight channel would result in a fully developed plane Poiseuille flow of mean velocity $U_{m} = \Rey \, h_c^2\,\Delta P/12 l_c$, where  
$\Delta P =P_A-P_B$ is the  pressure difference, and $h_c$ and $l_c=\sqrt{x_A^2+(h_s-y_B)^2}$ are the channel height and length. 
Assuming further that at both ends the induced flow is  localised at  points $A$ and $B$, the velocity vector is
\be
\bdelta\UU (\xx_{A}) = 
\bdelta\UU (\xx_{B}) 
= U_{m} \left( \begin{array}{c} \cos \theta \\ -\sin \theta \end{array} \right)
= \frac{U_{m} }{l_c}
\left( \begin{array}{c} |x_A| \\ h_s-y_B \end{array} \right),
\ee
where
$\theta>0$
is the angle between the channel axis and the horizontal $\ex$.
Taking the inner  product with the sensitivity yields the stochastic response reduction
\begin{align}
\delta E =
\frac{\Rey \, h_c^2\,\Delta P}{12 l_c^2} \left( \bnabla_{\UU_c} E(\xx_A) + \bnabla_{\UU_c} E(\xx_B) \right) 
\bcdot 
\left( \begin{array}{c} |x_A| \\ h_s-y_B \end{array} \right),
\label{eq:dE_smallchannel}
\end{align}
and a similar expression for harmonic gains $G_{k}^2(\omega)$.
This expression shows that
there is a trade-off  between pressure difference, channel length, channel angle, and sensitivity:
for instance, choosing $A$ far upstream increases  both $\Delta P$ and $l_c$, which have opposite effects on the channel velocity;  
similarly, choosing $A$ close to the step corner yields a larger sensitivity but  also increases unfavourably the angle $\theta$ between the control jet and the wall normal in $B$.

Figure~\ref{fig:smallchannel}$(b)$  shows the reduction in optimal harmonic gain  for $x_A=-1$, $y_B=0.75$ and $h_c=0.15$, as estimated with (\ref{eq:dE_smallchannel})  from geometry, pressure and sensitivity information only (solid line), and as obtained  from the  controlled non-linear flow with the channel included in the computational mesh (symbols).
In spite of strongly simplifying assumptions,  the estimated reduction has the correct order of magnitude.
Note that the channel velocity $U_m$ scales  like $h_c^2$, thus only a limited benefit can be expected when using narrow channels.

\section{Discussion}
\label{sec:discussion}

The  analysis of section~\ref{sec:sensit} for  $\Gamma=0.5$, $\Rey=500$, showed that the sensitivity of the stochastic gain was very similar to the sensitivity of the optimal harmonic gain at the optimal frequency $\omega_0$.
One can therefore reduce noise amplification using a steady volume or wall control configuration which reduces the amplification of harmonic forcing at $\omega_0$.
This \textit{a posteriori} observation implies that summation over frequencies and sub-optimals is not required in practice to predict the effect of steady control.

In this section, we investigate the following two aspects:
(i)~whether one can get an \textit{a priori} hint about the effect of steady control on the stochastic gain from the sensitivity of an alternative scalar quantity, the recirculation length;
(ii)~whether the optimal harmonic response at $\omega_0$ still dominates in different $\Gamma-\Rey$ configurations.

\subsection{Recirculation length}
\label{sec:recirclength}

In many flows, the length of the recirculation region is related to stability properties. 
It increases with Reynolds number as long as the flow is steady, while its mean value decreases when the flow becomes unsteady \citep{Sinha81,Armaly83}  as a consequence of mean flow corrections induced by Reynolds stresses.
Due to the presence of the recirculation region,  a shear layer forms, adjacent to the separatrix, which drives the strong convective inviscid instability underpinning the large harmonic/stochastic amplification, as detailed in section~\ref{sec:harm_resp}.

\begin{figure}
  \vspace{0.4cm}
  \centerline{
  	\hspace{0.2cm}
  	\begin{overpic}[width=13.5cm,tics=10]{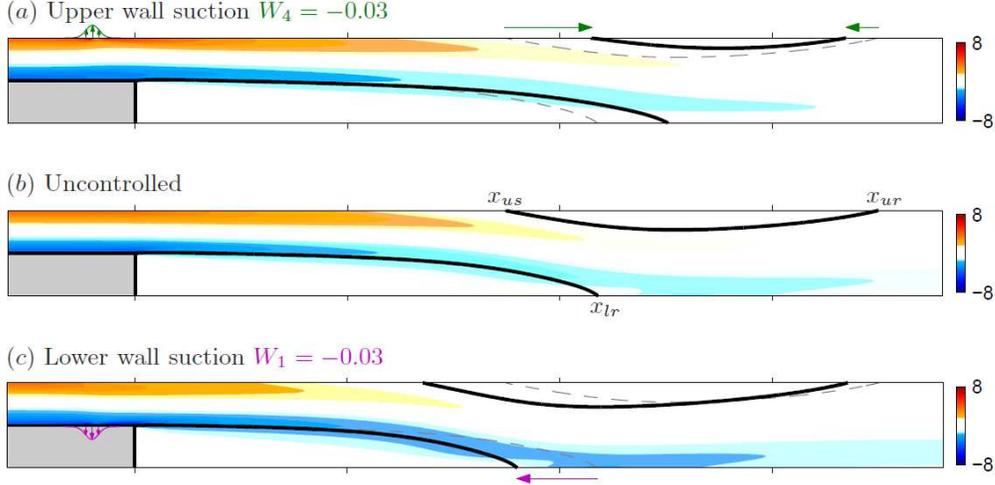}
  	\end{overpic}
  }
 \caption{
Effect of steady wall actuation: wall-normal suction at the 
upstream $(a)$ upper  or $(c)$ lower wall shortens the recirculation region at the upper or lower wall, respectively, compared to the $(b)$ uncontrolled flow  whose separatrices are reported as thin dashed lines in $(a)$ and $(c)$. 
Contours show spanwise vorticity $\partial_x V - \partial_y U$.
$\Gamma=0.5$, $\Rey=500$.
 }
   \label{fig:controlled_flows}
\end{figure}

In this context, it is natural to investigate whether there exists any relationship between recirculation length and stochastic gain when steady control is applied to the flow. 
To this aim, we focus on steady wall control, and compute the
effect of  blowing/suction at the  upstream channel walls (configurations 1 and 4 in figure~\ref{fig:wall_actuation_all}$(a)$)
on the length of the lower recirculation region
$l_l=x_{lr}$, and the length of the upper recirculation region
$l_u=x_{ur}-x_{us}$. 
Figure~\ref{fig:controlled_flows} shows vorticity contours in the controlled and uncontrolled flows. 
Upper wall suction (fig.~\ref{fig:controlled_flows}$(a)$) deflects the main flow upwards, which moves downstream the upper separation and lower reattachment points (compare dashed and solid separatrices without and with control).
As a result, the recirculation region on the upper wall is shortened while that on the lower wall is lengthened.
Conversely, lower wall suction (fig.~\ref{fig:controlled_flows}$(c)$) deflects the main flow downwards, which moves upstream the upper separation and lower reattachment points, resulting in a shorter recirculation region on the lower wall and a longer one on the upper wall.

\begin{figure}
  \vspace{0.2cm}
  \centerline{
	\hspace{0.3cm}
   	\begin{overpic}[width=4.cm,tics=10]{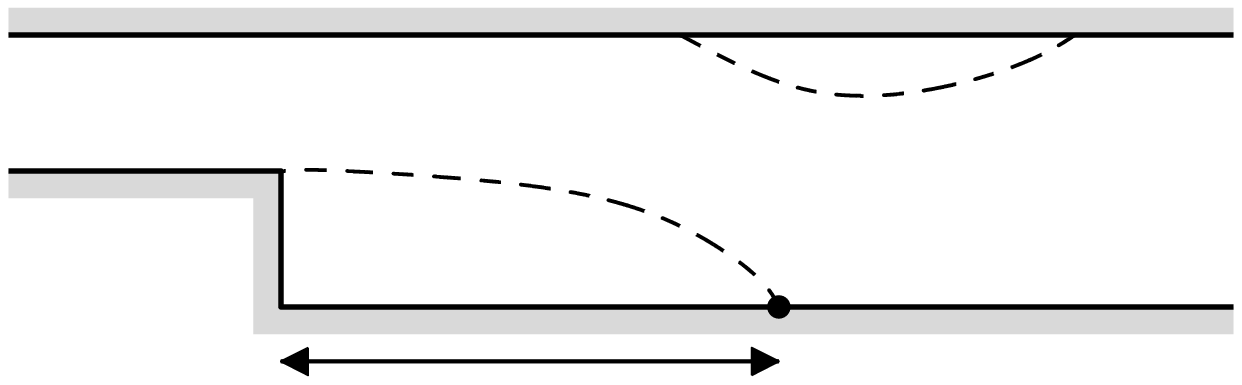}
   	    \put(-18,24) {$(a)$}
   		\put( 40,-5) {$l_l$}
		\put( 63, 8) {$x_{lr}$}	
  	\end{overpic}
  	\hspace{1.55cm}
   	\begin{overpic}[width=4.cm,tics=10]{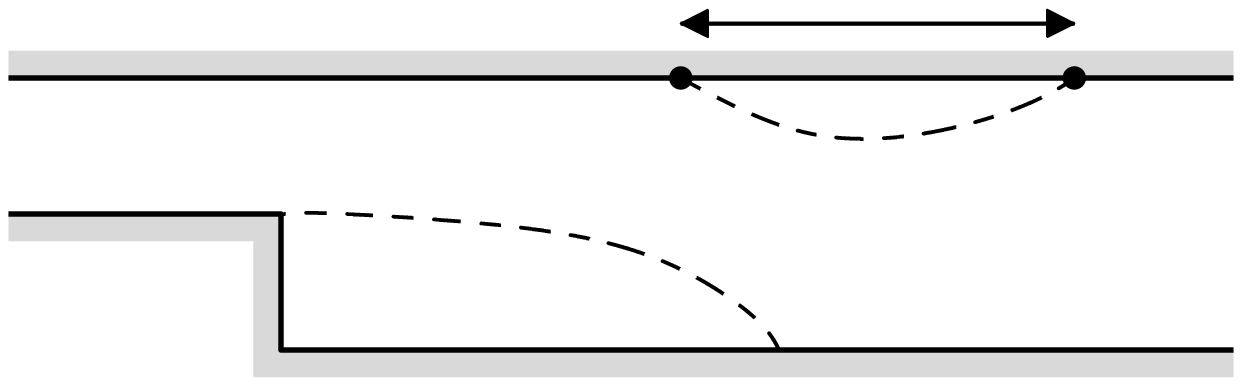}
	   	\put(-18,24)   {$(b)$}
   		\put( 67,34)   {$l_u$}
   		\put( 47,21.5) {$x_{us}$}
		\put( 85,21.5) {$x_{ur}$}
  	\end{overpic}
  }
  \vspace{0.28cm}
  \centerline{
    \psfrag{W}[t][]{flow rate $W$}
    \psfrag{lr}[r][][1][-90]{$l_l \,$}	
    \psfrag{lu}[r][][1][-90]{$l_u \,$}	
   	\begin{overpic}[height=5 cm,tics=10]{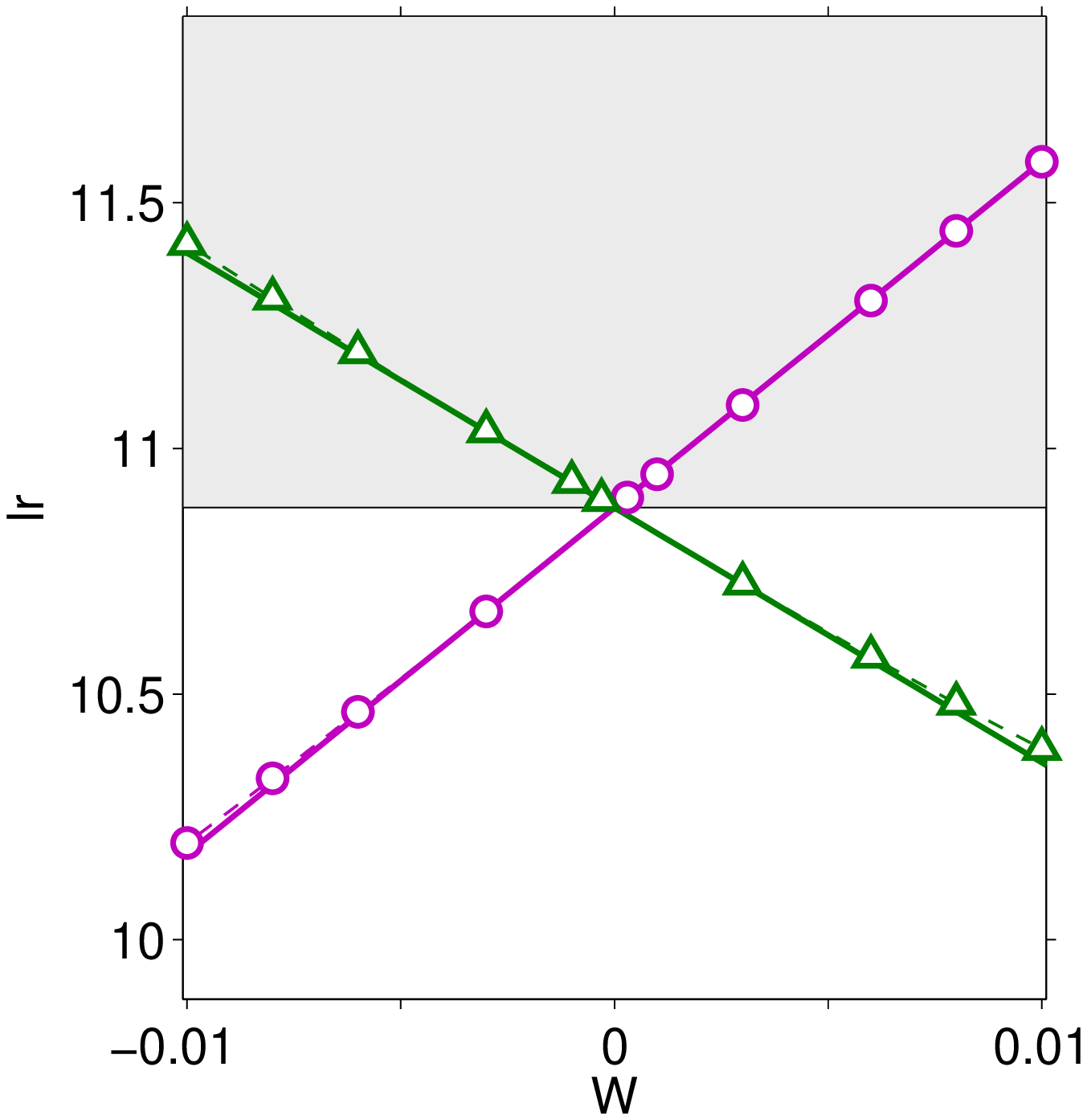}
		\put(22, 20)  {\footnotesize \textcolor[rgb]{0.75,0,   0.75}{1}}
   		\put(84.5, 28.){\footnotesize \textcolor[rgb]{0,   0.5, 0}   {4}}
   	\end{overpic}
   	\hspace{0.9cm}
  	\begin{overpic}[height=5 cm,tics=10]{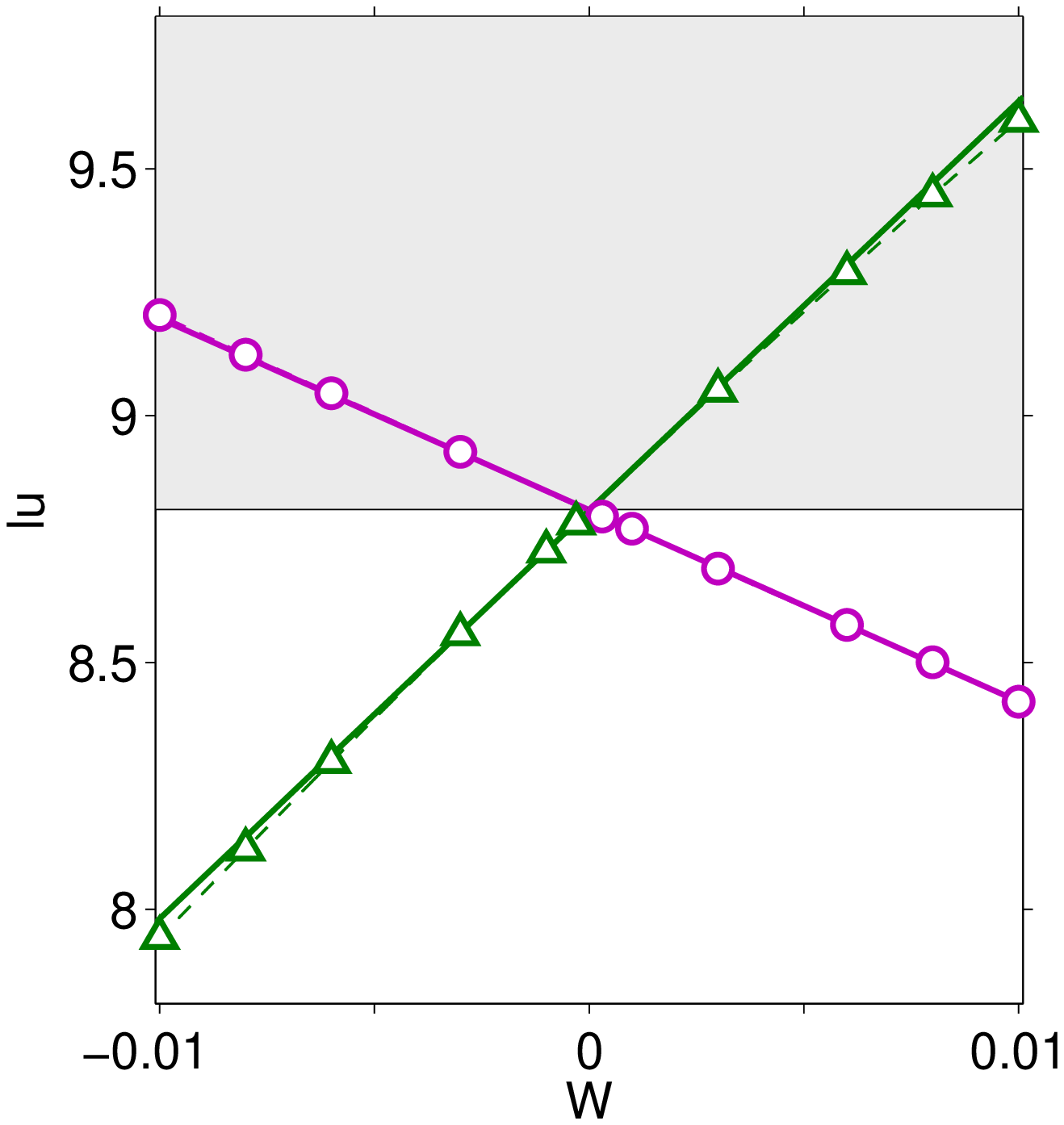}
   		\put(26,  20)  {\footnotesize \textcolor[rgb]{0,   0.5, 0}   {4}} 
 		\put(83., 30)  {\footnotesize \textcolor[rgb]{0.75,0,   0.75}{1}}
   	\end{overpic}
  } 
 \caption{Variation of  $(a)$ lower and  $(b)$ upper recirculation lengths with steady wall blowing/suction flow rate (configurations 1 and 4 shown in figure~\ref{fig:wall_actuation_all}).
Solid lines show predictions from sensitivity analysis, symbols are actual results from controlled flows. $\Gamma=0.5$, $\Rey=500$.
}
   \label{fig:wall_actuation_length}
\end{figure}

Figure~\ref{fig:wall_actuation_length} shows the variation in lower and upper recirculation length with control flow rate. Relatively small control flow rates have a significant impact on both recirculation lengths: 
$|\delta l| \simeq 5-10\%$ for $|W|=0.01$.
Applying sensitivity analysis to the recirculation lengths~\citep{Boujo14-JFM,Boujo14-PRSA}, we find that variations are almost linear in this range of flow rates:  non-linear results from full Navier--Stokes calculations (symbols) closely follow predictions from sensitivity analysis  (lines).

\begin{figure}
  \vspace{0.4cm}
  \centerline{  
  	\begin{overpic}[width=13.cm,tics=10]{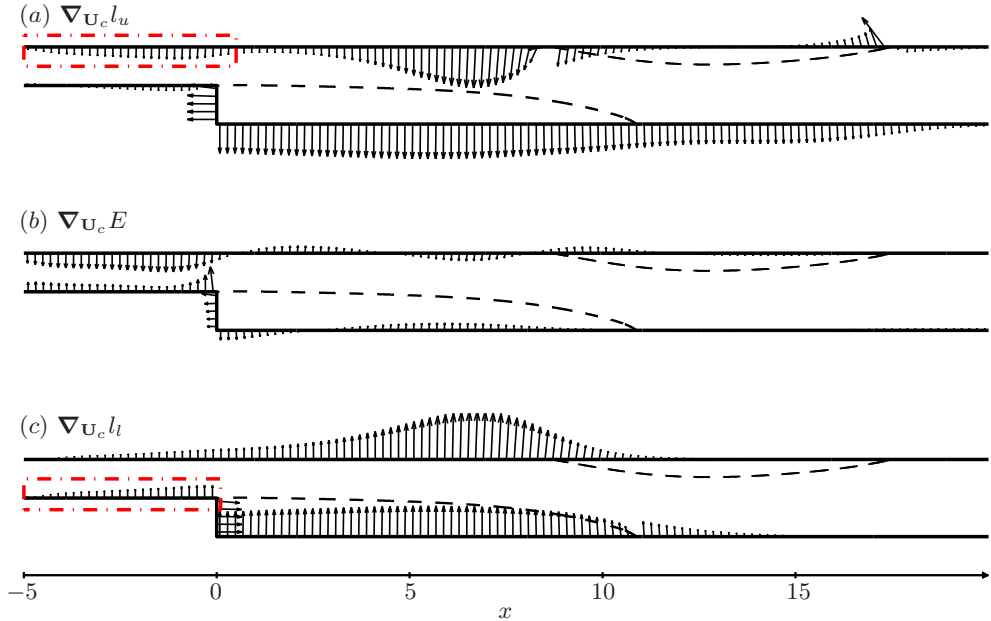} 
  	\put( 0,57)  {$(a)$ $\bnabla_{\UU_c} l_u$}
   	\put( 0,36)  {$(b)$ $\bnabla_{\UU_c} E$}
   	\put( 0,15)  {$(c)$ $\bnabla_{\UU_c} l_l$}
  	\put(-1.3, -2.) {\footnotesize$-5$}
    \put(19.5, -2.) {\footnotesize$ 0$}
    \put(39.4, -2.) {\footnotesize$ 5$}
    \put(58.5, -2.) {\footnotesize$10$}
    \put(78.5, -2.) {\footnotesize$15$} 	   	
   	\put(49,-4)  {$x$}
   	\end{overpic}
  } 
  \vspace{0.6cm}  
  \caption{
Sensitivity to wall control of
$(a)$ the upper recirculation length,
$(b)$ the stochastic gain (reproduced from figure~\ref{fig:DUw_gamma05_re500}$(b)$) and
 $(c)$ the lower recirculation length.
$\Gamma=0.5$, $\Rey=500$. 
Arrows point in the direction of positive sensitivity (i.e. increasing lengths and gain).
}
   \label{fig:zoom_compare}
\end{figure}

Figure~\ref{fig:zoom_compare} shows the sensitivity of the recirculation lengths  to wall control. 
The upper recirculation region (fig.~\ref{fig:zoom_compare}$(a)$) is most effectively shortened using wall suction on the upper wall upstream of $x_{us}$, or wall blowing on the vertical wall and the lower wall downstream of the step.
The sensitivity of the lower recirculation region (fig.~\ref{fig:zoom_compare}$(c)$) has a similar distribution but the opposite sign almost everywhere, indicating that any wall control has opposite effects on  $l_u$ and $l_l$.
As a side note, the sensitivity of $l_u$ is very large in the vicinity of $x_{us}$ and $x_{ur}$, and the sensitivity of $l_l$ is very large in the vicinity of $x_{lr}$.

If we compare now these two sensitivity maps to that of the stochastic gain (fig.~\ref{fig:zoom_compare}$(b)$, reproduced from fig.~\ref{fig:DUw_gamma05_re500}$(b)$), 
overall distributions are  different, and it seems unlikely at first sight that considerations about recirculation lengths can guide the design of a simple wall control configuration aiming at reducing noise amplification.
However, if we focus on the upstream channel ($x<0$) it appears that $E$ and $l_u$ (resp. $l_l$) have sensitivities of the same sign on the upper (resp. lower) walls.
This suggests that if $E$ is to be reduced with wall control in the upstream channel,  one can use wall control on the \textit{upper}  wall and aim at shortening the \textit{upper}  recirculation region, 
or use wall control on the \textit{lower} wall and aim at shortening the \textit{lower} recirculation region.

\subsection{On the predominance of the optimal harmonic response}
\label{sec:predom}

\begin{figure}
  \psfrag{om}[t][]{$\omega$}
  \psfrag{Gi}[][]{$G_{in,k}$}	
  \centerline{
	\begin{overpic}[width=6.5 cm,tics=10]{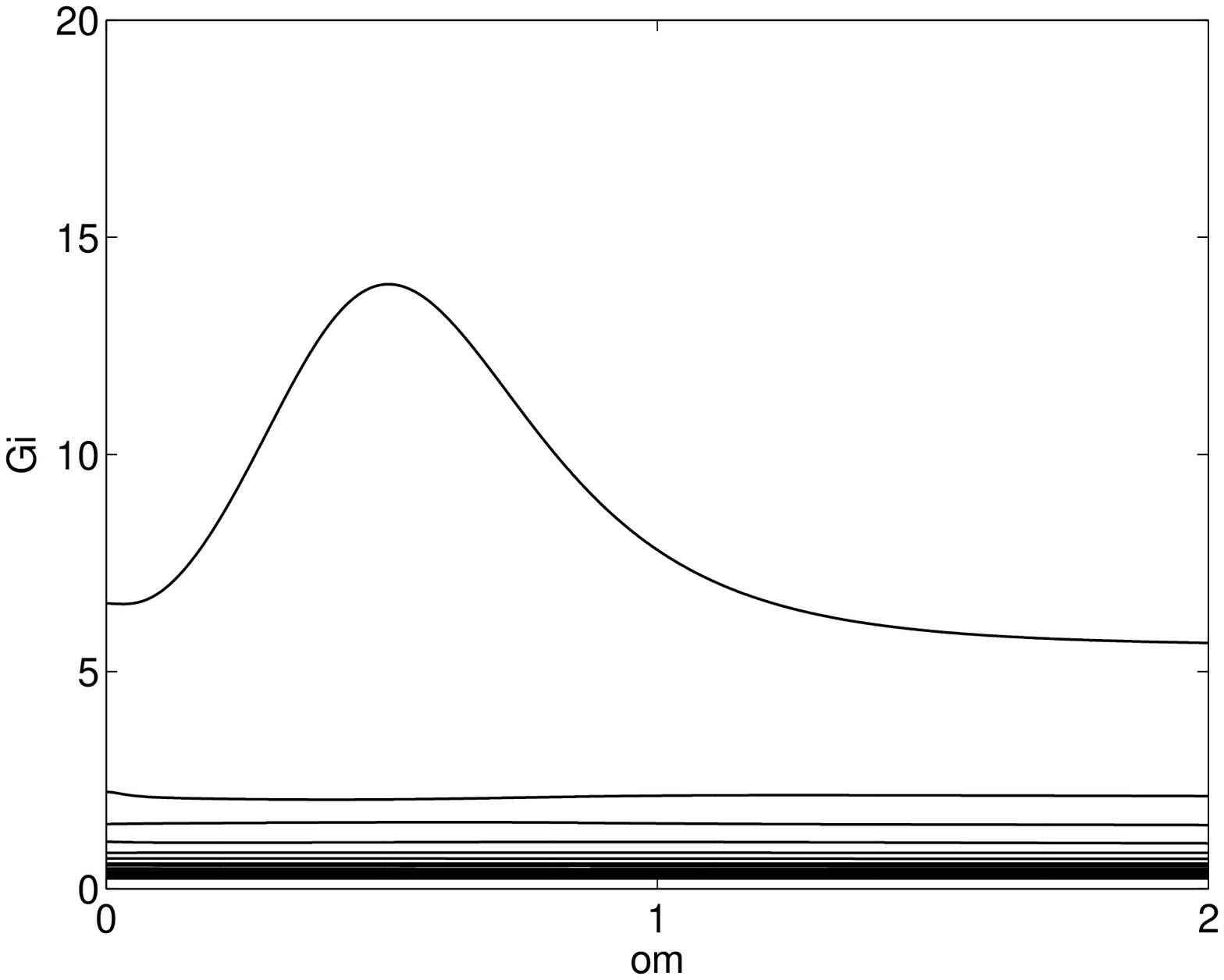}
   		\put(11,72){$(a)$}
   	\end{overpic}
   	\hspace{0.3cm}
   	\begin{overpic}[width=6.38 cm,tics=10]{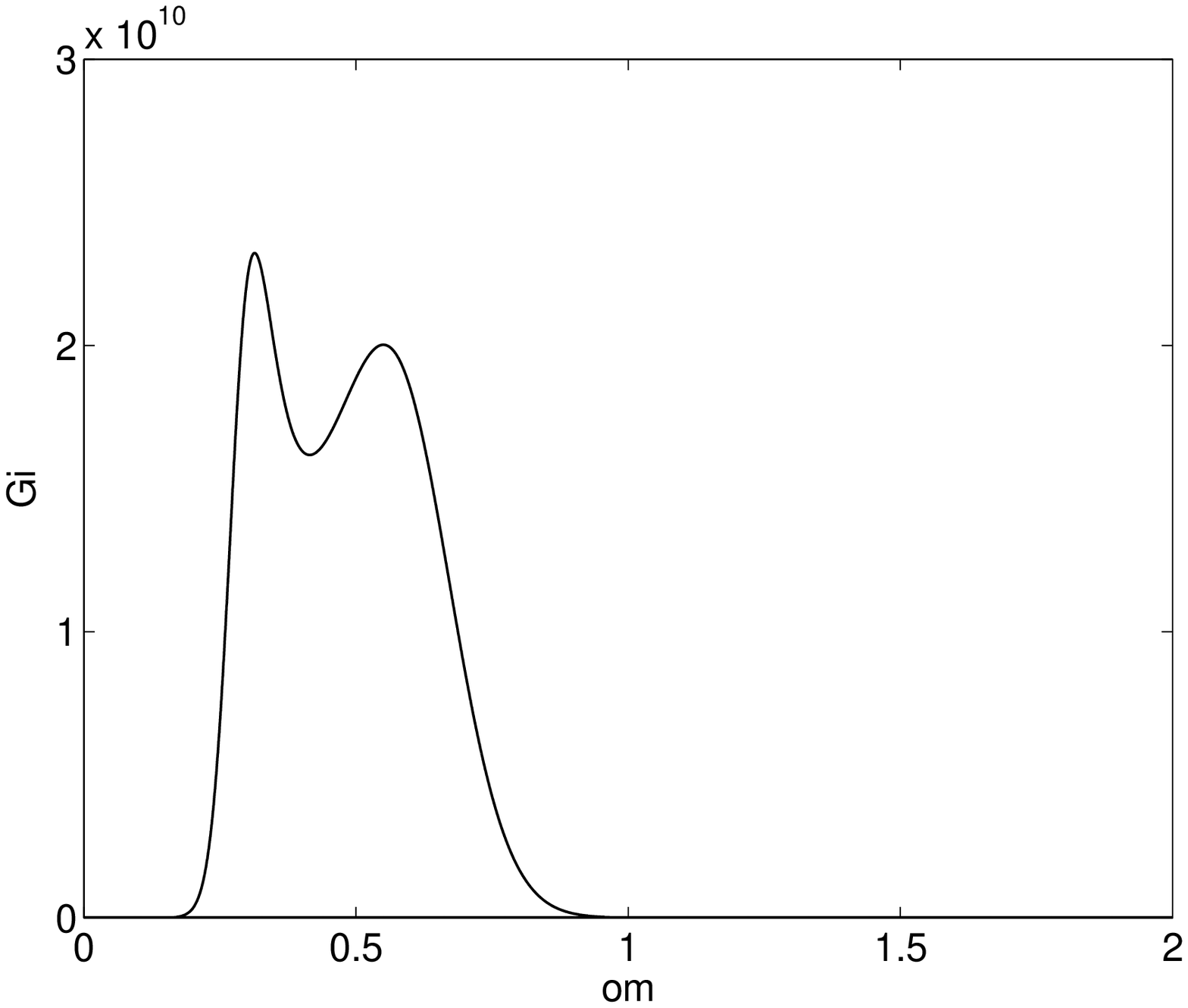}
   		\put(9.5,73){$(b)$}
   	\end{overpic}
  }
 \caption{Optimal harmonic gain and  sub-optimal gains for  inlet forcing.
  $(a)$~$\Gamma=0.5, \Rey=200$,  $(b)$~$\Gamma=0.3, \Rey=2800$.  
}
   \label{fig:G_others}
\end{figure}

We observed in section~\ref{sec:sensit}  that for  $\Gamma=0.5$, $\Rey=500$, the sensitivity of the stochastic gain $E$ was very similar to that of the optimal gain $G_{in,1}$ at the optimal frequency $\omega_0$.
In order to test the robustness of this observation, 
we now consider two other configurations where the sensitivity of $E$ is more likely to differ from the sensitivity of 
$G_{in,1}(\omega_0)$. 
In the first configuration we keep the same geometry ($\Gamma=0.5$) but decrease the Reynolds number to $\Rey=200$;
in the second one we use a smaller step with expansion ratio $\Gamma=0.3$ at the stable Reynolds number $\Rey=2800$  (recall that $\Rey_c>2900$ for this expansion ratio, as mentioned in section~\ref{sec:num_valid}).
From the harmonic gains shown in 
figure~\ref{fig:G_others} we can make the following remarks: 
(i)~reducing the Reynolds number  makes the peak of $G_{in,1}$ less marked, thus possibly increasing the relative importance of frequencies other than the optimal one, as well as the importance of sub-optimal forcings relative to the optimal one;
(ii)~a double peak appears when reducing the step height, which might result in contributions  of equal importance from two well separated  frequencies ($\omega_{0}=0.31$ and $\omega_{0}'=0.55$ in the present case).

\begin{figure}
  \centerline{
   	\begin{overpic}[width=11cm,tics=10]{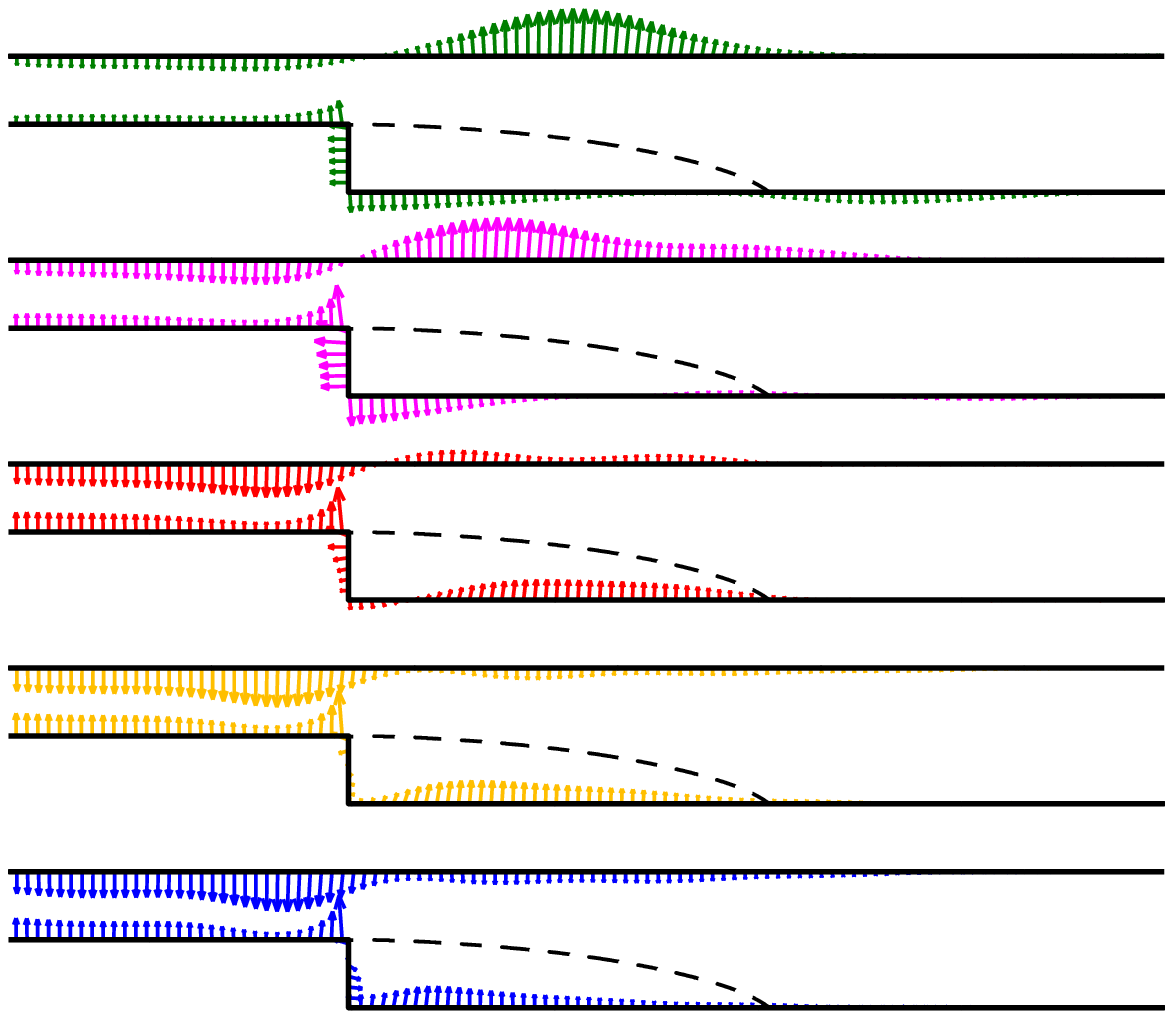}
   	\put( 0,85)  {$(a)$ $\bnabla_{\UU_c} G_{in,1}^2$}
   	\put(9,72.5)  {\footnotesize$\omega=0.1$}
   	\put(9,55. )  {\footnotesize$\omega=0.3$}
   	\put(9,37.5)  {\footnotesize$\omega=0.5$}
   	\put(9,20. )  {\footnotesize$\omega=0.7$}
   	\put(9, 2.5)  {\footnotesize$\omega=0.9$}
   	\put(70,72)   {\footnotesize$\max ||\bcdot||_2=1.6\times10^2$}
   	\put(70,54.5) {\footnotesize$\max ||\bcdot||_2=2.1\times10^3$}
   	\put(70,37.)  {\footnotesize$\max ||\bcdot||_2=5.3\times10^3$}
   	\put(70,19.5) {\footnotesize$\max ||\bcdot||_2=4.4\times10^3$}
   	\put(70, 2.)  {\footnotesize$\max ||\bcdot||_2=2.1\times10^3$}
   	\end{overpic}
  }
  \vspace{1.2cm}
  \centerline{
  	\begin{overpic}[width=11cm,tics=10]{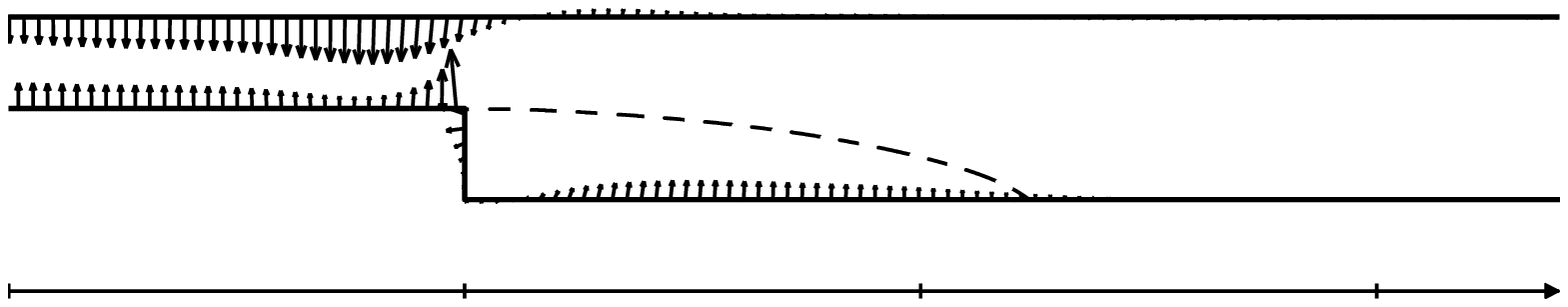}  	
   	\put( 0,20.5)  {$(b)$ $\bnabla_{\UU_c} E$}
   	\put(49,-4.5)  {$x$}
   	\put(70, 8)  {\footnotesize$\max ||\bcdot||_2=9.9\times10^2$}
    \put(-2,  -2.5) {\footnotesize$-5$}
    \put(28.7,-2.5) {\footnotesize$ 0$}
    \put(57.8,-2.5) {\footnotesize$ 5$}
    \put(86,  -2.5) {\footnotesize$10$}
   	\end{overpic}
  }    
 \vspace{0.5cm}
 \caption{Sensitivity of $(a)$ optimal harmonic gain
 and $(b)$ stochastic gain to wall control.  $\Gamma=0.5$, $\Rey=200$.
}
   \label{fig:DUw_gamma05_re200}
\end{figure}

\begin{figure}
  \centerline{
   	\begin{overpic}[width=13cm,tics=10]{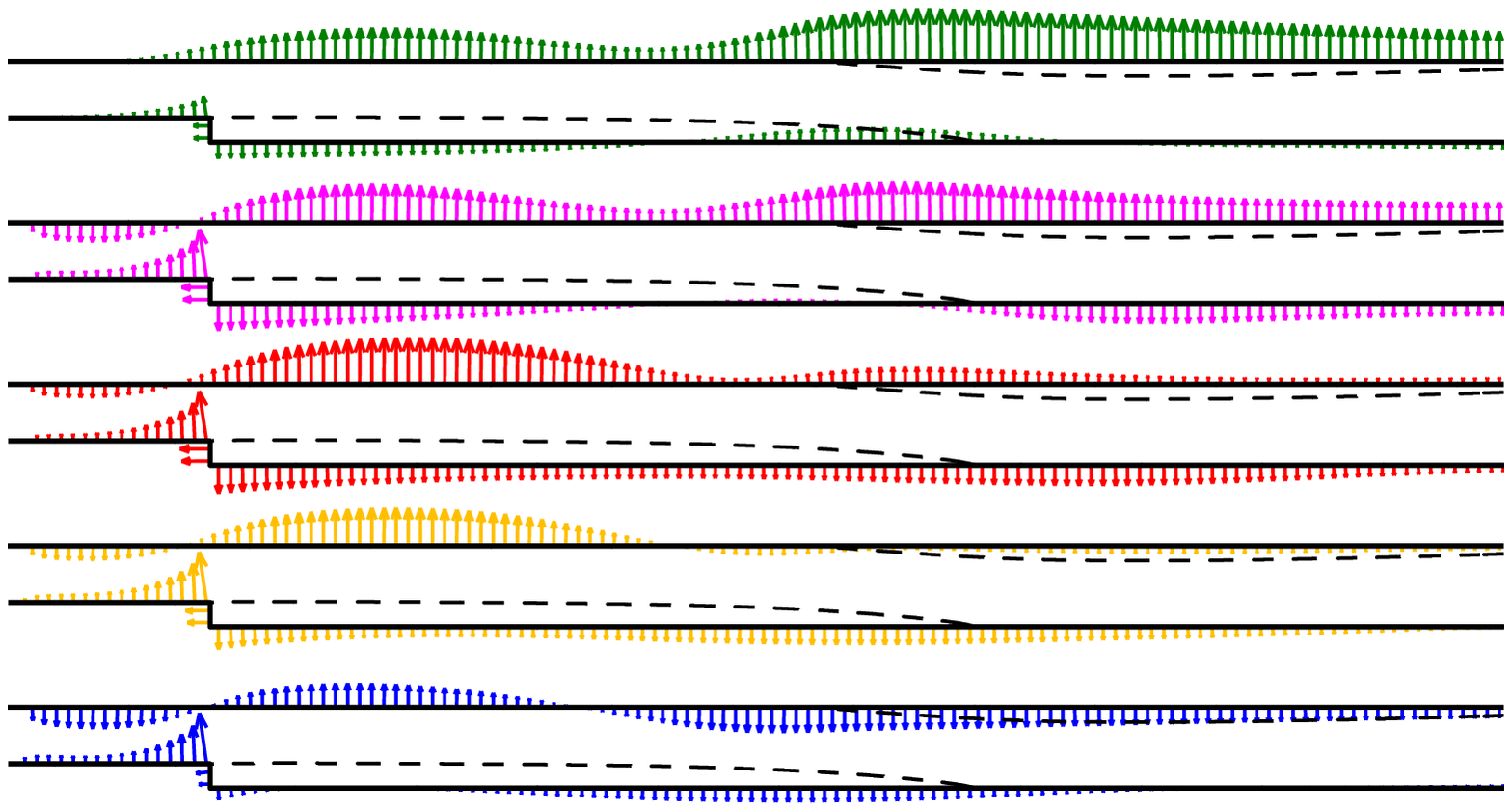}
   	\put( 0.,52.)  {$(a)$ $\bnabla_{\UU_c} G_{in,1}^2$}
   	\put(3,43.00) {\footnotesize$\omega=0.1$}
   	\put(3,32.25) {\footnotesize$\omega=0.3$}
   	\put(3,21.50) {\footnotesize$\omega=0.5$}
   	\put(3,10.75) {\footnotesize$\omega=0.7$}
   	\put(3, 0.0) {\footnotesize$\omega=0.9$}
   	\put(74,45)    {\footnotesize$\max ||\bcdot||_2=2.8\times10^{11}$}
   	\put(74,34.25) {\footnotesize$\max ||\bcdot||_2=1.5\times10^{23}$}
   	\put(74,23.5)  {\footnotesize$\max ||\bcdot||_2=1.7\times10^{23}$}
   	\put(74,12.75) {\footnotesize$\max ||\bcdot||_2=4.8\times10^{22}$}
   	\put(74, 2)    {\footnotesize$\max ||\bcdot||_2=1.5\times10^{19}$}
   	\end{overpic}
  }
  \vspace{1.2cm}
  \centerline{
  	\begin{overpic}[width=13cm,tics=10]{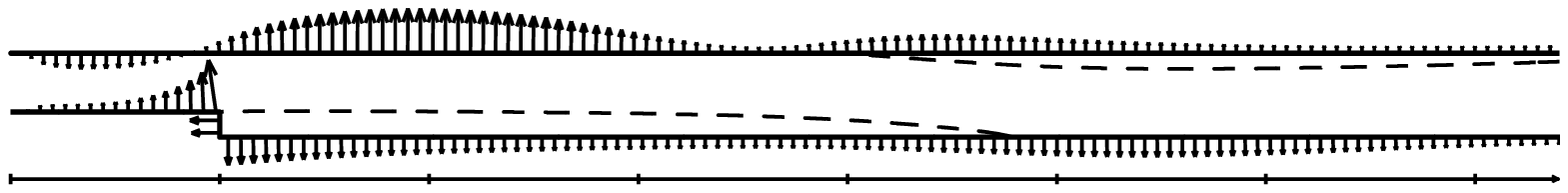}  	
   	\put( 0.,11)  {$(b)$ $\bnabla_{\UU_c} E$}
   	\put(49,-4.5)  {$x$}
   	\put(74, 4.25)  {\footnotesize$\max ||\bcdot||_2=2.1\times10^{22}$}
    \put(-1.5,-2.) {\footnotesize$-5$}    
    \put(13.2,-2.) {\footnotesize$ 0$}
    \put(26.5,-2.) {\footnotesize$ 5$}
    \put(39.2,-2.) {\footnotesize$10$}
    \put(52.8,-2.) {\footnotesize$15$}
    \put(66.2,-2.) {\footnotesize$20$}
    \put(79.5,-2.) {\footnotesize$25$}
    \put(93., -2.) {\footnotesize$30$}
   	\end{overpic}
  }    
 \vspace{0.5cm}
 \caption{Sensitivity of $(a)$ optimal harmonic gain  and $(b)$ stochastic gain
 to wall control. $\Gamma=0.3$, $\Rey=2800$.
}
   \label{fig:DUw_gamma03}
\end{figure}

Figures~\ref{fig:DUw_gamma05_re200} and \ref{fig:DUw_gamma03} show sensitivities to wall control for these two configurations.
Although slight differences can be noticed, the stochastic gain still has a sensitivity largely dominated by the sensitivity of the optimal harmonic  gain at the optimal frequency $\omega_0$. 
In particular, locations of maximal sensitivity of $E$ are captured robustly by the sensitivity of $G_{in,1}(\omega_0)$. 
Therefore, even though integrating over a range of frequencies (i.e.  performing a weighted average) should have a smoothing effect, the net result is almost unaffected by frequencies far from the optimal one.
Close inspection shows that all sensitivity fields (for base-flow modification, volume control and wall control) change continuously with $\omega$ around the optimal frequency, and slowly enough for their integral to be eventually dominated by 
$\bnabla_{\boldsymbol{*}} G_{in,1}^2(\omega_0)$. 
For instance, the two distinct peaks in 
$G_{in,1}$ for $\Gamma=0.3$, $\Rey=2800$ (fig.~\ref{fig:G_others}$(b)$), are actually associated with very similar sensitivity fields.
In addition, sub-optimals  appear to contribute only little. This is not surprising for $\Gamma=0.5$, $\Rey=500$, since sub-optimal gains $G_k$ for $k\geq2$ are smaller than $G_1$ by a factor of about two orders of magnitude in the range of most amplified frequencies (see fig.~\ref{fig:Gin_Gvol_gamma05_re500}). This is more surprising for $\Gamma=0.5$, $\Rey=200$, where this factor is reduced to about 5 (fig.~\ref{fig:G_others}$(a)$).
We  conclude that  sub-optimals  play a role only at even lower Reynolds numbers, where amplification mechanism are weak anyway.
Therefore,  sensitivity analysis and steady control design 
can be  conducted with good confidence on the optimal harmonic gain at the optimal frequency alone, rather than on the full stochastic response, thereby dramatically reducing the computational cost of the process.

These observations are limited to the backward-facing step flow and to expansion ratios and Reynolds numbers investigated in this paper, although we believe they might bear generality in other strong amplifier flows. 
In flows where perturbations undergo large amplification in well-separated frequency ranges, the relationship between local stability, transient growth, and response to harmonic/stochastic forcing might be more subtle.

\section{Conclusion}
\label{sec:conclu}

The response to time-harmonic and time-stochastic forcing in the two-dimensional flow past a backward-facing step was analysed.
For the expansion ratio $\Gamma=0.5$ and  Reynolds number $\Rey=500$, 
a global linear stability analysis predicts that the flow is stable 
but a large amplification of both harmonic and stochastic forcing was observed, typical of noise amplifier flows.
A local spatial stability analysis yielded very good agreement with global harmonic response, transient growth studies and direct numerical simulation, especially in terms of most amplified frequency ($\omega_0=0.5$) and streamwise location of maximum response ($x\simeq25-27$).

Sensitivity analysis was used to study in a systematic way the effect of small-amplitude steady control in the volume or at the wall.
A variational technique was used to derive analytical expressions of sensitivity,  extending existing  methods to stochastic forcing localised at the inlet.
For volume forcing, sensitivity is classically computed from the forcing and corresponding response, while for inlet forcing it is found from the response and from an intermediate adjoint velocity field.
In both cases, the sensitivity of the stochastic gain is expressed as a simple combination of sensitivities of optimal and sub-optimal  harmonic gains over all frequencies.

Sensitivity maps obtained from this analysis allowed us, without computing any controlled flows, to identify most sensitive regions where control can increase or decrease harmonic/stochastic gain the most efficiently.
In particular, passive control by means of a small cylinder decreases the gain in the main stream downstream of the step, and increases the gain in the shear layers at the edges of the recirculation regions.
Active control by means of wall blowing and suction is most effective  
on the vertical wall of the step and on the horizontal walls of the upstream channel.

For several  Reynolds numbers and expansion ratios, it was observed that the sensitivity of the stochastic response was dominated to a large extent by the sensitivity of the optimal harmonic response at the most amplified frequency.
This suggests that in this noise amplifier flow, and possibly in others, the design of open-loop control aiming at reducing noise amplification can be performed by targeting the optimal harmonic response at the optimal frequency only.

Possible extensions of the sensitivity analysis presented in this study include coloured noise, unsteady control and three-dimensional flows.
\\

Unpublished work \citep{Marquet10-BF2} dealing with the backward-facing step for $\Gamma=0.5$, $\Rey=500$,  was brought to our attention during the reviewing process. We thank these authors for interesting information and stimulating discussions.
We also wish to thank Xavier Garnaud, Cristobal Arratia and Eric Serre for their advice and comments, and the referees for valuable suggestions.
This work was supported by the Swiss National Science Foundation (grant no. 200021-130315) and  the French National Research Agency (project no. ANR-09-SYSC-001).

\appendix

\section{Influence of inlet length}
\label{sec:appA}

\begin{figure}
  \centerline{
  \psfrag{om}[t][]{$\omega$}
  \psfrag{Li}[t][]{$L_{in}$}
  \psfrag{Gv}     [][]{$G_{vol,1}$}	
  \psfrag{Gi}     [][]{$G_{in,1}$}	
  \psfrag{maxGvol}[r][][1][-90]{}	
  \psfrag{maxGin} [r][][1][-90]{}	
  \begin{overpic}[width=6.5 cm,tics=10]{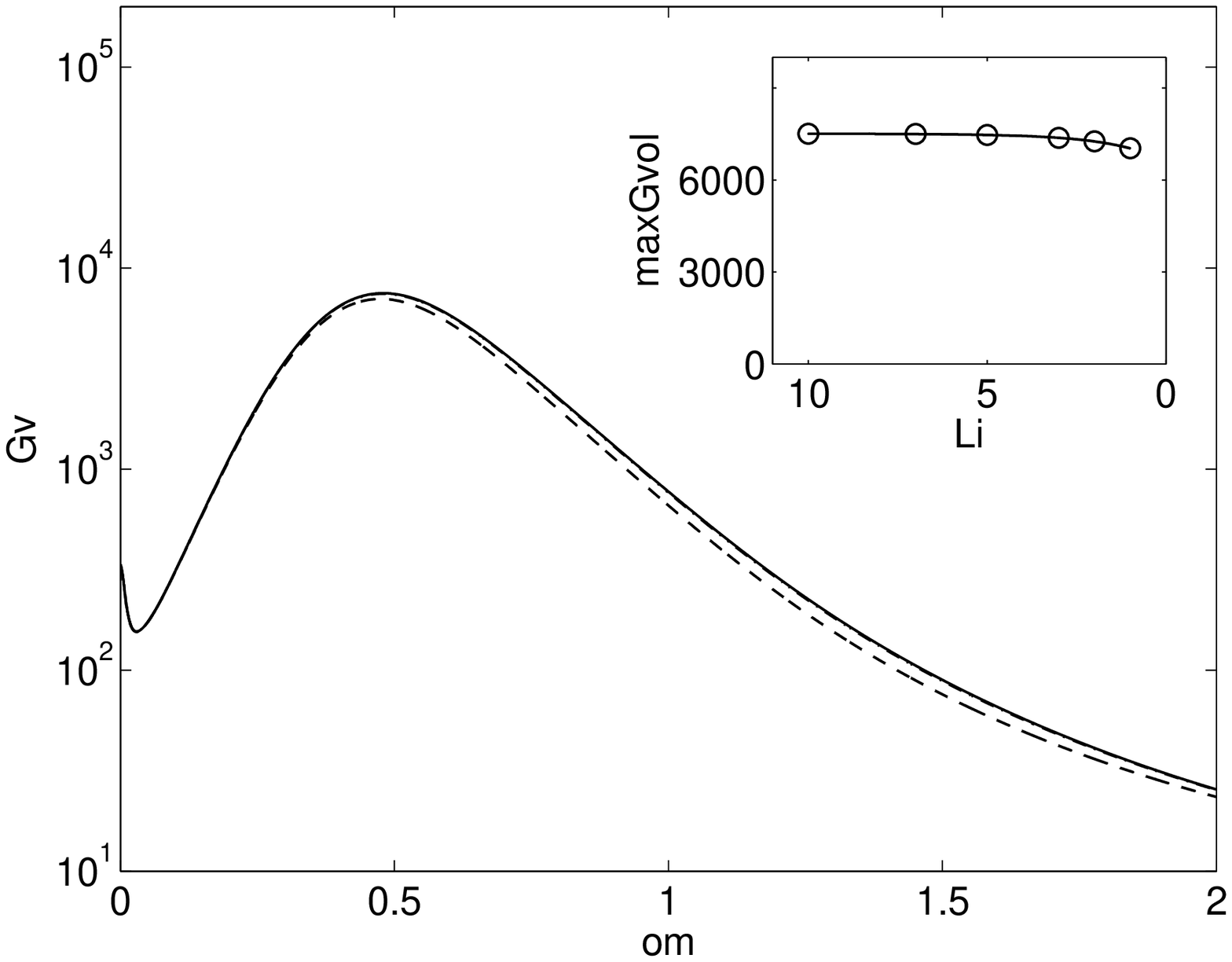}
   	\put(39,70){$\displaystyle \max_{\omega}G_{vol,1}$}
	\put(12,72){$(a)$}
   	\end{overpic}   
   	\hspace{0.2cm}
   	\begin{overpic}[width=6.5 cm,tics=10]{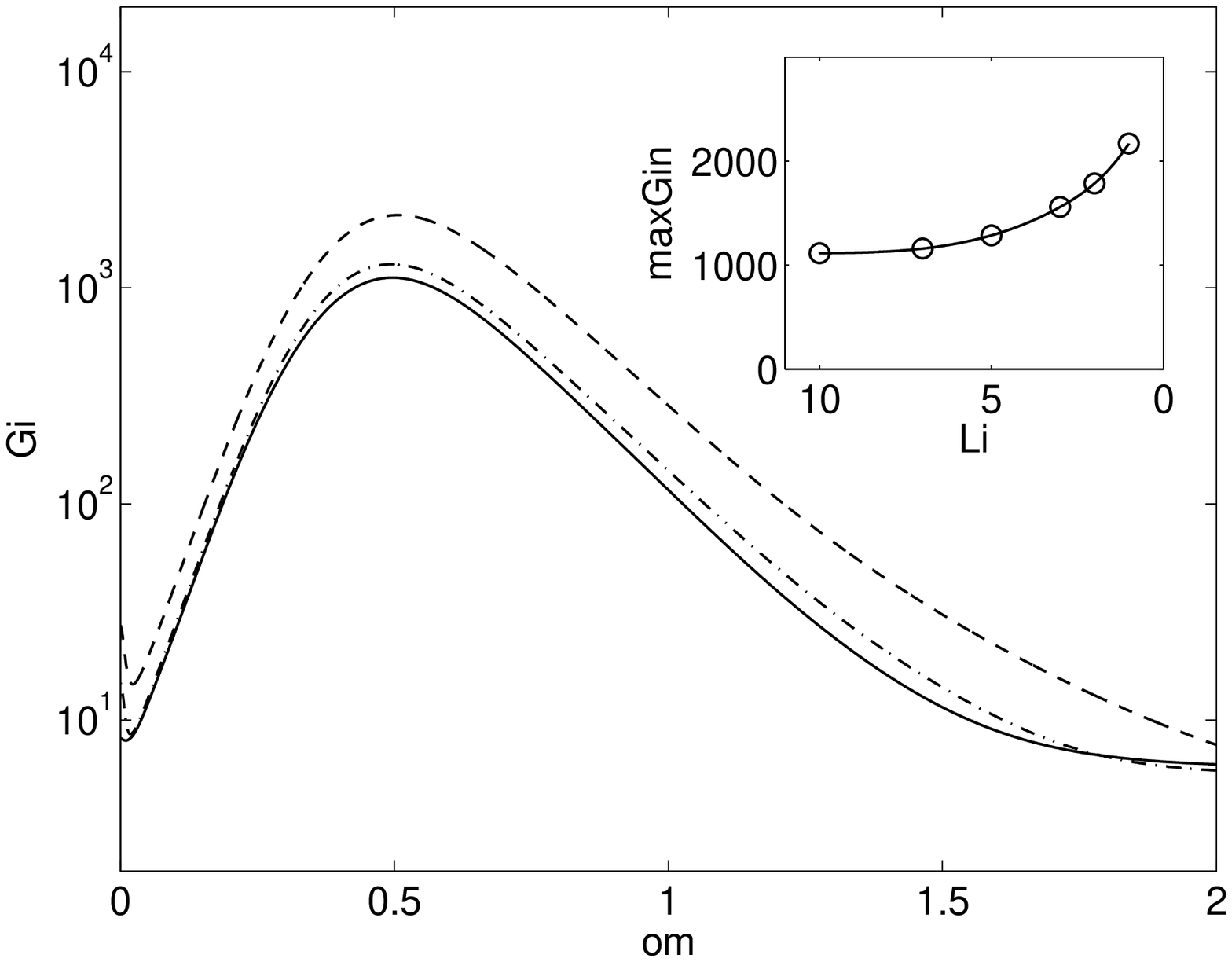}
   	\put(41,70){$\displaystyle \max_{\omega} G_{in,1}$}
   	\put(12,72){$(b)$}
   	\end{overpic}   	
  }
 \caption{Optimal harmonic gain for  $(a)$ volume and  $(b)$ inlet forcing for different inlet lengths: $L_{in}=1$ (dashed line),  $L_{in}=5$ (dash-dotted line) and  $L_{in}=10$ (solid line). Insets show the convergence of the maximum gain value with increasing inlet length. $\Gamma=0.5$, $\Rey=500$.
 }
   \label{fig:Li_effect}
\end{figure}

\begin{figure}
  \psfrag{x}[t][]{$x$}
  \psfrag{y}[][][1][-90]{$y$}	
  \centerline{
   	\begin{overpic}[width=13cm,tics=10]{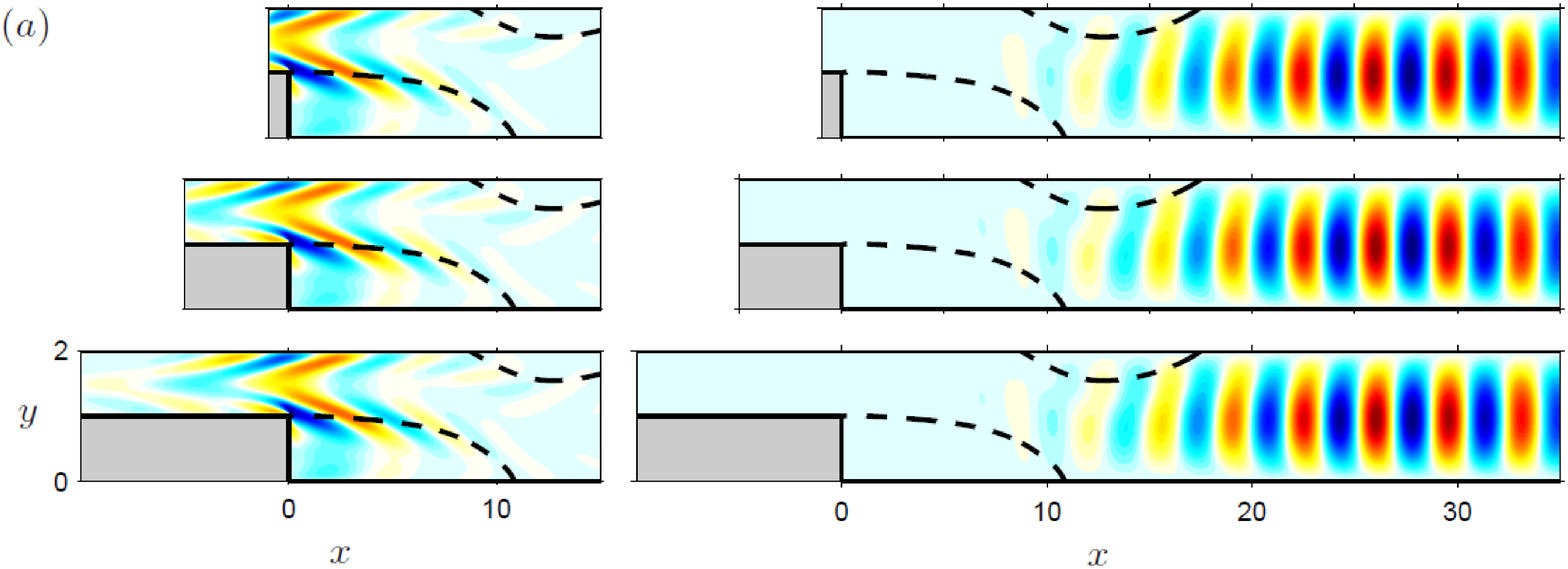}
   	\end{overpic}
  }
  \centerline{
   	\begin{overpic}[width=13cm,tics=10]{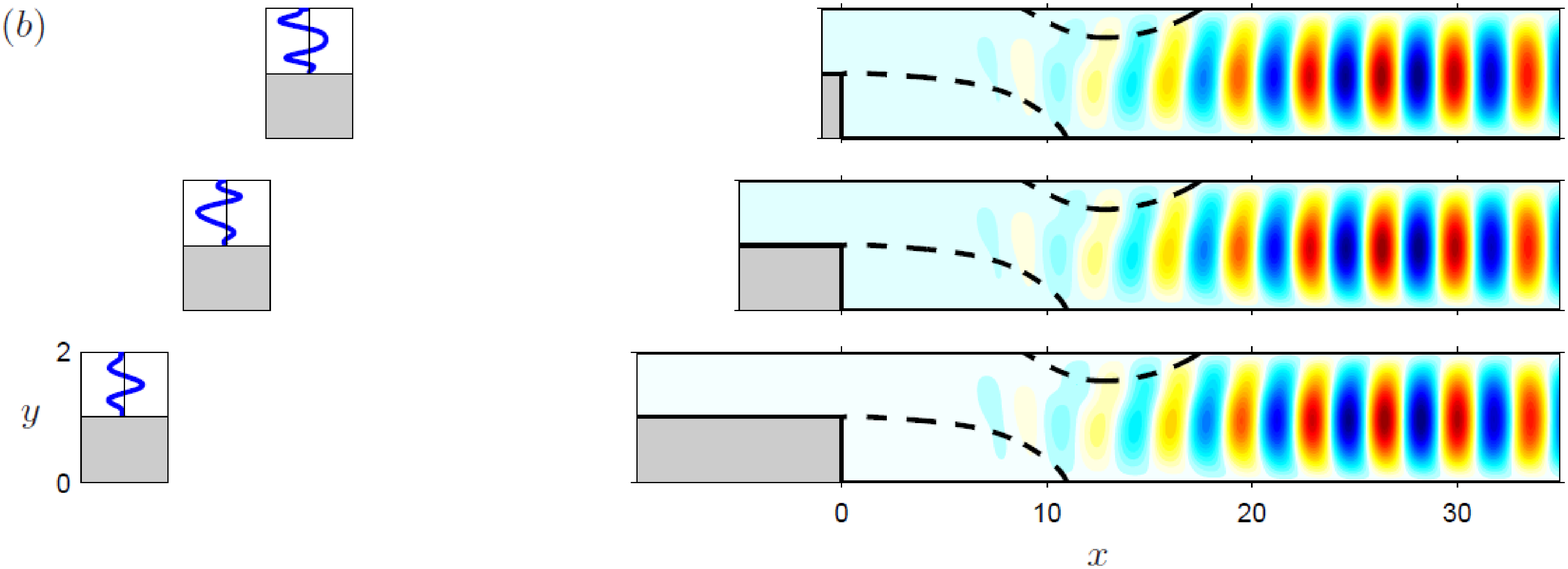}
   	\end{overpic}
  }
 \caption{Influence of inlet length ($L_{in}=1$, 5 and 10 from top to bottom) on optimal harmonic forcing (left; real part of streamwise component $\ff_1\bcdot\ex$) and  optimal harmonic response (right; real part of cross-stream component $v_1$):
$(a)$ volume forcing and corresponding response, $(b)$ inlet forcing and corresponding response. $\Gamma=0.5$, $\Rey=500$, $\omega=0.5$.}
   \label{fig:Li_effect_opt_forc_resp}
\end{figure}

Figures~\ref{fig:Li_effect} and \ref{fig:Li_effect_opt_forc_resp} show how the optimal harmonic  gain, forcing and response depend on the inlet length.
In the case of volume forcing (fig.~\ref{fig:Li_effect}$(a)$), the optimal gain $G_{vol,1}$ is not much affected by the inlet length, since the optimal forcing is well localised near the step corner and only a small amount of energy is introduced in the upstream inlet region. This is  illustrated for $\omega=0.5$ in figure~\ref{fig:Li_effect_opt_forc_resp}$(a)$.
In the case of inlet forcing (fig.~\ref{fig:Li_effect}$(b)$), the optimal gain varies significantly when the inlet length is increased 
up to $L_{in}\simeq 5$, consistent with the observation of \cite{Gar13}. 
Here $G_{in,1}$  decreases with $L_{in}$ due to viscous effects which smooth out perturbations when they enter the flow farther away upstream of the step corner, i.e. upstream of the locally unstable region.
Figure~\ref{fig:Li_effect_opt_forc_resp}$(b)$ shows that 
the optimal response keeps the same spatial structure
although the  inlet optimal forcing does vary with $L_{in}$ due to a phase effect (as mentioned in section~\ref{sec:harm_resp}, the inlet optimal forcing is similar  to the profile of the volume optimal forcing close to $x=L_{in}^+$, and here we fix the phase at $x=0$, $y=1.5$ for all cases).

\section{Influence of cut-off frequency}
\label{sec:appB}

\begin{figure}
  \psfrag{re}[t][]{$\Rey$}
  \psfrag{omc}[t][]{$\omega_c$}
  \psfrag{E}[r][][1][-90]{$E$}	
  \psfrag{En}[r][][1][-90]{$\displaystyle\widetilde E$}	
  \centerline{
   	\begin{overpic}[width=6.5cm,tics=10]{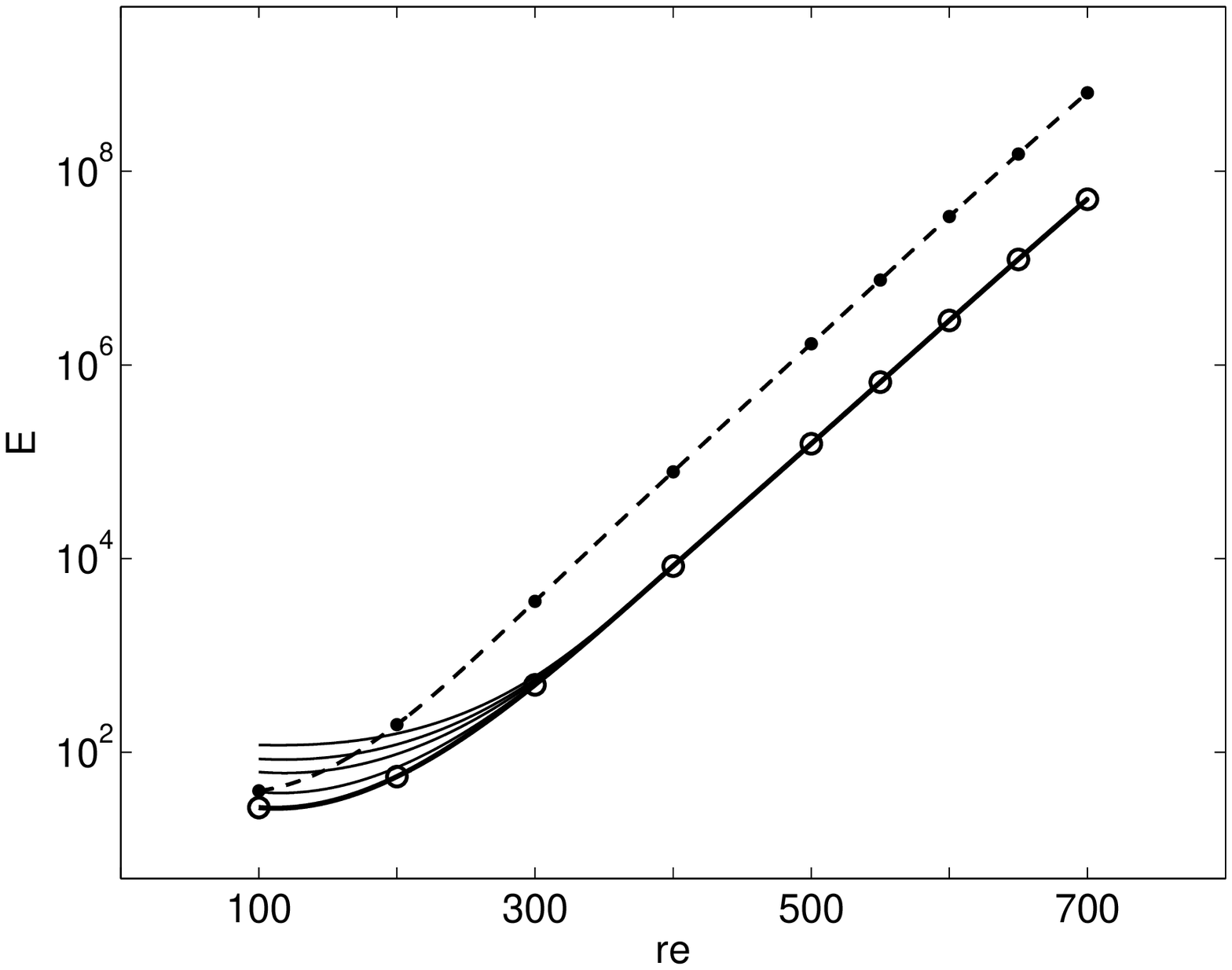}
   	\put(56,64)  {\footnotesize  $\displaystyle\max_\omega G_{in,1}^2$}
	\put(13.8,20.5)   {\footnotesize $\omega_c=10$}
	\put(13.8,9.5)   {\footnotesize $\omega_c=2$}
   	\end{overpic}
  }
 \caption{
 Stochastic gain $E$  vs $ \Rey$ (solid lines) for  cut-off frequencies $\omega_c=2$ (thick line) and $\omega_c=3,$ 5, 7, 10 (thin lines).
 The maximum harmonic optimal gain is also shown for reference (dashed line).
}
   \label{fig:wc}
\end{figure}

Figure~\ref{fig:wc} shows the effect of the cut-off frequency  in the integral evaluation of the stochastic response (see section~\ref{sec:num_valid}).
Increasing $\omega_c$ yields a slight increase in $E$ because
the optimal harmonic gain $G_{in,1}(\omega)$ 
does not decrease to zero at large frequencies $\omega \geq 2$ but instead saturates to a finite value.
However for large Reynolds numbers this finite value is negligible compared to the peak value $\max_{\omega} G_{in,1}$ at $\omega_0=0.5$, and  $E$ is unaffected by the exact value of the cut-off frequency provided $\omega_c$ is sufficiently larger than $\omega_0$.

\section{Influence of $k$ and $\Rey$}
\label{sec:appC}

\begin{figure}
  \centerline{
  \psfrag{k*}[t][]{$k^*$}
  \psfrag{gamma=0.5 msh2 Re=100...700}[t][]{ }
  \psfrag{sum of Ik, k=1...k*}[r][][1][-90]{$\displaystyle\frac{1}{E} \sum_{k=1}^{k^*} I_k \,\,$}	
   	\begin{overpic}[width=6.5 cm,tics=10]{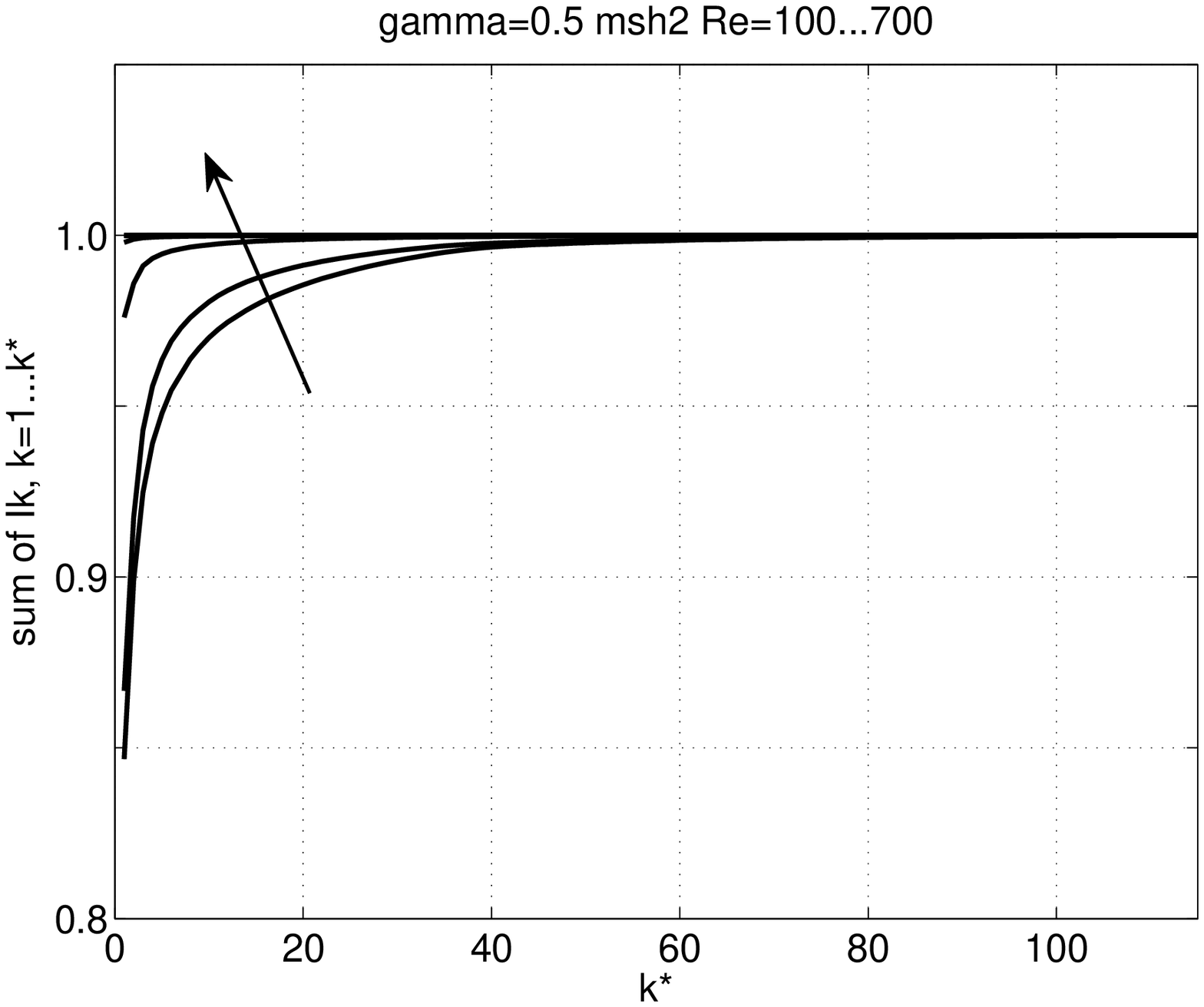}
   	\put(16, 70)  {\footnotesize $\Rey$}
   	\end{overpic}
  }
 \caption{Convergence  of the partial sum $\sum_{k} I_k$ (\ref{eq:E_G}) towards the full stochastic gain $E$. Reynolds number $\Rey$=100, 200, $\ldots$, 700.
}
   \label{fig:k}
\end{figure}

As mentioned in sections~\ref{sec:sensit}-\ref{sec:discussion}, the stochastic gain $E$ and its sensitivities 
$\bnabla_{\boldsymbol{*}} E$ 
are  mostly influenced by optimal 
harmonic quantities  $G_1^2$ and 
$\bnabla_{\boldsymbol{*}} G_1^2$,  especially at larger Reynolds numbers.
Figure~\ref{fig:k} quantifies this phenomenon.
At $\Rey=300$, the contribution from the optimal harmonic gain alone reaches more than 97\% of $E$, while that of the first sub-optimal gain $k=2$ amounts to a mere 1\%.
Even at a Reynolds number as low as $\Rey=100$,  
$G_{in,1}$ contributes for 85\%,
$G_{in,2}$ for 5\%, and 25 sub-optimals are enough to reach 99\% of $E$.

\vspace{1cm}
\bibliographystyle{jfm}
\bibliography{biblio_stochastic}

\end{document}